# Baryons in the Nambu Jona-Lasinio models

Eric Blanquier
*12 Boulevard des capucines, F-11 800 Trèbes, France*
*E-mail: ericblanquier@hotmail.com*

**Abstract**

The octet and decuplet $SU(3)_f$ baryons are studied with the Polyakov, Nambu and Jona-Lasinio model, in a diquark-quark approach, via the resolution of the Bethe-Salpeter equation. This work includes estimations of their masses according to the temperature and the baryonic density, with or without the isospin symmetry. Several improvements of this modeling are proposed and tested: the treatment of the baryons unstable regime, the inclusion of the diquark momentum dependence, the possibility to model baryons without the static approximation, the description of the octet baryons via scalar and axial flavor components, the use of a Nambu and Jona-Lasinio diquark propagator instead of a structureless one. The effects of the color superconductivity on the nucleon mass are also investigated. Certain of the mentioned improvements can be applied simultaneously. The advantages and limitations of each of them are underlined.

## I. INTRODUCTION

The baryons play a central role in the particle, hadronic and nuclear physics. Therefore, it appears important to understand their behavior according to external parameters, like the temperature, the chemical potential and the density. This remark is notably relevant in the framework of the compact stars [1] or high-energy colliders. In the RHIC [2] and the LHC [3], high temperatures are involved. One objective is, e.g., to form a quark gluon plasma from heavy-ion collisions. The NICA [4-6] and FAIR [7] are designed to explore zones of the Quantum Chromodynamics (QCD) phase diagram located at higher densities [8]. Consequently, to model the baryons in such cases, quark degrees of freedom must be taken into account. The Lattice Quantum Chromodynamics (LQCD) [9, 10] appears as an interesting source of data in this framework. However, the fermion sign problem [11] constitutes nowadays a limitation of LQCD to describe finite densities.

The Nambu and Jona Lasinio (NJL) model [12, 13] is an effective quark model based on the QCD [14-25]. It constitutes a relevant alternative to describe quark physics, at finite temperatures via the Matsubara formalism [26], and at finite chemical potentials or densities. It is not affected by the fermion sign problem. In addition, it shares symmetries with QCD. It allows modeling, e.g., the restoration of the chiral symmetry. In the NJL description, the quarks are studied at low energies. Their interactions are supposed to be local, and described via effective terms [27]. In other words, massive gluons are considered and their degrees of freedom are *frozen* [22]. A direct consequence is the quark confinement is absent. Therefore, free quarks or diquarks can be observed in this description. In order to correct this unphysical feature, the NJL model is completed by the inclusion of a Polyakov loop [28]. The goal is to mimic a mechanism of confinement. The obtained PNJL model has become very popular to study quark physics [29-36]. Several evolutions of this model have been proposed like the quark back-reaction to the gluonic sector [37] (labeled as μPNJL in [38, 39]) or the EPNJL [40, 41] that introduces an entanglement between the chiral condensate and the Polyakov loop. The objective is to improve the reliability of the obtained data, in comparison to some LQCD results at vanishing densities.

In addition, the NJL approach and its various versions are able to build composite objects made with the quarks and antiquarks. A modeling of the light mesons constitutes one of the successes of the NJL description [23, 42]. Rapidly, it had been imagined to describe baryons with this model. Starting from the first terms of the Faddeev equations [43, 44], a baryon can be seen as a diquark-quark bound state [22, 45-49]. This explains why hypothetical particles like diquarks are considered in this framework [39, 50]. The diquark-quark picture has permitted to model baryons with different models, notably via the resolution of the Bethe-Salpeter equation [51, 52]. Since the mesons and diquarks can be modeled via this equation in the NJL model [22], this has suggested to adapt this method to the NJL baryons, e.g. [53-59]. In fact, the Bethe-Salpeter equation in a diquark-quark picture is not the only ways to reach this objective. It is also possible to solve a Dyson-Swinger equation as in [60]. Moreover, Ref. [61] has also imagined NJL baryons as a clusters of three quarks.

However, it is often considered that the treatment of the PNJL baryons stays incomplete, notably in the quark-diquark picture. Indeed, on the one hand, it is known for its limitations. The difficulties to calculate the baryon mass at high temperatures or densities, in its *unstable regime*, is one of them. Indeed, it is observed in the devoted papers that the curve of the baryon mass stops when its mass becomes equal to the ones of its constituents [53, 55, 56, 62, 63]. It appears crucial to propose a solution to overcome this limitation.

In addition, the mesons and baryons are sometimes treated separately in PNJL works, in order to gain in precision. In practice, the constants used in the model are tuned on the one hand to fit the masses of the mesons, and modified on the other hand to obtain the ones of the baryons. Nonetheless, modeling the mesonic and baryonic sectors at the same time presents real advantages. This may concern, e.g., cross-section calculations [64] or a dynamical study [65] involving these particles together. This second Ref. has shown that the calculation of their masses according to the temperature and the densities allows modeling an action at a distance between the considered particles. This constitutes an additional motivation to estimate the masses of the baryons according to the mentioned parameters.

Moreover, the isospin symmetry is usually applied in such works. It considers that the masses and the chemical potentials of the light quarks $u$ and $d$ are equal, i.e. $m_u = m_d \equiv m_q$ and $\mu_u = \mu_d \equiv \mu_q$. Even if this allows evident simplifications in the descriptions, it could be relevant to perform some calculations without it, to evaluate their precision and reliability. In other words, can the PNJL baryon modeling correctly reproduce the experimental baryonic spectrum?

On the other hand, approximations are frequently used in the PNJL baryon description. It should be relevant to identify these ones, to investigate their influence on the results, and then to see the possibilities to avoid them. The static approximation is certainly the most known of them [48]. It considers "infinitely" massive exchanged quarks in the baryon modeling. This allows simplifying the related equations in a non-negligible way. In fact, this is sometimes mandatory to allow their resolution. However, this certainly has great consequences on the results, even if Ref. [48] has predicted a rather modest effect on the masses. Nevertheless, Refs. [55, 56] point out one of its negative aspects: an unphysical mass inversion for the proton and neutron when $m_u \neq m_d$, with a proton heavier than a neutron. Correcting this anomaly appears to be a priority. However, is it necessarily required to abandon the static approximation to fix this problem?

Another simplification considers diquarks at rest to build baryons, e.g. [53]. This constitutes also a drastic approximation in the framework of an NJL approach. Indeed, a diquark momentum-dependence should be taken into account in the modeling. Therefore, it should be investigated how to overcome this approximation. In other words, what are the elements of the devoted equations that should be updated? In addition, a crucial point is to identify its effects in the results. Is this approximation severely alter the reliability of the baryon modeling?

In the baryon description found in the PNJL literature, the $SU(3)_f$ octet baryons are frequently described via a scalar flavor component and decuplet ones with an axial flavor one [63]. The axial flavor component of the octet baryons is thus neglected. Consequently, it should be relevant to investigate the contribution of this component on the behavior of these baryons [66, 67]. Is this contribution low enough to justify neglecting it? Can the inclusion of the axial flavor component solve the proton-neutron mass inversion mentioned above?

Moreover, in the PNJL quark-diquark picture associated with a Bethe-Salpeter equation, the use of the quark and diquark propagators are required to build the baryon. In (P)NJL works, e.g. [55, 63], the propagator of the diquark is the one of a structureless particle (free particle propagator), instead of the NJL diquark propagator available in this model (notably used to evaluate the diquark mass). Indeed, this choice greatly facilitates the establishment of the equations to be solved. However, this approximation certainly has non-trivial consequences. It could be interesting to identify them and to evaluate their importance. Since this constitutes a major modification, will the results stay comparable to the ones found with the structureless propagator?

In addition, at low temperatures and high densities, the quarks are expected to form couples, comparable to the electronic Cooper pairs. This is the color superconductivity (CSC) phenomenon [68-70]. The (P)NJL model is able to work in these conditions and can describe the CSC [25]. This one is characterized by several phases [71, 72], which depend on the quark-quark pairings. For example, in the 2-flavor color superconducting phase (2SC) [73], *ud* pairs are created. In the color-flavor-locked phase (CFL) [74], there is formation of *ud*, *us*, and *ds* pairs. As studied in [75], these phases can be in competition with the meson condensation [76, 77], associated with the creation of meson condensates. Moreover, the color superconductivity has an influence on various observables, and notably the masses. This statement concerns the quarks [38], and also the mesons and diquarks [39, 78]. This should have consequences on the baryons masses. They should be estimated in the same way, e.g., in the conditions explored by the FAIR and NICA colliders [79]. Is the effect of the color-superconductivity on the baryons low enough to neglect this phenomenon at moderate densities, as done habitually?

Moreover, the determination of the meson coupling constant is well known is the NJL model [42]. By extension, this is also the case for the diquarks, e.g., [55]. In contrast, concerning the baryons, this is not trivial because of their fermionic nature. In addition, with the various improvements suggested above, is it possible to propose a unique method to estimate the baryon coupling constants? Indeed, such data appear essential in the framework of cross section calculations involving baryons [64].

In order to answer these questions, this paper is organized as follow. In a first step, the Sec. II recalls the main ideas of the (P)NJL models to describe the quarks, with the inclusion of the Polyakov loop. Then, the Sec. III focuses on a description of the (P)NJL composite particles: the mesons, diquarks and baryons. Even if the mesons are not studied numerically in this paper, the similarities between the mesons and diquarks modeling are underlined. In addition, this section recalls the method used in previous works to model PNJL baryons as a quark-diquark bound state, via the Bethe-Salpeter equation. In contrast, the Sec. IV analyses the various approximations performed in this modeling, and proposes then improvements to avoid them. Firstly, this concerns the inclusion of the complex numbers to describe the instability of the baryons. Then, it is proposed to include the diquark momentum dependence. The next improvement is the abandon of the static approximation. In parallel, the inclusion of the axial flavor component into the octet baryon modeling is detailed. Afterward, the replacement of the structureless diquark propagator by an NJL one is explained. Then, the required updates to describe color-superconductivity are presented, in order to propose a baryon modeling in this regime. Moreover, a method to estimate the baryon coupling constant is detailed. Except for the inclusion of the Polyakov loop, these propositions are novelties in the framework of such PNJL description. For each of them, the results are presented and compared in the Sec. V. The conclusion and outlook are in Sec. VI. Five appendices complete these sections. The Appendix A recalls the flavor wavefunctions and the Feynman diagrams of the octet and decuplet baryons in the diquark-quark picture. The Appendix B details calculations of the Sec. IV. The Appendix C gives extra explanations about the inclusion of the axial flavor component of the octet baryons. The Appendix D indicates the mutual compatibility of the presented evolutions. The Appendix E shows a comparison of the found baryons masses, at zero temperature and zero density.

## II. DESCRIPTION OF THE (P)NJL QUARK MODEL.

### A. The PNJL Lagrangian density.

In the framework of the PNJL model, the associated Lagrangian density is [22, 25]

$$\begin{aligned}
\mathcal{L} = & \sum_{f=u,d,s} \bar{\psi}_f \left( i \slashed{D} - m_{0f} + \gamma_0 \mu_{0f} \right) \psi_f - U(T, \Phi, \bar{\Phi}) \\
& + G \sum_{a=0}^{8} \left[ \left( \bar{\psi} \tau_a \psi \right)^2 + \left( \bar{\psi} i \gamma_5 \tau_a \psi \right)^2 \right] \\
& - G_V \sum_{a=0}^{8} \left[ \left( \bar{\psi} \gamma_\mu \tau_a \psi \right)^2 + \left( \bar{\psi} \gamma_\mu i \gamma_5 \tau_a \psi \right)^2 \right] \\
& - K \left\{ \det_f \left[ \bar{\psi}(1+\gamma_5)\psi \right] + \det_f \left[ \bar{\psi}(1-\gamma_5)\psi \right] \right\} \\
& + G_{DIQ} \sum_{a=2,5,7} \sum_{a'=2,5,7} \left( \bar{\psi} i\gamma_5 \tau_a \lambda_{a'} \psi^C \right)\left( \bar{\psi}^C i\gamma_5 \tau_a \lambda_{a'} \psi \right) \\
& + G_{DIQ} \sum_{a=2,5,7} \sum_{a'=2,5,7} \left( \bar{\psi} \tau_a \lambda_{a'} \psi^C \right)\left( \bar{\psi}^C \tau_a \lambda_{a'} \psi \right) \\
& + \frac{G_{DIQ}}{4} \sum_{a=2,5,7} \sum_{a'=2,5,7} \left( \bar{\psi} i\gamma_5 \gamma_\mu \tau_a \lambda_{a'} \psi^C \right)\left( \bar{\psi}^C i\gamma_5 \gamma_\mu \tau_a \lambda_{a'} \psi \right) \\
& + \frac{G_{DIQ}}{4} \sum_{a=0,1,3,4,6,8} \sum_{a'=0,1,3,4,6,8} \left( \bar{\psi} \gamma_\mu \tau_a \lambda_{a'} \psi^C \right)\left( \bar{\psi}^C \gamma_\mu \tau_a \lambda_{a'} \psi \right)
\end{aligned} \qquad (1)$$

The quark fields are noted as $\psi$, whose naked masses and bare chemical potentials are, respectively, $m_{0f}$ and $\mu_{0f}$. Also, $\slashed{D} = \gamma^\mu D_\mu$, where

$$D_\mu = \partial_\mu - i A_\mu \ . \qquad (2)$$

The $A_\mu$ and the Polyakov loop potential $U(T, \Phi, \bar{\Phi})$ [34, 36] are due to the inclusion of the Polyakov loop, as detailed in the Subsec. II.D. Moreover, $\tau_j/2$ and $\lambda_k/2$ are the eight generators of, respectively, $SU(3)_{flavor}$ with $f = u, d, s$, and $SU(3)_{color}$. So, one considers the $N_c = 3$ colors, i.e. r,g,b. In addition, $\tau_0 = \lambda_0 = \sqrt{2/3}\, I_{d3}$, where $I_{d3}$ is the $3 \times 3$ identity matrix. The superscript $C$ corresponds to the charge conjugation operator. The constant $G$ is placed in front of the scalar/pseudoscalar quark-antiquark interaction terms; $G_V$ is for vector/axial quark-antiquark ones. $K$ is the 't Hooft term [23]. These three constants notably intervene in the meson modeling. In the same way, $G_{DIQ}$ concerns, respectively, the scalar, pseudoscalar, vectorial and axial quark-quark interaction terms. $G_{DIQ}$ is relevant for the diquark description. Even if these constants are linked by relations, they are frequently treated as more or less independent parameters in the (P)NJL literature. Since the model is not renormalizable, a cutoff is required in the numerical calculations. It corresponds to the upper bound of the integrals upon the momentum. The quark naked masses, the mentioned constants and the cutoff constitute a parameter set. In this paper, the two sets used in Refs. [38, 39, 55, 56, 64, 65] are considered again, TABLE I. The first set (ISO) respects the isospin symmetry: it considers that the *u* and *d* quarks have the same naked mass $m_{0q}$. In contrast, the second set (NISO) does not use this symmetry.

**TABLE I.** (P)NJL parameter sets. The quark naked masses $m_{0f}$ and the cutoff $\Lambda$ are expressed in MeV. The constants $G$, $G_V$ and $G_{DIQ}$ are in MeV$^{-2}$ and $K$ is in MeV$^{-5}$.

| Parameter set | $m_{0u}$ | $m_{0d}$ | $m_{0s}$ | $G\Lambda^2$ | $G_V$ | $K\Lambda^5$ | $G_{DIQ}$ | cutoff $\Lambda$ |
|---|---|---|---|---|---|---|---|---|
| ISO | 4.75 | 4.75 | 147.0 | 1.922 | 0.310 $G$ | 10.00 | 0.705 $G$ | 708.0 |
| NISO | 4.00 | 6.00 | 120.0 | 1.922 | 0.295 $G$ | 10.00 | 0.705 $G$ | 708.0 |

### B. The gap equations.

The Lagrangian density Eq. (1) is then rewritten in the framework of the mean field approximation. If the meson condensation and the color superconductivity are not taken into account, this leads to [25, 70, 80-82]:

$$\mathcal{L}_{MF} = \sum_{f=u,d,s} \bar{\psi}_f \left[ i\slashed{\partial} - I_{d4}\left(m_{0f} - 4G\sigma_f + 2K\sigma_j\sigma_k\right) + \gamma_0\left(\mu_{0f} - 4G_V\rho_f\right) \right] \psi_f \\ - U(T, \Phi, \bar{\Phi}) - 2G \sum_{f=u,d,s} \sigma_f^2 + 4K\sigma_u\sigma_d\sigma_s + 2G_V \sum_{f=u,d,s} \rho_f^2 \quad , \quad (3)$$

where $I_{d4}$ is the $4\times 4$ identity matrix, $\sigma_f$ is the flavor *f* chiral phase condensate,

$$\sigma_f = \langle\langle \bar{\psi}_f \psi_f \rangle\rangle = \frac{N_c}{\beta} \sum_n \int \frac{d^3p}{(2\pi)^3} \text{Tr}\left[ S_f^+(i\omega_n, \vec{p}) \right] , \quad (4)$$

and $\rho_f$ is the density of the flavor *f* quarks,

$$\rho_f = \langle\langle \psi_f^+ \psi_f \rangle\rangle = \frac{N_c}{\beta} \sum_n \int \frac{d^3p}{(2\pi)^3} \text{Tr}\left[ \gamma_0 S_f^+(i\omega_n, \vec{p}) \right] . \quad (5)$$

In these relations, $S_f^+$ is the propagator of a flavor *f* quark, $\vec{p}$ is a momentum and Tr the trace upon the Dirac matrices $\gamma$. Also, $\beta$ is the inverse of the temperature *T*. It appears when the Matsubara formalism is employed via the substitution:

$$\int \frac{d^4p}{(2\pi)^4} \rightarrow \frac{i}{\beta} \sum_{n=-\infty}^{+\infty} \int \frac{d^3p}{(2\pi)^3} , \quad (6)$$

which leads to perform the summation upon the fermionic Matsubara frequencies $\omega_n = (2n+1)\pi T$, with $n \in \mathbb{Z}$. Moreover, in the first line of Eq. (3), the $I_{d4}$ term leads to write the gap equations for the quark masses:

$$m_f = m_{0f} - 4G\sigma_f + 2K\sigma_j\sigma_k \Big|_{\substack{f=u,d,s \\ f\neq j \text{ and } f\neq k}} , \quad (7)$$

where $m_f$ is the dressed mass of the flavor *f* quarks. Another gap equation is found with the $\gamma_0$ term of Eq. (3), for the quark chemical potentials:

$$\mu_f = \mu_{0f} - 4G_V \rho_f , \quad (8)$$

where $\mu_f$ refers to an *effective* chemical potential used in the NJL model. The modifications leaded by the inclusion of the Polyakov loop are explained in Subsec. II.E. In practice, $m_f$ is used to build the composite (P)NJL particles. So, in order to work in the $T, \rho_B$ plane, with $\rho_B = 2\rho_q/3$ and

$\rho_q = \rho_u = \rho_d$, the Eqs. (7) and (5) constitutes a set of coupled equations to be solved. In this work, the standard nuclear density is $\rho_0 = 0.16\,\mathrm{fm}^{-3}$.

### C. Expression of the PNJL grand potential.

The PNJL grand potential $\Omega$ is obtained from $\mathcal{L}_{MF}$, and is expressed as [83]:

$$\Omega = U(T,\Phi,\bar{\Phi}) + \Omega_M + 2G \sum_{f=u,d,s} \sigma_f^2 - 4K \sigma_u \sigma_d \sigma_s - 2G_V \sum_{f=u,d,s} \rho_f^2 , \tag{9}$$

with the $\Omega_M$ thermodynamical potential [25]:

$$\Omega_M = -\frac{T}{2} \int \frac{d^3 p}{(2\pi)^3} \sum_n \mathrm{Tr}\left\{ \ln\left[ \beta \tilde{S}^{-1}(i\omega_n, \vec{p}) \right] \right\} . \tag{10}$$

In this relation, $\tilde{S}^{-1}$ is a matrix gathering the inverse propagators of the quarks and antiquarks. In the Nambu-Gorkov basis $\Psi = (\psi, \psi^C)^T$, $\tilde{S}^{-1}$ is written on the diagonal form [70]:

$$\tilde{S}^{-1}(i\omega_n, \vec{p}) = \begin{bmatrix} \left[S^+(i\omega_n, \vec{p})\right]^{-1} & 0 \\ 0 & \left[S^-(i\omega_n, \vec{p})\right]^{-1} \end{bmatrix} , \tag{11}$$

where $S^+$ and $S^-$ concerns, respectively, the quarks and the antiquarks. More precisely,

$$\left[S^{\pm}(\slashed{p})\right]^{-1} = \begin{bmatrix} \slashed{p} \pm \gamma_0 \tilde{\mu}_u - m_u & 0 & 0 \\ 0 & \slashed{p} \pm \gamma_0 \tilde{\mu}_d - m_d & 0 \\ 0 & 0 & \slashed{p} \pm \gamma_0 \tilde{\mu}_s - m_s \end{bmatrix} . \tag{12}$$

The expression of the $3\times 3$ matrices $\tilde{\mu}_f$ is detailed in the Subsec. II.E for the NJL and PNJL models.

### D. The Polyakov loop potential.

In order to include the Polyakov loop into the NJL description, the Polyakov line $L$ and its conjugate $L^\dagger$ are considered:

$$L(\vec{x}) = \mathcal{P} \exp\left( i \int_0^\beta A_4(\vec{x}, \tau) \, d\tau \right) , \tag{13}$$

and so on for $L^\dagger$, where $\mathcal{P}$ is the path ordering operator and $A_4 = iA_0$ is the temporal component of the Euclidian gauge field $A$ [36]. It is frequently defined as $\beta A_4 = \phi_3 \lambda_3 + \phi_8 \lambda_8$ [33], with $\phi_3, \phi_8 \in \mathbb{R}$. Indeed, the Gell-Mann matrices $\lambda_3, \lambda_8$ allows obtaining a diagonal matrix [30],

$$A_4 = \mathrm{diag}\left[ A_{4(11)}, A_{4(22)}, A_{4(33)} \right] . \tag{14}$$

In the mean field approximation, Eq. (13) is simplified. This gives, for $L$ and $L^\dagger$:

$$L = \exp(i\beta A_4), \quad L^\dagger = \exp(-i\beta A_4) . \tag{15}$$

In the PNJL model, they intervene in a rewriting of $\Omega_M$,

$$\Omega_M = -2N_c \int \frac{d^3 p}{(2\pi)^3} \sum_{f=u,d,s} E_f \\ + \frac{1}{N_c} \mathrm{Tr}_c \left\{ T \ln\left[ 1 + L^\dagger \exp\left( -\frac{E_f - \mu_f}{T} \right) \right] + T \ln\left[ 1 + L \exp\left( -\frac{E_f + \mu_f}{T} \right) \right] \right\} , \tag{16}$$

where $\text{Tr}_c$ is a trace over the colors and $E_f = \sqrt{\vec{p}^2 + m_f^2}$. On the other hand, the average of $L$ and $L^\dagger$ are defined as

$$\Phi = \text{Tr}_c(L)/N_c, \quad \bar{\Phi} = \text{Tr}_c(L^\dagger)/N_c . \tag{17}$$

When the color superconductivity is not included in the modeling, $\Phi$ and $\bar{\Phi}$ are found with the resolution of $\partial\Omega/\partial\Phi = 0$ and $\partial\Omega/\partial\bar{\Phi} = 0$, in which they are treated as real and independent variables [32]. Physically, $\Phi$ is an order parameter of the $\mathbb{Z}_3$ symmetry in pure gauge calculations: $\Phi$ is equal to zero in a confined regime, and $\Phi \to 1$ in a deconfined one. Since the $\mathbb{Z}_3$ symmetry breaking is no longer exact in the presence of dynamical quarks, the Polyalov loop only proposes a *simulation of confinement* in such a quark model [36]. In the PNJL calculations, $\Phi$ and $\bar{\Phi}$ intervene in the expression of the Polyakov loop potential $U(T, \Phi, \bar{\Phi})$ mentioned in Eq. (1). The following expression is considered in this paper [32, 33],

$$\frac{U(T, \Phi, \bar{\Phi})}{T^4} = -\frac{a(T)}{2}\Phi\bar{\Phi} + b(T)\ln\left[1 - 6\Phi\bar{\Phi} + 4(\Phi^3 + \bar{\Phi}^3) - 3(\Phi\bar{\Phi})^2\right], \tag{18}$$

where

$$a(T) = a_0 + a_1(T_0/T) + a_2(T_0/T)^2 \text{ and } b(T) = b_3(T_0/T)^3, \tag{19}$$

and $a_0, a_1, a_2, b_3$ and $T_0$ are defined TABLE II.

**TABLE II.** Parameters of the PNJL model.

| $a_0$ | $a_1$ | $a_2$ | $b_3$ | $T_0$ |
|---|---|---|---|---|
| 3.51 | −2.47 | 15.2 | −1.75 | 270 MeV |

The Eq. (18) allows keeping $\Phi$ lower than 1. In pure gauge calculations, $T_0$ is the critical temperature of the deconfinement transition [30]. Even if a $\mu$-dependence via $T_0$ is proposed in the recent PNJL version [37], named the "$\mu$PNJL model" in Refs. [38, 39], a constant $T_0$ is used in this paper. Indeed, the found masses with $\mu$PNJL and NJL are rather close.

### E. Influence of the Polyakov loop on the Fermi-Dirac statistics.

In the NJL model described in this work, the chemical potential of a flavor $f$ quark does not depend on its color $c$. In Eq. (12), $\tilde{\mu}_f = \mu_f I_{d3}$, i.e.

$$\mu_f^r = \mu_f^g = \mu_f^b \equiv \mu_f . \tag{20}$$

In contrast, when the Polyakov loop is included, the $A_4 = iA_0$ carried out by $\slashed{D}$, Eq. (2), and the $\gamma_0$ term in the first line of Eq. (3) are merged. This leads to define $\tilde{\mu}_f = \mu_f I_{d3} - iA_4$, i.e.

$$\mu_f^r = \mu_f - iA_{4(11)}, \quad \mu_f^g = \mu_f - iA_{4(22)}, \quad \mu_f^b = \mu_f - iA_{4(33)} . \tag{21}$$

The chemical potentials defined Eq. (21) intervenes in the Fermi-Dirac statistics, which are derived from Eq. (16) [38]. A color averaging is performed, in order to define

$$f_\Phi^\pm(E_f \mp \mu_f) = \frac{1}{N_c}\sum_{c=r,g,b}\frac{1}{1+\exp\{\beta[E_f \mp \mu_f^c]\}} . \tag{22}$$

In the PNJL model, Eq. (22) is frequently written on the form [34, 36], for the quarks,

$$f_\Phi^+(E_f - \mu_f) = \frac{\{\bar{\Phi} + 2\Phi\exp[-\beta(E_f - \mu_f)]\}\exp[-\beta(E_f - \mu_f)] + \exp[-3\beta(E_f - \mu_f)]}{1 + 3\{\bar{\Phi} + \Phi\exp[-\beta(E_f - \mu_f)]\}\exp[-\beta(E_f - \mu_f)] + \exp[-3\beta(E_f - \mu_f)]}, \tag{23}$$

and, for the antiquarks,

$$f_\Phi^-(E_f + \mu_f) = f_\Phi^+(E_f - \mu_f)\Big|_{\substack{\Phi \leftrightarrow \bar{\Phi} \\ \mu_f \leftrightarrow -\mu_f}} . \quad (24)$$

Consequently, $f_\Phi^\pm$ replace the "standard" Fermi-Dirac statistics of the quarks and antiquarks in the PNJL approach.

### III. MODELING COMPOSITE PARTICLES IN THE NJL MODELS

#### A. General description.

The modeling of a composite particle formed by a particle/antiparticle pair is schematized in the Figure 1 [23]. Its mass is *m* and is four-momentum *K*. In this description, the interactions between the particle/antiparticle are described via simple vertices, i.e. this concerns (effective) instantaneous interactions.

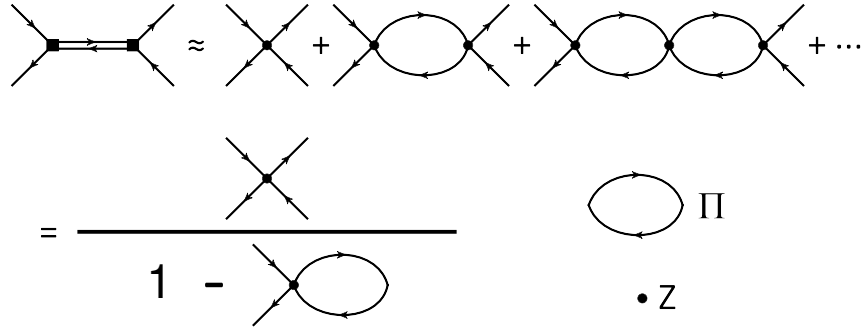

**Figure 1. Illustration of the employed method, in the random phase approximation.**

With the Bethe-Salpeter equation [22, 84, 85], the composite particle is described by the transition matrix *T*, which is expressed as:

$$T = Z + Z\Pi T , \quad (25)$$

where *Z* is the interaction kernel. It gathers the terms at the level of the vertex. Moreover, this relation introduces the irreducible polarization function $\Pi$ [23, 42, 86], i.e. the *loop function* that describes the particle/antiparticle pair. In this framework, *Z* and $\Pi$ are scalars or matrices. The Eq. (25) leads to the geometric progression:

$$T = Z + Z\Pi Z + Z\Pi Z\Pi Z + \ldots = \frac{Z}{1 - Z\Pi} , \quad (26)$$

assuming $\lim_{n \to \infty}(Z\Pi)^n = 0$. This expression can be written on the equivalent forms $Z(1 - \Pi Z)^{-1}$ or $(1 - Z\Pi)^{-1} Z$. The *Z* in the numerator of Eq. (26) is the first (resp. last) vertex in the chain $Z\Pi Z\Pi Z\ldots Z$. If the composite particle is a spin-0 boson, $(1 - Z\Pi)^{-1} Z$ and the Klein-Gordon propagator $1/(K^2 - m^2)$ are expected to have the same behavior at the vicinity of the pole $K^2 = m^2$, i.e. to diverge. This leads to the relation:

$$\frac{Z}{1 - Z\Pi(k_0, \vec{k})}\bigg|_{K^2 = m^2} \propto \frac{1}{K^2 - m^2}\bigg|_{K^2 = m^2} , \quad (27)$$

where $k_0$ and $\vec{k}$ are, respectively, the energy and the momentum of the particle. In the case of a spin-1 boson, Eq. (27) stays valid. Indeed, whatever the gauge choice for the propagator, the pole is also $K^2 = m^2$. Throughout this paper, the left hand side of Eq. (27) is named the *NJL propagator*

$S^{NJL}$ and the right hand side the *quantum field theory propagator* $S$, or $S^{QFT}$ if coupling constants are included. The particle type, i.e. Meson, Diquark or Baryon (in the fermionic case), is specified in subscript via its initial. In order to obtain the divergence of $(1-Z\Pi)^{-1}Z$ at the pole, the matrix $1-Z\Pi$ should not be invertible when $K^2 = m^2$, i.e. one should have:

$$\det\left[1-Z\Pi\left(\sqrt{m^2+\vec{k}^2},\vec{k}\right)\right]=0 , \qquad (28)$$

The Eq. (28) is the general relation to be solved to obtain the mass $m$ of a composite particle. The descriptions performed in this Sec. III concerns the NJL model, but the extension to PNJL is explained in the Subsubsec. IV.A.1.

### B. Mesons and diquarks.

#### 1. Mesons.

The spin-0 or spin-1 NJL meson modeling corresponds to the description performed above. The loop $\Pi_M$ is composed by a quark $q$ going towards to the future, and $q'$ going towards the past, Figure 2. According to the Feynman point of view, $q'$ is an antiquark. For an uncoupled meson made by the $q\bar{q}'$ quark-antiquark pair, Eq. (27) becomes $S_M^{NJL}\big|_{K^2=m_M^2} = S_M^{QFT}\big|_{K^2=m_M^2}$, i.e.

$$\left.\frac{Z_M}{1-f Z_M \Pi_M(k_0,\vec{k})}\right|_{K^2=m_M^2} = \left.\frac{-g_{q\bar{q}'}^2}{K^2-m_M^2}\right|_{K^2=m_M^2} , \qquad (29)$$

where $g_{q\bar{q}'}$ is the coupling constant between the meson and the $q\bar{q}'$ pair. For a spin-1 meson, a possibility is to consider the 't Hooft-Feynman gauge [87, 88] in the QFT propagator. Also, $Z \equiv f Z_M$ where $f = 2$ is a flavor factor. $Z_M$ depends on the treated meson. Thanks to the NJL assumptions, i.e. static gluons with an effective mass, this is consistent to translate the quark/antiquark interactions $Z$ via a momentum independent term associated with the vertex mentioned above [27, 53, 56]. Its various expressions are found in the NJL literature [23, 42]. The meson irreducible polarization function $\Pi_M$ is written as:

$$-i\Pi_M\left(i\nu_m,\vec{k}\right) = -N_C \frac{i}{\beta}\sum_n \int \frac{d^3p}{(2\pi)^3} \text{Tr}\left[S_q(i\omega_n,\vec{p})\Gamma S_{q'}(i\omega_n - i\nu_m,\vec{p}-\vec{k})\Gamma\right] , \qquad (30)$$

where $\Gamma = i\gamma_5$ for pseudoscalar mesons, $\Gamma = I_{d4}$ for scalar mesons, $\Gamma = \gamma_\mu$ for vectorial and $\Gamma = \gamma_\mu i\gamma_5$ for axial ones. $S_q(\slashed{p}) = (\slashed{p}+\gamma_0\mu_q - m_q I_{d4})^{-1}$ is the propagator of the $q$ quark, and so on for $S_{q'}$. With Eq. (29), the mass $m_M$ of an uncoupled meson of momentum $\vec{k}$ is found via:

$$1-2Z_M \Pi_M\left(\sqrt{m_M^2+\vec{k}^2},\vec{k}\right)=0 . \qquad (31)$$

The method to treat coupled ones ($\eta,\eta'$, $\pi_0$, etc.) is presented in Refs [23, 42, 55, 56].

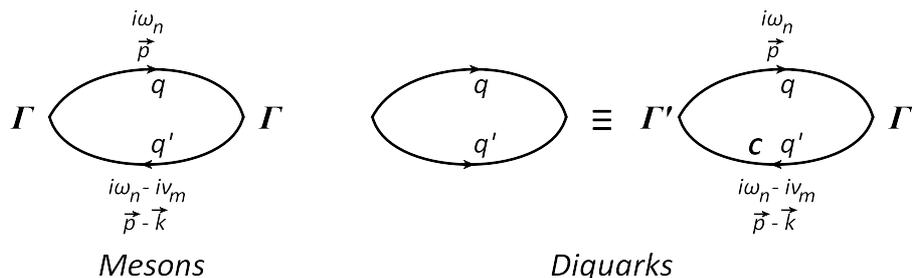

Mesons          Diquarks

**Figure 2. Irreducible polarization functions of the mesons and diquarks.**

*2. Diquarks.*

The description of the NJL diquarks consists in applying the charge conjugation on $q'$, Figure 2. Therefore, $q'$ is a charge conjugate antiquark, and *mimics* the behavior of a quark. This trick allows keeping the loop structure described above to model a diquark as a quark-quark pair. The irreducible polarization function $\Pi_D$ of a diquark $[qq']$ is:

$$-i\Pi_D(iv_m, \vec{k}) = -\frac{i}{\beta}\sum_n \int \frac{d^3p}{(2\pi)^3} \text{Tr}\left[S_q(i\omega_n, \vec{p})\Gamma' S_{q'}^C(i\omega_n - iv_m, \vec{p} - \vec{k})\Gamma'\right], \quad (32)$$

where $S_{q'}^C(\slashed{p}) = (\slashed{p} - \gamma_0 \mu_{q'} - m_{q'} I_{d4})^{-1}$ is the propagator of $q'$. The Eq. (32) is invariant by the exchange $q \leftrightarrow q'$ [55, 56]. The scalar diquarks corresponds to $\Gamma' = i\gamma_5$, $\Gamma' = I_{d4}$ for pseudo-scalar, $\Gamma' = \gamma_\mu$ for axial and $\Gamma' = \gamma_\mu i\gamma_5$ for vectorial ones. The names are reversed in comparison with the mesons because of the charge conjugation $C = i\gamma_0\gamma_2$ [22, 51]. Moreover, $Z = 2G_{DIQ}$ for scalar and pseudoscalar diquarks, $Z = G_{DIQ}/2$ for axial and vectorial ones [22]. The $S_D^{NJL}$ and $S_D^{QFT}$ scalar diquark propagators satisfy the relation:

$$\left.\frac{4G_{DIQ}}{1 - 2G_{DIQ}\Pi_D(k_0, \vec{k})}\right|_{K^2 = m_D^2} = \left.\frac{-g_{qq'}^2}{K^2 - m_D^2}\right|_{K^2 = m_D^2}. \quad (33)$$

Consequently, the mass $m_D$ of a scalar diquark of momentum $\vec{k}$ is obtained by:

$$1 - 2G_{DIQ}\Pi_D\left(\sqrt{m_D^2 + \vec{k}^2}, \vec{k}\right) = 0. \quad (34)$$

As done in Refs. [14, 42] for the mesons, $g_{qq'}$ is found via Eq. (33):

$$g_{qq'} = \sqrt{\frac{4E_D}{\left.\frac{\partial}{\partial k_0}\Pi_D(k_0, \vec{k})\right|_{k_0^2 = E_D^2}}}, \quad (35)$$

where $E_D = \sqrt{m_D^2 + \vec{k}^2}$ is the diquark energy. The reasoning is similar for the axial diquarks if the 't Hooft-Feynman gauge is considered, or if the $-k^\mu k^\nu/m_D^2$ term in the QFT (Proca) propagator is neglected with the unitary gauge, as in [63].

### C. Baryons with the static approximation.

*1. The static approximation.*

The modeling of the baryons is feasible via a simplification of the Faddeev equations [43, 44], which leads to consider baryons as quark-diquark bound states. In practice, scalar and axial diquarks are good candidates to reach this objective, because of their stability [55, 56]. The description performed in Subsec. III.A can be adapted to composite fermionic particles, which logically include baryons. In order to come back to a loop structure, the static approximation is applied in the NJL approaches [48]. It considers that the mass $m_{qe}$ of the exchanged quark $q_e$ is strong enough to neglect its four-momentum $\slashed{p}$ (and its chemical potential $\mu_{qe}$), Figure 3.

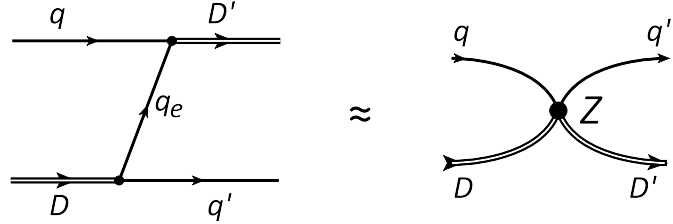

**Figure 3.** Schematization of the static approximation.

This gives a simplified expression of the exchanged quark propagator $S_{qe}$:

$$S_{qe} = \frac{1}{\slashed{p} + \gamma_0 \mu_{qe} - m_{qe} I_{d4}} \approx \frac{1}{-m_{qe}} I_{d4}. \tag{36}$$

In Figure 3, the Z materializes the effective vertex obtained via this approximation. This *constant* term gathers the simplified propagator Eq. (36), coupling constants *g*, flavor factors *f*, and the matrices translating the diquark-quark interactions. With a scalar diquark interaction, this leads to:

$$Z = g_{q'qe} \, f \, i\gamma_5 \frac{-1}{m_{qe}} g_{qqe} \, f' \, i\gamma_5. \tag{37}$$

In the case where $D' = D$ and $q' = q$, one has $g_{q'qe} = g_{qqe}$. For an axial diquark interaction, $i\gamma_5$ are replaced by $\gamma_\mu$.

## *2. The loop functions.*

This description presents some similarities with the diquark modeling: the composite particle is described by a loop of two particles, in which one of them is charge conjugate. With the baryons, this can be the quark or the diquark. So, as proposed in the Refs. [55, 56] and then employed in Ref. [89], two terms are taken into account in the baryon loop function $\Pi_B = 1/2\left(\Pi_B^{(1)} + \Pi_B^{(2)}\right)$, Figure 4 and Figure 5. The numbering of these two terms is reversed in comparison with these Refs.

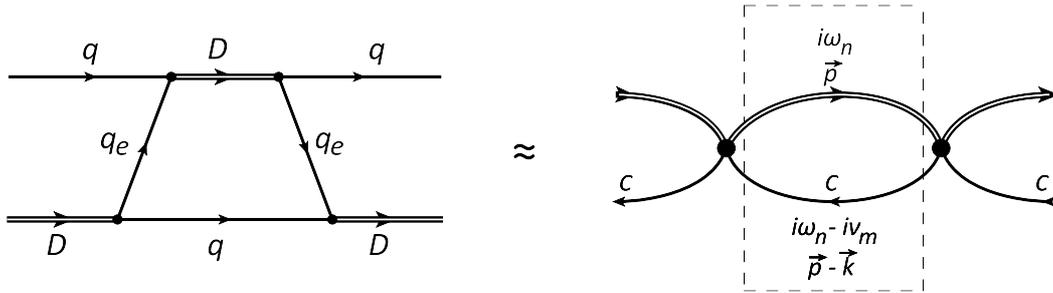

**Figure 4.** Static approximation: first term. The loop function corresponds to the dotted rectangle.

The fist term is expressed as:

$$-i\Pi_B^{(1)}\left(iv_m, \vec{k}\right) = -\frac{i}{\beta} \sum_n \int \frac{d^3 p}{(2\pi)^3} S_D\left(i\omega_n, \vec{p}\right) S_q^C\left(i\omega_n - iv_m, \vec{p} - \vec{k}\right), \tag{38}$$

where $S_D$ is the diquark propagator:

$$S_D\left(i\omega_n, \vec{p}\right) = \frac{1}{\left(i\omega_n + \mu_D\right)^2 - E_D^2}, \tag{39}$$

with $E_D^2 = \vec{p}^2 + m_D^2$, and $S_q^C$ is the charge conjugate quark propagator:

$$S_q^C(i\omega_n - i\nu_m, \vec{p} - \vec{k}) = \frac{\gamma_0(i\omega_n - i\nu_m - \mu_q) - \vec{\gamma}\cdot(\vec{p}-\vec{k}) + m_q I_{d4}}{(i\omega_n - i\nu_m - \mu_q)^2 - E_q^2} , \qquad (40)$$

where $E_q^2 = (\vec{p} - \vec{k})^2 + m_q^2$.

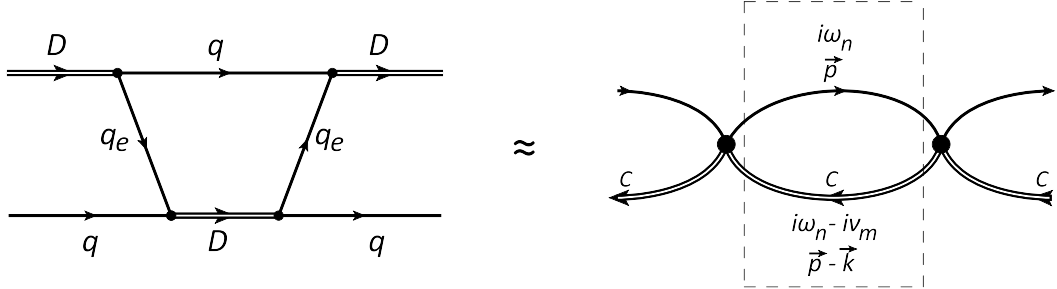

**Figure 5. Static approximation: second term.**

The second term is

$$-i\Pi_B^{(2)}(i\nu_m, \vec{k}) = -\frac{i}{\beta}\sum_n \int \frac{d^3 p}{(2\pi)^3} S_q(i\omega_n, \vec{p}) S_D^C(i\omega_n - i\nu_m, \vec{p} - \vec{k}) , \qquad (41)$$

where $S_q$ is the quark propagator:

$$S_q(i\omega_n, \vec{p}) = \frac{\gamma_0(i\omega_n + \mu_q) - \vec{\gamma}\cdot\vec{p} + m_q I_{d4}}{(i\omega_n + \mu_q)^2 - E_q^2} , \qquad (42)$$

with $E_q^2 = \vec{p}^2 + m_q^2$, and $S_D^C$ is the charge conjugate diquark propagator:

$$S_D^C(i\omega_n - i\nu_m, \vec{p} - \vec{k}) = \frac{1}{(i\omega_n - i\nu_m - \mu_D)^2 - E_D^2} , \qquad (43)$$

with $E_D^2 = (\vec{p} - \vec{k})^2 + m_D^2$.

As detailed in the appendix B.2, the terms $\Pi_B^{(1)}$ and $\Pi_B^{(2)}$ allows obtaining a final expression of $\Pi_B$ that exclusively depends on $I_{d4}$, and not on $\gamma_0$ or $\vec{\gamma}$. In other words, this expression is similar to the one of Refs. [55, 56]. In contrast, the Refs. [62, 63] have to handle a $\Pi_B$ with $\gamma_0$ and $I_{d4}$ because they only consider $\Pi_B^{(1)}$, and not $\Pi_B^{(2)}$.

*3. Equation to be solved.*

These descriptions focus on baryons with *constant D, q* and $q_e$, i.e. described via one unique state $D+q$, like the nucleon scalar flavor component. However, the reasoning can be generalized. For example, the baryon $\Lambda$ scalar flavor component is modeled by a linear combination of the three states $[ds]+u$, $[us]+d$ and $[ud]+s$. The static approximation allows calculating all the loop functions separately, and gathering them in the matrix [53, 55, 56],

$$\Pi_B = \mathrm{diag}(\Pi_u, \Pi_d, \Pi_s) , \qquad (44)$$

where $\Pi_u$, $\Pi_d$ and $\Pi_s$ refer, respectively, to the loop functions that describe these states, appendix A. The reasoning is similar for the vertex *Z*, which becomes

$$Z = \begin{bmatrix} 0 & -g_{ds}g_{us}/m_s & 2g_{ds}g_{ud}/m_d \\ -g_{us}g_{ds}/m_s & 0 & 2g_{us}g_{ud}/m_u \\ 2g_{ud}g_{ds}/m_d & 2g_{ud}g_{us}/m_u & 0 \end{bmatrix}, \qquad (45)$$

Whatever the treated baryon, the relation to be solved to obtain its mass $m_B$ is

$$\det\left[1 - Z\, \Pi_B\left(\sqrt{m_B^2 + \vec{k}^2}, \vec{k}\right)\right] = 0 . \qquad (46)$$

For example, this equation becomes $1 - 2\dfrac{-2g_{ud}^{\,2}}{m_d}\Pi_B\left(\sqrt{m_B^2 + \vec{k}^2}, \vec{k}\right) = 0$ for a proton modeled exclusively by its scalar flavor component. The expression for each octet and decuplet baryon is available in the Refs. [55, 56].

## IV. APPROXIMATIONS AND IMPROVEMENTS OF THE BARYON MODELING

### A. Inclusion of complex numbers into the calculations.

#### 1. The Polyakov loop.

The irreducible polarization functions $\Pi_{M/D}$ of the mesons and diquarks are proportional to:

$$\Pi_{M/D}(k_0, \vec{k}) \propto A(m_q, \mu_q) + A(m_{q'}, \pm\mu_{q'}) + F\, B_0(k, m_q, \mu_q, m_{q'}, \pm\mu_{q'}, \operatorname{Re}(k_0)), \qquad (47)$$

where the plus sign concerns mesons and the minus the diquarks. For pseudo scalar mesons and scalar diquarks, the $F$ term is:

$$F = (m_q - m_{q'})^2 - (k_0 + \mu_q - (\pm\mu_{q'}))^2 + \vec{k}^2. \qquad (48)$$

For vectorial mesons and axial diquarks, it becomes:

$$F = m_q^2 + m_{q'}^2 - 4 m_q m_{q'} - (k_0 + \mu_q - (\pm\mu_{q'}))^2 + \vec{k}^2. \qquad (49)$$

Also,

$$A(m_f, \mu_f) = \frac{16\pi^2}{\beta}\sum_n \int \frac{d^3 p}{(2\pi)^3}\frac{1}{(i\omega_n + \mu_f)^2 - E_f^2} = \frac{4\pi^2}{m_f N_c}\langle\langle \bar{\psi}_f \psi_f \rangle\rangle, \qquad (50)$$

$$B_0(k, m_q, \mu_q, m_{q'}, \mu_{q'}, i\nu_m)$$
$$= \frac{16\pi^2}{\beta}\sum_n \int \frac{d^3 p}{(2\pi)^3}\frac{1}{\left[(i\omega_n + \mu_q)^2 - E_q^2\right]\left[(i\omega_n - i\nu_m + \mu_{q'})^2 - E_{q'}^2\right]}, \qquad (51)$$

are generic functions proposed by Refs. [42, 86]. They constitute a reference in NJL numerical calculations at finite $T$ and $\mu$. The substitution $\mu_f \to \mu_f^c$ Eq. (21) in $F$, carried out by the PNJL description, is strictly exact for the mesons, because $\mu_q^c - \mu_{q'}^c = \mu_q - \mu_{q'}$. Consequently, the inclusion of the Polyakov loop into the mesons description is exclusively done via the replacement of the quark/antiquark Fermi-Dirac statistics by Eqs. (23) and (24) [36]. This replacement is also performed for the diquarks. However, one has to admit the approximation $\mu_q^c + \mu_{q'}^{c'} \approx \mu_q + \mu_{q'}$ in the diquark $F$ terms, as noted in Ref. [39]. These aspects intervene in the same way for the baryons modeling, as in Ref. [55] and throughout this paper. Clearly, even if the presented baryon equations are NJL ones, the $\mathrm{NJL} \to \mathrm{PNJL}$ transition can be done in appendix B via the modified quark/antiquark Fermi-Dirac statistics Eqs. (23) and (24).

### 2. Treatment of the baryon instability.

Moreover, in Eq. (51), after the Matsubara summation, the analytic continuation $i\nu_m \to \text{Re}(k_0)$ is performed to use $k_0$ as the energy of the modeled composite particle. With Eqs. (31) and (34), $k_0$ allows finding the meson/diquark mass $m_{M/D}$. The complex part of $B_0$ is estimated via the Sokhotski–Plemelj theorem:

$$\lim_{\varepsilon \to 0^+} \frac{1}{x \pm i\varepsilon} = P\left(\frac{1}{x}\right) \mp i\pi\delta(x), \tag{52}$$

where $P$ denotes the Cauchy principal value, and $\delta$ is the Dirac delta function. In the particle unstable regime, $k_0$ and $m_{M/D}$ are complex. The mesons and diquarks equations Eqs (31) and (34) take into account $\text{Im}(k_0)$ via the $F$ term. The absence of $\text{Im}(k_0)$ as $B_0$ input argument leads to a slight underestimation of the particle width $\Gamma = -2\,\text{Im}(m_{M/D})$ [39, 55, 56].

The Eq. (51) presents similarities with the baryon loop functions Eqs. (38) and (41). This suggests using a slightly modified version of $B_0$ to calculate them, as in Refs. [53-56, 64, 65]. Nevertheless, the $F$ term does not intervene in the baryon modeling described Sec. III. Consequently, $\text{Im}(k_0)$ (and so the baryon width $\Gamma$) is not taken into account at all in Eq (46). This explains why the quoted Refs. do not explore the baryon unstable regime. In order to solve this problem, one uses a version of $B_0$ that considers $k_0 \in \mathbb{C}$ as an input argument (via an analytic continuation that allows $i\nu_m \to k_0$) and does not need Eq. (52). It also allows including $\text{Im}(m_D)$ in the modeling. This $B_0$ version has already been tested and mentioned in Refs. [39, 56] to study mesons and diquarks.

### B. Momentum dependence.

In the modeling of the baryon described in the Sec. III, $g$ and $m_D$ are the values for a diquark at rest, i.e. $\vec{p} = \vec{0}$ and $P^2 = p_0^2$ in the relation for scalar diquarks

$$\left.\frac{4G_{DIQ}}{1 - 2G_{DIQ}\,\Pi_D(p_0, \vec{p})}\right|_{P^2 = m_D^2} = \left.\frac{-g_{qqe}^2}{P^2 - m_D^2}\right|_{P^2 = m_D^2}. \tag{53}$$

In fact, both depend on the momentum $p$ in the NJL description, Eqs. (35) and (34), Figure 6. This $p$-dependence has an incidence on the Eq. (53). Indeed, as visible in the Figure 7, the pole approximation is well satisfied when they are p-dependent. More precisely, $m_D(p)$ guaranties the correct value of the pole, and $g_{qqe}(p)$ enhances the reliability of Eq. (53) at its vicinity.

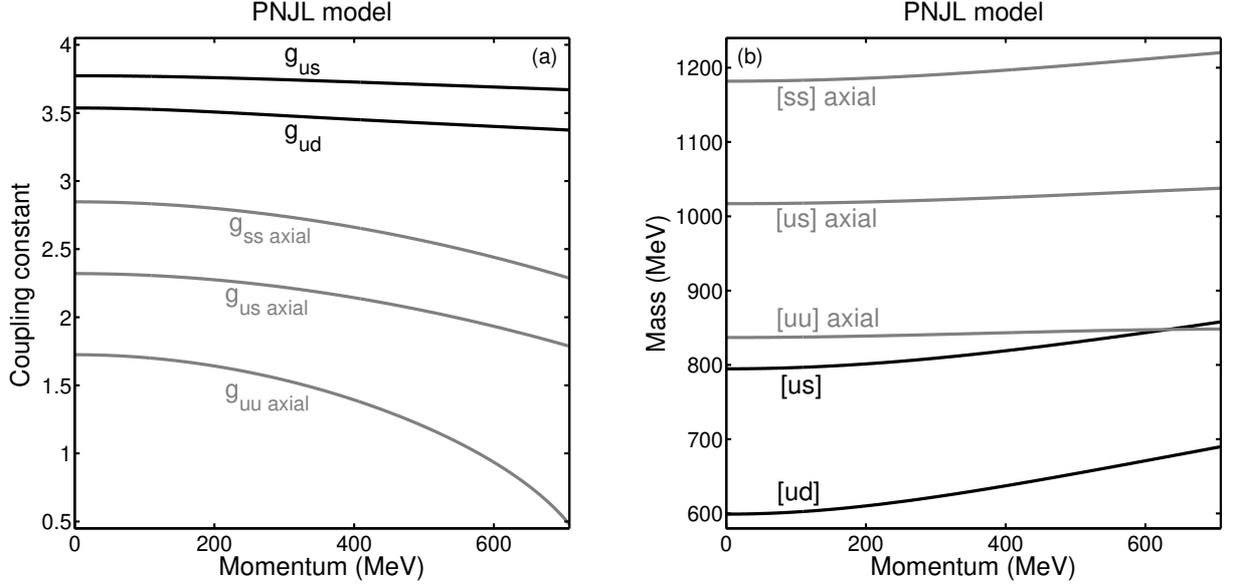

**Figure 6.** (a) Coupling constants and (b) masses of the scalar and axial diquarks according to the momentum, with $T = 10 \text{ MeV}$ and $\rho_B = 0$ (ISO parameter set).

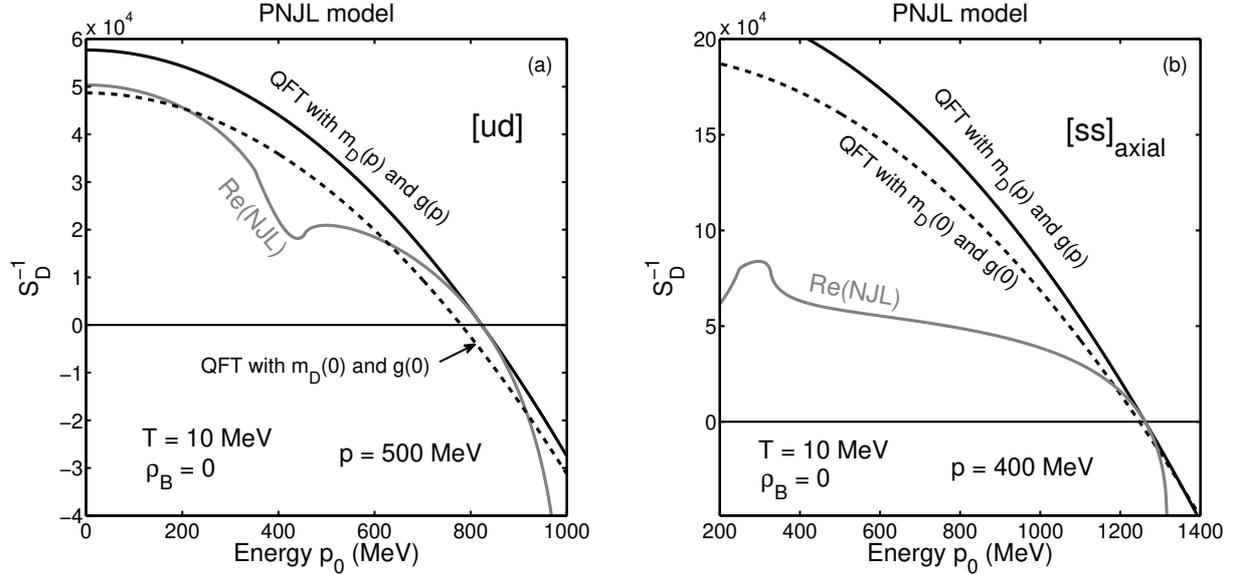

**Figure 7.** Inverse QFT and NJL propagators of a (a) scalar $[ud]$ diquark and of an (b) axial $[ss]$ diquark, according to its energy $p_0$ (ISO parameter set). The poles are visible at the level of the roots of $S_D^{-1}$.

Consequently, the use of constant $g_{qqe}$ in Eq. (37) and $m_D$ in the integrals upon $p$ in Eq. (38) and Eq. (41) constitutes *a priori* a non-negligible approximation. This suggests to consider instead $g_{qqe}(p)$ and $m_D(p)$. As a consequence, $g_{qqe}(p)$ is extracted from Z, Eq. (37), and included in the quoted integrals, which becomes:

$$-i\Pi_B^{(1)}\left(iv_m,\vec{k}\right) = -\frac{i}{\beta}\sum_n \int \frac{d^3p}{(2\pi)^3}\left[g_{qqe}(p)\right]^2 S_D(i\omega_n,\vec{p})S_q^C(i\omega_n - iv_m,\vec{p}-\vec{k}), \quad (54)$$

$$-i\Pi_B^{(2)}\left(iv_m,\vec{k}\right) = -\frac{i}{\beta}\sum_n \int \frac{d^3p}{(2\pi)^3}\left[g_{qqe}(\|\vec{p}-\vec{k}\|)\right]^2 S_q(i\omega_n,\vec{p})S_D^C(i\omega_n - iv_m,\vec{p}-\vec{k}), \quad (55)$$

with the presence of $m_D(p)$ in the expression of $E_D$ in $S_D$, and $m_D(\|\vec{p}-\vec{k}\|)$ in $S_D^C$. An indirect consequence of the p-dependence of $m_D$ is the differentiation $E_D^2 = m_D^2 + p^2 \Rightarrow p\,dp = E_D\,dE_D$ is no longer exact. Therefore, it cannot be employed to perform the substitution $\int d^3p \to \int dE$, as done in Ref. [86]. This justifies the use of $p$ as integration variable throughout this paper.

### C. Beyond the static approximation.

In order to avoid the static approximation, the two terms presented Eqs. (38) and (41) are redrawn in Figure 8 and Figure 9. Both figures recall what particles are charge conjugate, in order to determine the four-momentum of the exchanged quark. Indeed, the charge conjugation (as used in this work) acts on the particle charge, not on its momentum.

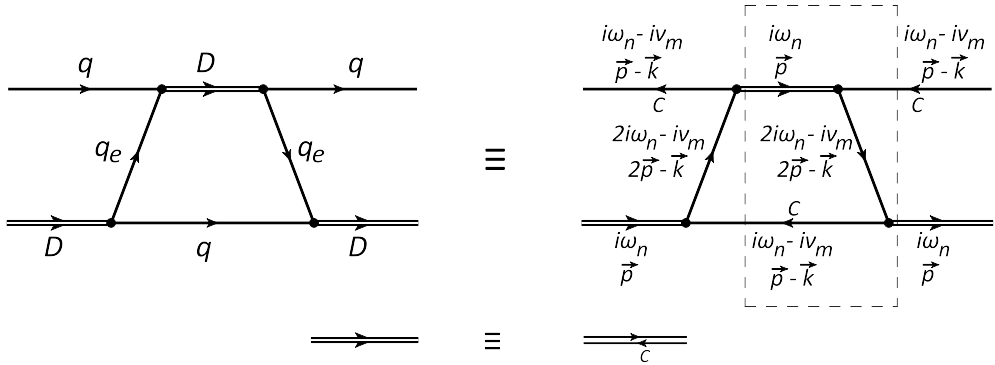

Figure 8. First term of the "loop function" (dotted rectangle) without the static approximation.

Keeping the same notations as in the Sec. III (even the name of "loop function"), the fist term becomes:

$$-i\Pi_B^{(1)}(iv_m,\vec{k})$$
$$= -\frac{i}{\beta}\sum_n \int \frac{d^3p}{(2\pi)^3} S_D(i\omega_n,\vec{p}) i\gamma_5 S_{qe}(2i\omega_n - iv_m, 2\vec{p}-\vec{k}) i\gamma_5 S_q^C(i\omega_n - iv_m, \vec{p}-\vec{k}) \quad (56)$$

with

$$S_{qe}(2i\omega_n - iv_m, 2\vec{p}-\vec{k}) = \frac{\gamma_0(2i\omega_n - iv_m + \mu_{qe}) - \vec{\gamma}\cdot(2\vec{p}-\vec{k}) + m_{qe} I_{d4}}{(2i\omega_n - iv_m + \mu_{qe})^2 - E_{qe}^2}, \quad (57)$$

and $E_{qe}^2 = (2\vec{p}-\vec{k})^2 + m_{qe}^2$.

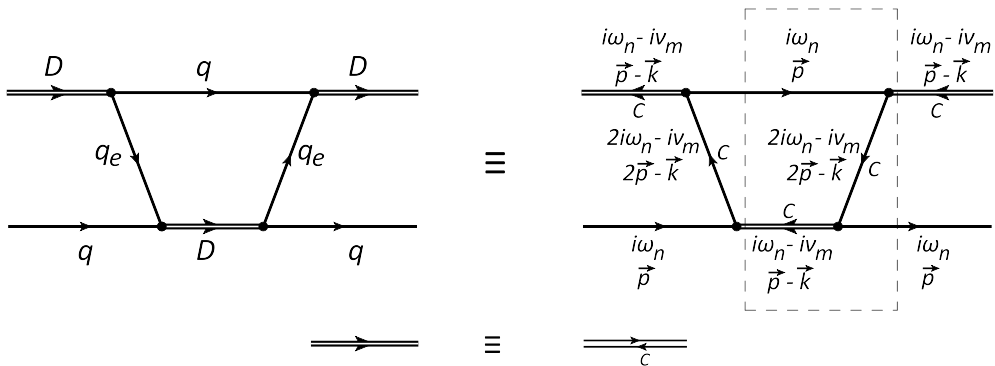

Figure 9. Second term of the "loop function" without the static approximation.

Concerning the second term, one obtains in the same way:

$$-i\Pi_B^{(2)}(iv_m,\vec{k})$$
$$= -\frac{i}{\beta}\sum_n \int \frac{d^3p}{(2\pi)^3} S_q(i\omega_n,\vec{p}) i\gamma_5 S_{qe}^C(2i\omega_n - iv_m, 2\vec{p}-\vec{k}) i\gamma_5 S_D^C(i\omega_n - iv_m, \vec{p}-\vec{k}),$$ (58)

with

$$S_{qe}^C(2i\omega_n - iv_m, 2\vec{p}-\vec{k}) = \frac{\gamma_0(2i\omega_n - iv_m - \mu_{qe}) - \vec{\gamma}\cdot(2\vec{p}-\vec{k}) + m_{qe} I_{d4}}{(2i\omega_n - iv_m - \mu_{qe})^2 - E_{qe}^2}.$$ (59)

In the framework of the Bethe-Salpeter equation, the exchanged quark $q_e$ in Eqs. (56) and (58) translates the interaction between the constituent quark-diquark couple. The expression of $Z$, Eq. (37), is updated in order to remove $-1/m_{qe}$ and the $i\gamma_5$ (or $\gamma_\mu$) matrices. More precisely, $Z$ and the "loop function" has been merged together into $\Pi_B^{(1)}$ or $\Pi_B^{(2)}$. The calculations are available in the appendix B.3. Moreover, the p-dependence corrections introduced in Subsec. IV.B are fully compatible with a description that does not consider the static approximation, Appendix D. Nevertheless, if the static approximation allows modeling baryons with several states, like $\Lambda$, the presented approach focuses on one state baryons, like the nucleons. Indeed, the pattern visible in Figure 8 and Figure 9 can be reproduced and juxtaposed *ad infinitum* only if $D, q, qe$ do not change. In the case of baryons like $\Lambda$, this would require treating the three states together in a same $\Pi_B$, but at the price of an increased complexity of the modeling.

### D. Axial flavor component of the octet baryons.

As described in [66], the inclusion of the axial flavor components in the octet baryons modeling is performed via an update of the $Z$ and $\Pi$ terms. With the proton, they are expressed as:

$$Z = \begin{bmatrix} -2\dfrac{2g_{ud}^S g_{ud}^S}{m_d} & -2\sqrt{2}\dfrac{\gamma_5 g_{ud}^S \gamma_\mu g_{uu}^A}{m_u} & -2\dfrac{\gamma_5 g_{ud}^S \gamma_\mu g_{ud}^A}{m_d} \\ -2\sqrt{2}\dfrac{\gamma^\mu g_{uu}^A \gamma^5 g_{ud}^S}{m_u} & 0 & -2\sqrt{2}\dfrac{4g_{uu}^A g_{ud}^A}{m_u} \\ -2\dfrac{\gamma^\mu g_{ud}^A \gamma^5 g_{ud}^S}{m_d} & -2\sqrt{2}\dfrac{4g_{ud}^A g_{uu}^A}{m_u} & -2\dfrac{4g_{ud}^A g_{ud}^A}{m_d} \end{bmatrix},$$ (60)

where the Figure 10 illustrates the various couplings gathered in this matrix, and

$$\Pi_B = \text{diag}(\Pi_u^S, \Pi_d^A, \Pi_u^A),$$ (61)

where the superscripts *S* and *A* stand for, respectively, scalar and axial. $\Pi_u^S$ refers to the $[ud]^S + u$ loop function, $\Pi_d^A$ to $[uu]^A + d$ and $\Pi_u^A$ to $[ud]^A + u$.

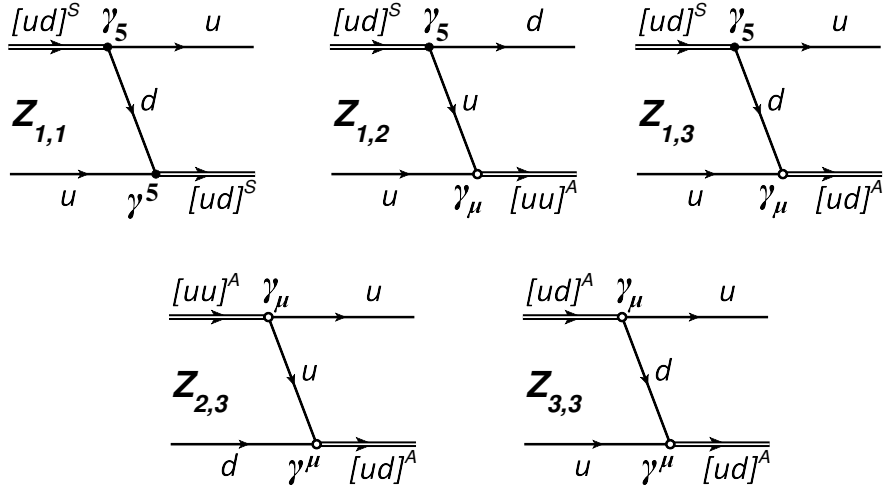

Figure 10. Graphical description of the $Z_{i,j}$ matrix elements of Eq. (60).

The Eq. (46) is still the equation that allows obtaining the proton mass. The modeling is similar for the neutron, $\Sigma^{\pm}$, $\Xi^0$ and $\Xi^-$, with or without the isospin symmetry. More details about the associated calculations and the modeling of $\Lambda$ and $\Sigma^0$ are available in Appendix C. Whatever the baryon, the momentum dependence corrections are fully usable in this description of the octet baryons. In contrast, this requires applying the static approximation, as recalled in the Appendix D.

### E. NJL diquark propagator in the baryon description.

#### 1. Presentation of the employed method.

The use of a QFT diquark propagator in Eq. (38) and Eq. (41) wants to say that the diquark is treated as a *structureless* particle in the baryon modeling, in which $m_D$ is found with Eq. (34) for a given temperature, density and momentum. Consequently, a possible evolution of the modeling could be to replace the QFT scalar diquark propagator by an NJL one. So, Eq. (53), which establishes a link between $S_D^{NJL}$ and $S_D^{QFT}$, is considered again:

$$\left.\frac{4G_{DIQ}}{1-2G_{DIQ}\Pi_D(p_0,\vec{p})}\right|_{P^2=m_D^2} = \left.\frac{-g_{qqe}^{\,2}}{P^2-m_D^2}\right|_{P^2=m_D^2}. \tag{62}$$

Concretely, Eq. (39) becomes

$$S_D(i\omega_n,\vec{p}) = \frac{1}{(i\omega_n+\mu_D)^2-E_D^2} \to \frac{\dfrac{4G_{DIQ}}{-g_{qqe}^{\,2}}}{1-2G_{DIQ}\Pi_D(i\omega_n+\mu_D,\vec{p})}. \tag{63}$$

In the same way, Eq. (43) is rewritten as

$$S_D^C(i\omega_n-i\nu_m,\vec{p}-\vec{k}) = \frac{1}{(i\omega_n-i\nu_m-\mu_D)^2-E_D^2} \to \frac{\dfrac{4G_{DIQ}}{-g_{qqe}^{\,2}}}{1-2G_{DIQ}\Pi_D(i\omega_n-i\nu_m-\mu_D,\vec{p}-\vec{k})}. \tag{64}$$

The substitution is feasible for axial diquarks in a similar way, with the same precautions as in Subsubsec. III.B.2 with the Proca propagator. The term $-g_{qqe}^{\,-2}$ visible in these two Eqs. disappears with the $-g_{qqe}^{\,2}$ found in the expression of $Z$, Eq. (37). Therefore, the diquark coupling constant $g_{qqe}$ does not intervene in this evolution. This remark stays relevant when the p-dependence corrections are applied: $m_D(p)$ and $m_D(\|\vec{p}-\vec{k}\|)$ are still included in the calculations, but not $g_{qqe}(p)$ and $g_{qqe}(\|\vec{p}-\vec{k}\|)$ that disappear. In this case, the Figure 7 reveals that the $S_D^{NJL}$ and $S_D^{QFT}$ diquark propagators are very close near to the pole, but have visible differences elsewhere. This statement is particularly true for the axial diquarks.

*2. Diquarks and antidiquarks.*

During the Matsubara summation described in Appendix B, the Eq. (62) recalls that $S_D^{NJL}$ and $S_D^{QFT}$ must have the same poles: $i\omega_n = -\mu_D \pm E_D$ for Eq. (63) and $i\omega_n = \mu_D \pm E_D + i\nu_m$ for Eq. (64). Moreover, this appendix underlines that the replacements done in Eqs. (63) and (64) are feasible with or without the static approximation.

In fact, in the left hand side of Eq. (62), $S_D^{NJL}$ describes a diquark when $p_0 > 0$. In this case, $E_D$ is found with its mass. In contrast, $S_D^{NJL}$ models an antidiquark at negative $p_0$, and $E_D$ is calculated with this antidiquark mass. At zero baryonic density, this observation is without consequence because they have the same mass $m_D$. However, this affirmation is incorrect otherwise, Figure 11(a). In other words, in Figure 11(b), $S_D^{NJL}$ (gray curve) is not there an even function according to $p_0$, in contrast with $S_D^{QFT}$ ("QFT diquark" curve). This feature concerns both terms $\Pi_B^{(1)}$ and $\Pi_B^{(2)}$.

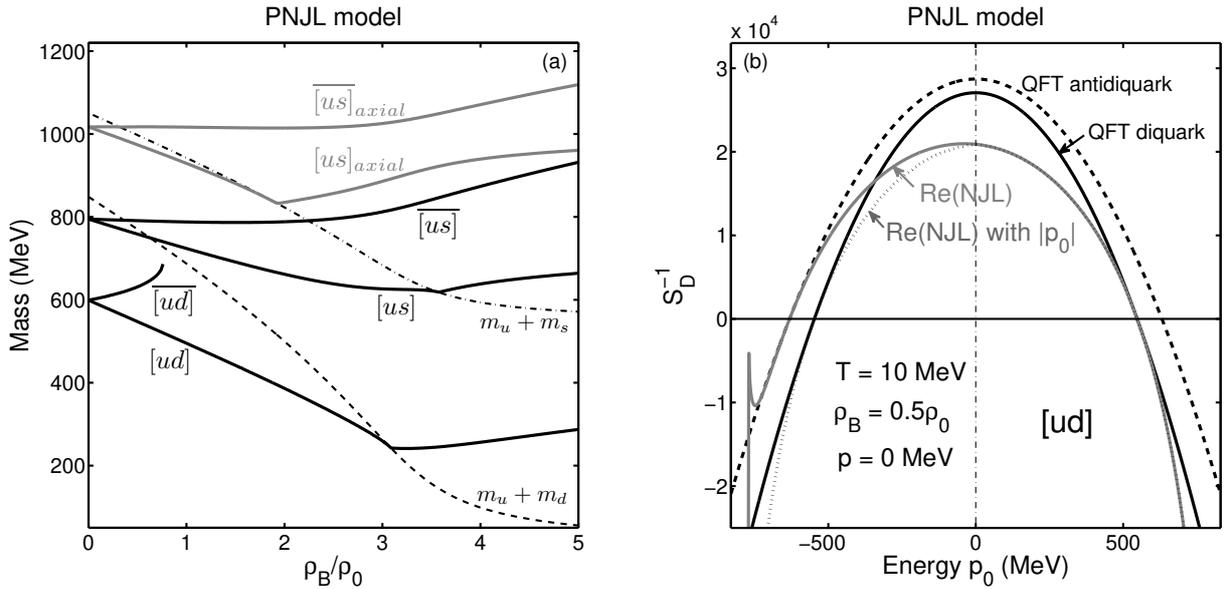

**Figure 11.** (a) Scalar and axial diquarks/antidiquarks (ISO parameter set). The antidiquark $\overline{[ud]}_{axial}$ cannot be modeled when $\rho_B > 0.04\rho_0$. Also, $[ss]$ and $\overline{[ss]}$ are degenerate in the whole $T, \rho_B$ plane. (b) Inverse QFT and NJL propagators of the scalar $[ud]$ and inverse QFT propagator of the scalar $\overline{[ud]}$ ("QFT antidiquark" curve).

In practice, two solutions are possible to handle $S_D^{NJL}$ in the baryon modeling. The first one considers both diquark and antidiquark masses according to the sign of $p_0$, as explained above. Since the scalar antidiquark $\overline{[ud]}$ cannot be modeled for a wide range of $\rho_B$, this limits the possibilities to study nucleons at moderate and high densities. The second solution applies an absolute value to $\text{Re}(p_0)$. So, it exclusively considers diquarks and not antidiquarks ("Re(NJL) with $|p_0|$" curve). This choice permits $\Pi_D$ to be an even function in the whole $T, \rho_B$ plane, and leads to some simplifications in appendix B.3 and B.5.

## F. Color superconductivity.

When the color superconductivity and the meson condensation are taken into account, the Eqs. (3) and (9) include additional terms. These ones notably concern the energy gaps $\Delta$ of the pairs that appear during these phenomena. The energy gaps of the meson condensates also intervene in the non-diagonal terms of the matrix Eq. (12), and the ones of the diquark condensates in those of the matrix Eq. (11). At zero strangeness, only the 2-flavor color-superconducting phase (2SC) and the pion condensation are considered. The 2SC phase corresponds to the formation of pairs involving a red $u$ quark with a green $d$ quark, or a green $u$ quark with red $d$ one. The energy gap $\Delta_{ud}$ is the binding energy of these pairs, and is defined as:

$$\Delta_{ud} = -2 G_{DIQ} \left\langle\left\langle \bar{\psi}^C \gamma_5 \tau_2 \lambda_2 \psi \right\rangle\right\rangle. \tag{65}$$

This phase is present when $\Delta_{ud} \neq 0$. In contrast, all the energy gaps are equal to zero in the normal quark phase (NQ). Moreover, the pion condensation requires a strong asymmetry between $\mu_u$ and $\mu_d$, which cannot be reached when $\rho_B$ is employed [39]. Consequently, the $T, \rho_B$ plane is divided between the NQ and the 2SC phases, Figure 12(a). The 2SC phase affects some observables. This notably concerns the masses of the quarks, mesons and diquarks, the chemical potentials and the Polyakov fields [38, 39]. The increase in the mass and width of the scalar $[ud]$ diquarks whose color is $rg$ is particularly strong, Figure 12(b, c). An interpretation of this behavior is proposed in Ref. [39]: the $[ud]^{rg}$ cease to act as free and independent "particles", in order to form the diquark condensate.

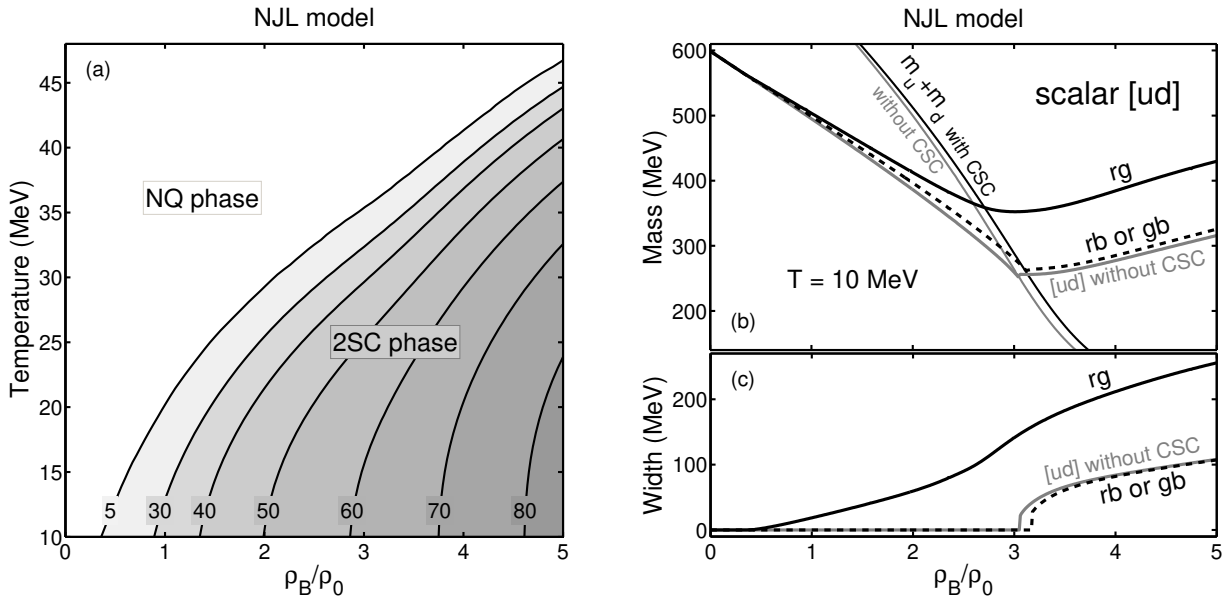

Figure 12. (a) Phase diagram in the $T, \rho_B$ plane, from [38]. In the level curves, the values of the gap $\Delta_{ud}$ are in MeV. (b) Mass and (c) width of the scalar $[ud]$ diquarks according to the baryonic density, including or not the influence of the color superconductivity (CSC), from [39].

A complete treatment of the (P)NJL baryon modeling in a color superconducting regime is out of the scope of this paper. However, the quarks and diquarks described in the 2SC phase are usable to build baryons, following the method described above. Obviously, this constitutes an approximation, but this can bring information about the behavior of baryons like the nucleons in such conditions. In addition, since the used modeling of the 2SC phase only considers scalar diquarks at rest, this forbids applying the p-dependence corrections of Subsec. IV.B in this case.

### G. Expression of the baryon coupling constant.

In this subsection, a method is proposed to estimate the baryon coupling constant $g_{Dq}$ that stays valid in all the various evolutions described above. This can be obtained via a pole approximation, as done in Eq. (33) that becomes Eq. (35), but adapted to a fermionic composite particle. For a spin-1/2 baryon, this leads to consider both propagators:

$$S_B^{NJL} = \frac{Z}{1 - \bar{Z} \Pi_B(k_0, \vec{k})} \tag{66}$$

and

$$S_B^{QFT} = \frac{-g_{Dq}^2}{K^2 - m_B^2} . \tag{67}$$

The expression of $\Pi_B$ depends on the used evolution (p-dependence corrections, etc.). This has an influence on the vertex term $\bar{Z}$. This one is equal to the $Z$ in the numerator only in the description of Sec. III. In this paper, two approximations are applied at this stage. Firstly, $Z$ is expected to be constant, and is defined in Eq. (37) in all the cases. Secondly, at the vicinity of the pole $K^2 = m_B^2$, the equality between Eqs. (66) and (67) is supposed to be found when a trace is applied:

$$\text{Tr}\left[\frac{Z}{1 - \bar{Z}\Pi_B(k_0, \vec{k})}\right]\bigg|_{K^2 = m_B^2} = \text{Tr}\left[\frac{-g_{Dq}^2 (\gamma_0 k_0 - \vec{\gamma} \cdot \vec{k} + m_B I_{d4})}{k_0^2 - k^2 - m_B^2}\right]\bigg|_{K^2 = m_B^2} . \tag{68}$$

When a QFT diquark propagator is employed in the modeling, $\Pi_B$ only depends on $I_{d4}$. For a baryon at rest, this gives:

$$\frac{Z}{1 - \bar{Z} \Pi_B(k_0, \vec{0})}\bigg|_{k_0 = m_B} = -g_{Dq}^2 \frac{m_B}{k_0^2 - m_B^2}\bigg|_{k_0 = m_B} . \tag{69}$$

This relation allows extracting the coupling constant:

$$g_{Dq} = \sqrt{\frac{2Z}{\bar{Z}\frac{\partial \Pi_B(k_0, \vec{0})}{\partial k_0}\bigg|_{k_0 = m_B}}} . \tag{70}$$

This expression can be compared to the one of [56, 64]. When an NJL diquark propagator is used, $\Pi_B$ can be written on the form $\Pi_B = \gamma_0 \Pi_1 + I_{d4} \Pi_2$. Injected into the Eq. (46), this leads to consider these two relations, for a baryon at rest:

$$1 - \bar{Z}\left[-\Pi_1(m_B, \vec{0}) + \Pi_2(m_B, \vec{0})\right] = 0 \tag{71}$$

and

$$1 - \bar{Z}\left[\Pi_1(m_B, \vec{0}) + \Pi_2(m_B, \vec{0})\right] = 0 . \tag{72}$$

In practice, the baryon mass $m_B$ satisfies one of them. It can be shown that:

$$g_{Dq} = \sqrt{\frac{Z}{\bar{Z}\frac{\partial\left[\mp\Pi_1(k_0,\vec{0})+\Pi_2(k_0,\vec{0})\right]}{\partial k_0}\bigg|_{k_0=m_B}}} \quad , \tag{73}$$

where the minus (resp. plus) sign is applied when Eq. (71) (resp. (72)) is verified.

For a spin-3/2 baryon, like $\Delta$, the Dirac propagator Eq. (67) should be replaced by the Rarita-Schwinger one [90]. On the one hand, the denominator $K^2 - m_B^2$ (and its associated pole) is unchanged during this replacement, e.g., [91, 92]. This confirms the possibility to use the Eq. (46) in the modeling of spin-3/2 baryons with no restriction. On the other hand, the *free isobar propagator* visible in [91] could undergo the application of Eq. (68), in order to obtain a result equivalent to Eqs. (70) and (73). However, the use of the trace appears to be a cruder approximation in this case in comparison with spin-1/2 baryons.

## V. RESULTS

Throughout this section, a first objective is to compare the masses obtained at zero temperature and zero density with the experimental data collected in TABLE III. The baryon masses are calculated with the ISO and NISO parameter sets, TABLE I. A global comparison between the NISO parameter set results and the experimental ones is available in Appendix E, for each evolution proposed in this work. This NISO parameter set gives very good results with pseudo-scalar mesons [55]. Therefore, it will not be modified in order to obtain a better agreement with be baryons. Indeed, the goal is to be able to use the same parameter set to describe the mesons and baryons (even if the mesonic results, strictly identical to the ones of [55, 56], are not included in this paper).

A second objective is to plot the variations of the masses according to the temperature *T* and the baryonic density $\rho_B$. In this case, the ISO parameter set is employed. Since it respects the isospin symmetry, this reduces the number of curves to be represented and avoids overloading the graphs. Anyway, the ISO and NISO are rather similar. As a consequence, their results have a rather close dependence according to *T* and $\rho_B$.

TABLE III. Actual masses of the octet and decuplet baryons, in MeV [93].

| Proton | Neutron | $\Lambda$ | $\Sigma^+$ | $\Sigma^0$ | $\Sigma^-$ | $\Xi^0$ | $\Xi^-$ |
|---|---|---|---|---|---|---|---|
| 938.27 | 939.57 | 1115.68 | 1189.37 | 1192.64 | 1197.45 | 1314.86 | 1321.71 |

| $\Delta^{++}\ \Delta^+\ \Delta^0\ \Delta^-$ | $\Sigma^{*+}$ | $\Sigma^{*0}$ | $\Sigma^{*-}$ | $\Xi^{*0}$ | $\Xi^{*-}$ | $\Omega^-$ |
|---|---|---|---|---|---|---|
| 1209 to 1211 | 1382.83 | 1383.7 | 1387.2 | 1531.80 | 1535.0 | 1672.45 |

### A. First results.

In the (P)NJL literature, a composite particle is *stable* when its mass is lower than its constituents, *unstable* otherwise. This instability corresponds to the possibility of its disintegration into its constituents. In practice, the stability may be obtained at low and moderate temperatures and densities. The octet and decuplet baryons have been studied in previous papers, e.g. [53-56, 62, 63], but only as stable particles. These results are in agreement with those presented Figure 13 to Figure 15, at least in their stable regime. In contrast, the description of their instability is rather new. This one is observed at higher temperatures or densities. The stable-unstable transition is also named Mott transition. It corresponds to the angular points drawn by the baryon curve, at a critical temperature (or Mott temperature) or critical density, at its intersection with the curve of its constituents. The fact to obtain a Mott temperature for the nucleon (266 MeV) slightly lower than the one of the scalar $[ud]$ diquark (267 MeV, Figure 16(a) ) is not in agreement with [57]. Indeed, this Ref. indicates a temperature range in which the nucleon is stable when its diquark constituent is unstable, so a Borromean state. The angular point visible in this Ref. for the nucleon is not a Mott transition, in contrast with Figure 13. Since this Ref. uses a rather similar way to study the nucleon, a different parameterization of the model could explain this disagreement.

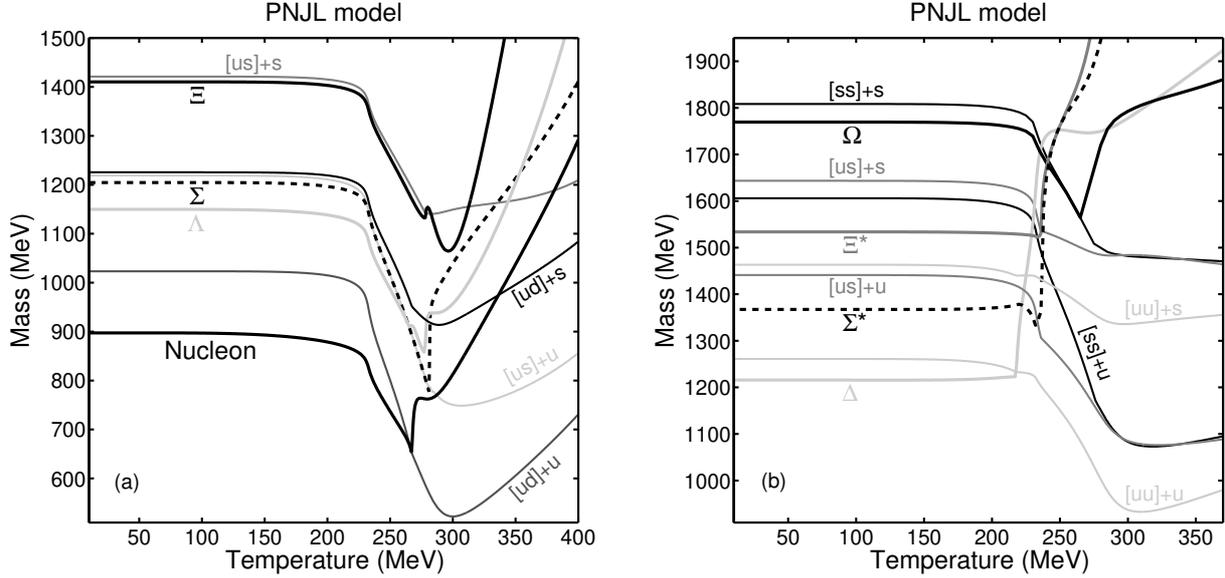

**Figure 13. Masses of the (a) octet and (b) decuplet baryons according to the temperature, at zero density (ISO parameter set). The octet baryons are modeled via scalar diquarks, and the decuplet via axial ones. The baryons $\Sigma$ and $\Sigma^0$ are degenerate, and so on for $\Sigma^*$ and $\Sigma^{0*}$.**

The unstable regime of the mesons and diquarks is characterized by a continuous increase in their mass [23, 42, 53, 55]. This affirmation should be modified for certain baryons. With the nucleon, a brutal increasing is observed just after the stable-unstable transition, followed by a smooth variation or even a local maximum, and then the mass increases again. This behavior is due to the diquark instability. As a whole, the critical temperature of a baryon is lower than or equal to the one(s) of its diquark constituent(s), a so on for its critical density. When the diquark becomes unstable, its mass $m_D$ and coupling constant $g_{qqe}$ used in Eq. (46) become complex numbers, Figure 16(a, b). The non-monotonic variations of the baryon mass come from this coupling constant. Moreover, the use of $\text{Im}(m_D)$ into the baryon modeling leads to reduce the increasing in the baryon mass (real and imaginary parts) in its unstable regime, Figure 16(c). Even if this figure focuses on the nucleon, the other baryons follow more or less this behavior. However, the brutal fall of the $\Xi$ mass in its unstable regime, according to $\rho_B$, constitutes an exception.

The case of the $\Lambda$ baryon is particularly interesting. Indeed, it is modeled by two states $[qs]+q$ and $[ud]+s$ with the isospin symmetry. Therefore, this baryon presents two Mott transitions according to the temperature, and according to the baryonic density. This feature had been identified in [55, 56]. However, the local minimum visible between the two transitions according to the temperature, Figure 13, had not been detected in these references.

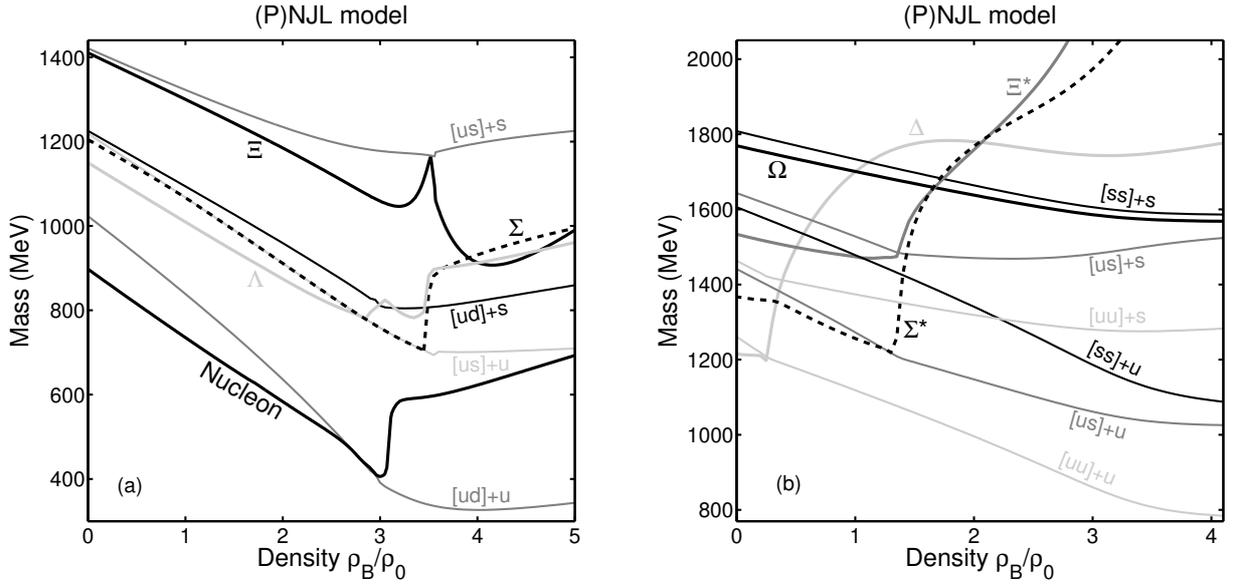

**Figure 14. Masses of the (a) octet and (b) decuplet baryons according to the baryonic density, at $T = 10$ MeV (ISO parameter set). The diquarks are (a) scalar and (b) axial.**

On the other hand, Figure 15 exhibits the nucleon mass in the $T, \rho_B$ and in the $T, \mu_B$ planes, where $\mu_B = 3\mu_q$ is the baryonic chemical potential. No discontinuity is found in the first one. The strong mass variations reported above are still visible in this graph. They allow identifying easily the whole frontier between the stable and the unstable regimes of the nucleon. In contrast, the first order chiral phase transition explains the shape of the plotted surface in the $T, \mu_B$ plane, at low temperatures and $\mu_B \approx 1200$ MeV. Such graphs recall the ones plotted in refs. [38, 39] for the quarks, mesons and diquarks. In other words, it is possible to distinguish the stable, metastable and unstable states linked to this first order phase transition in the $T, \mu_B$ plane [38], but not in the $T, \rho_B$ one. Since the main goal of this work is to study baryons, one focuses on graphs according to the temperature and the baryonic density.

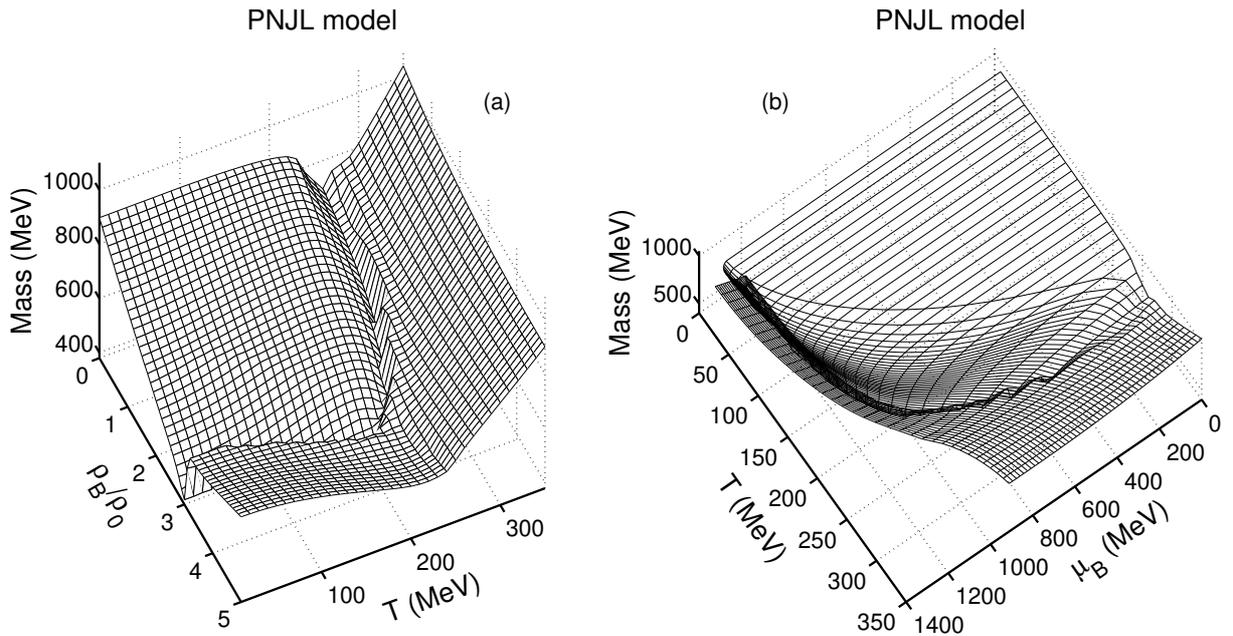

**Figure 15. Nucleon mass in the (a) $T, \rho_B$ plane and in the (b) $T, \mu_B$ plane (ISO parameter set).**

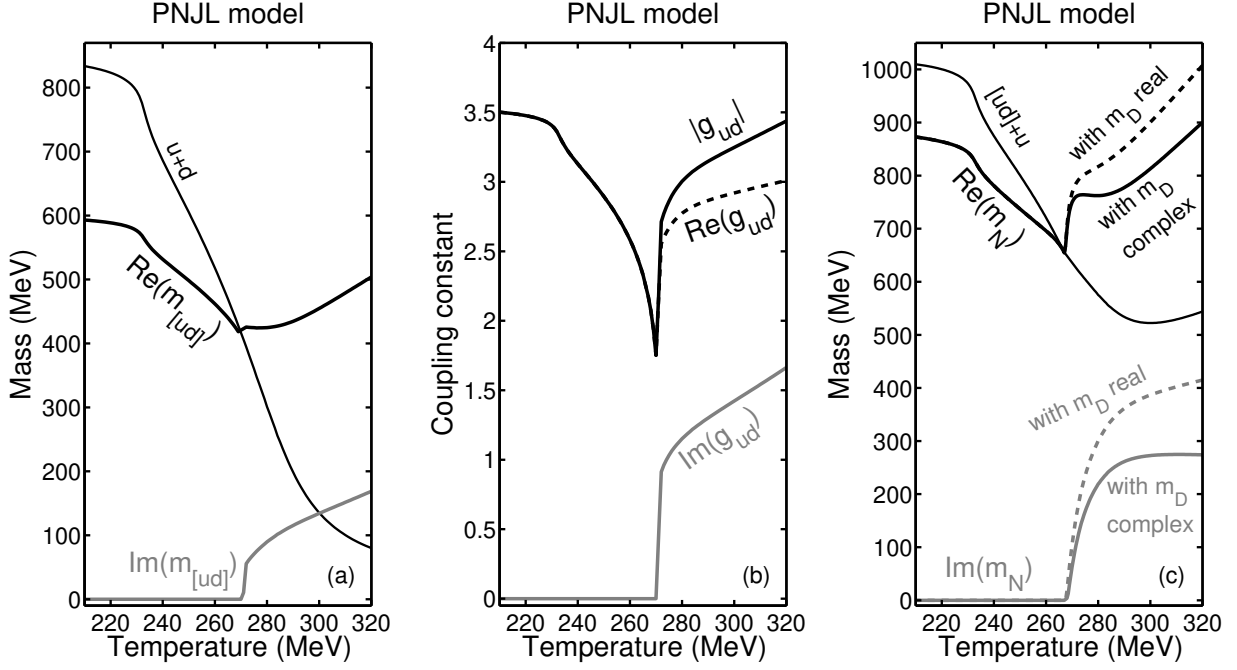

Figure 16. (a) Mass and (b) coupling constant of the scalar $[ud]$ diquark, and (c) mass of the nucleon according to the temperature, at zero density (ISO parameter set).

**B. Influence of the momentum dependence.**

The TABLE IV and TABLE V allow a first comparison to estimate the effect of the p-dependence corrections on the baryon masses. The ones found without momentum dependence are in agreement with the refs. [55, 56], except minor differences. As a whole, the inclusion of these corrections leads to an increase in the masses. However, its importance depends on the baryon. The effect is strong with the nucleons, with or without the isospin symmetry. Their masses without correction were too weak; the corrected ones are closer to the experimental values. In contrast, the effect is moderate for other baryons, e.g., $\Xi$ or $\Omega$. In these cases, the increase in mass leads them to move away from the experimental data.

**TABLE IV.** Masses of the octet and decuplet baryons at $T=0$ and $\rho_B=0$, in MeV (ISO parameter set). The data of the row "No p" do not include p-dependence corrections; the "p" row includes them.

|       | Nucleon | $\Lambda$ | $\Sigma$ | $\Xi$ | $\Delta$ | $\Sigma^*$ | $\Xi^*$ | $\Omega$ |
|-------|---------|-----------|----------|-------|----------|------------|---------|----------|
| No p  | 897.4   | 1149.9    | 1204.6   | 1410.4 | 1215.6  | 1367.3     | 1533.8  | 1769.4   |
| p     | 955.7   | 1184.2    | 1215.2   | 1419.5 | 1260.9  | 1434.9     | 1592.6  | 1798.9   |

**TABLE V.** Same legend as **TABLE IV** with the NISO parameter set.

|       | Proton | Neutron | $\Lambda$ | $\Sigma^+$ | $\Sigma^0$ | $\Sigma^-$ | $\Xi^0$ | $\Xi^-$ |
|-------|--------|---------|-----------|------------|------------|------------|---------|---------|
| No p  | 887.0  | 884.7   | 1105.1    | 1148.8     | 1151.1     | 1153.4     | 1332.3  | 1335.3  |
| p     | 944.7  | 943.9   | 1140.2    | 1163.8     | 1166.3     | 1168.9     | 1340.0  | 1342.6  |

|       | $\Delta^{++}$ | $\Delta^+$ | $\Delta^0$ | $\Delta^-$ | $\Sigma^{*+}$ | $\Sigma^{*0}$ | $\Sigma^{*-}$ | $\Xi^{*0}$ | $\Xi^{*-}$ | $\Omega^-$ |
|-------|---------------|------------|------------|------------|---------------|---------------|---------------|------------|------------|------------|
| No p  | 1211.6        | 1212.5     | 1213.4     | 1214.2     | 1338.5        | 1339.2        | 1339.9        | 1478.8     | 1480.3     | 1674.8     |
| p     | 1249.6        | 1250.0     | 1253.0     | 1255.8     | 1395.9        | 1398.1        | 1400.3        | 1531.8     | 1533.9     | 1700.0     |

When the isospin symmetry is not applied, the p-dependence corrections do not modify the mass hierarchy visible, e.g., with the $\Delta$: the $\Delta^{++}$ stays the lightest and $\Delta^-$ the heaviest. This affirmation also includes the proton and neutron. The corrections lead to reduce the mass gap between them.

Nevertheless, the proton stays heavier than the neutron in this description… Therefore, the p-dependence alone does not fix the mass inversion problem mentioned above.

The Figure 17 studies the masses according to the temperature. As a whole, the shape of the curves is not really modified by the corrections. The fluctuations of the octet baryon masses in their unstable regime seem to be reduced. This is particularly visible for $\Xi$: at $T \approx 290$ MeV, the depth of the minimum is decreased. However, differences between the curves applying or not these corrections are observed at high temperatures, in particular with $\Delta$ and $\Omega$. Moreover, at low temperatures, the gap between the nucleon mass and the one of its constituents is lowered by the momentum dependence. This gap corresponds to the nucleon binding energy. Even with its reduction, the Mott temperature of the nucleon is unchanged. This observation can be generalized: the momentum dependence corrections seem to have a negligible effect on the Mott temperatures, except for $\Omega$.

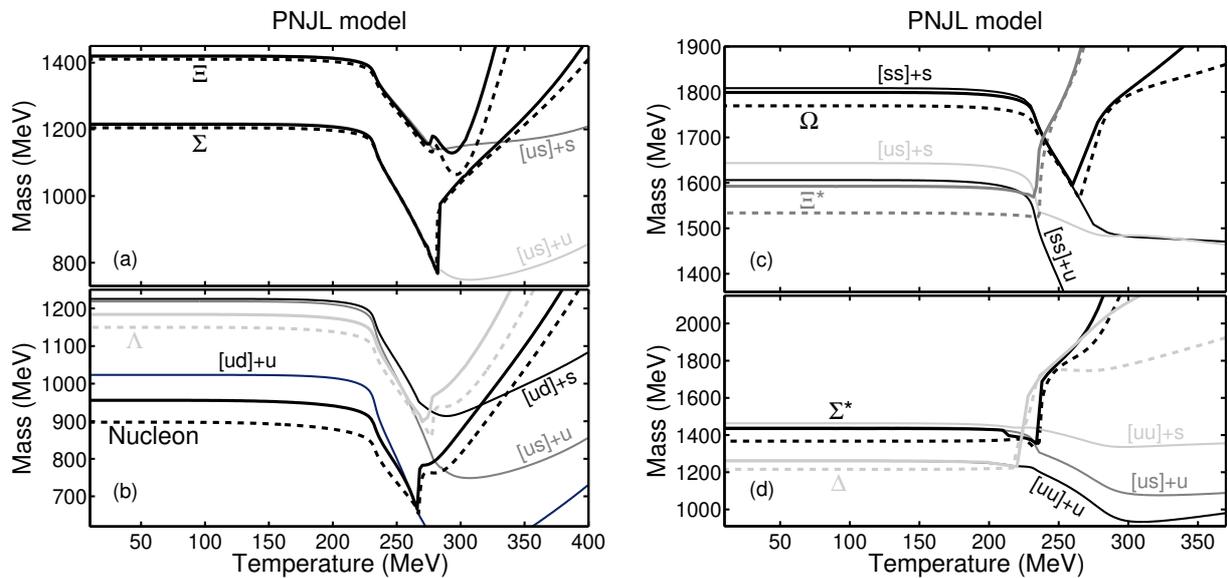

**Figure 17.** Masses of the (a, b) octet and (c, d) decuplet baryons according to the temperature, at zero density (ISO parameter set). The p-dependence corrections appear in the baryon solid curves; the dotted ones do not include them. The diquarks are (a, b) scalar and (c, d) axial.

In the Figure 18, the masses are plotted according to the baryonic density. On the one hand, the shape of the curves is not altered by the momentum dependence, as with the temperature. With the nucleon, the local maximum located near the stable-unstable transition is reinforced. On the other hand, the critical densities are reduced by these corrections. The gaps are limited with the octet baryons, but not with the decuplet ones.

Concerning $\Sigma^*$, a slight mass decrease is visible in Figure 17(d) and Figure 18(d) at, respectively, $T \approx 220$ MeV and $\rho_B \approx 0.3\rho_0$, with the p-dependence curves. This phenomenon is due to the axial $[uu]$ diquark, which becomes unstable at these values. It is invisible without the p-dependence corrections.

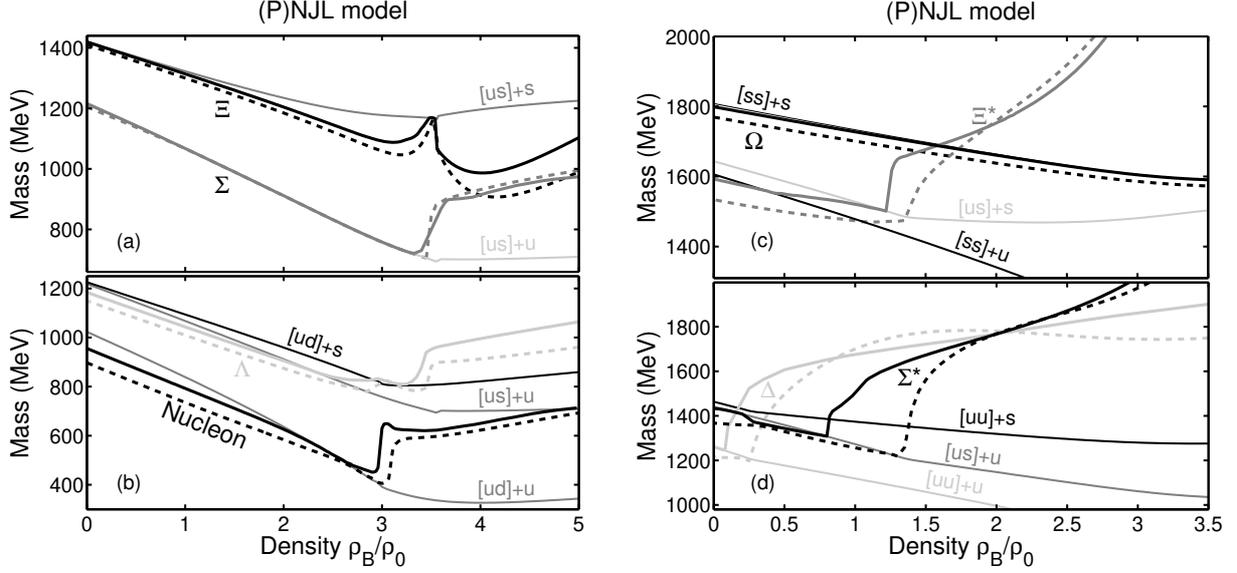

**Figure 18.** Masses of the (a, b) octet and (c, d) decuplet baryons according to the baryonic density, at $T = 10 \text{ MeV}$ (ISO parameter set). The p-dependence corrections appear in the baryon solid curves; the dotted ones do not include them. The diquarks are (a, b) scalar and (c, d) axial.

### C. Results without the static approximation.

The Refs. [55, 56] explain how the static approximation leads to the unwanted mass inversion for the proton and neutron. If this approximation is not applied, the problem should disappear. As attested by TABLE VII, this is actually the case. Indeed, the neutron is found to be heavier than the proton, even if their mass difference is higher than expected. This property stays valid with or without the p-dependence corrections. In addition, the estimated masses are rather close to the experimental ones when these corrections are applied. This observation can be extended to the other baryons modeled in TABLE VII, Appendix E. Nevertheless, the reduced size of TABLE VI and TABLE VII recalls that only the baryons described by a single state $D + q$ are studied in this subsection.

**TABLE VI.** Masses (in MeV) of octet and decuplet baryons at $T = 0$ and $\rho_B = 0$, without the static approximation, with the ISO parameter set. The "No p" row does not include p-dependence corrections; the "p" row includes them.

|  | Nucleon | $\Sigma$ | $\Xi$ | $\Delta$ | $\Omega$ |
|---|---|---|---|---|---|
| No p | 903.2 | 1186.2 | 1411.5 | 1229.4 | 1759.9 |
| p | 942.7 | 1197.3 | 1416.7 | 1238.7 | 1771.4 |

**TABLE VII.** Same legend as **TABLE VI** with the NISO parameter set.

|  | Proton | Neutron | $\Sigma^+$ | $\Sigma^-$ | $\Xi^0$ | $\Xi^-$ | $\Delta^{++}$ | $\Delta^-$ | $\Omega^-$ |
|---|---|---|---|---|---|---|---|---|---|
| No p | 891.5 | 894.9 | 1139.1 | 1144.5 | 1331.7 | 1334.0 | 1218.3 | 1225.3 | 1664.7 |
| p | 930.9 | 934.2 | 1150.9 | 1156.4 | 1336.8 | 1339.2 | 1226.8 | 1234.3 | 1674.5 |

Upon a physical point of view, the effect of the static approximation is expected to be weak when the exchanged quark is massive, maybe with a dressed strange quark. This could concern $\Sigma$ and $\Omega$. In addition, $m_s$ stays strong at high temperatures. Therefore, the effect of this approximation could stay reduced whatever the temperature [53]. This may explain why the two curves of $\Sigma$ are rather close in Figure 19. However, it leads to major modifications at the level of the $\Omega$ Mott transition.

Moreover, its effect should be important when a light quark is exchanged, i.e. with the nucleons, $\Xi$ and $\Delta$. This last statement is confirmed by the nucleons and $\Delta$ with p-dependence corrections,

Appendix E. Indeed, their masses at zero temperature and density are modified in a non-negligible way when the static approximation is no longer applied. In addition, the light quark masses $m_q$ tend towards their naked values $m_{0q}$ at high temperatures [22, 94]. Consequently, the static approximation should have non-negligible consequences in these conditions. This is a possible explanation of the divergence between the two curves of $\Xi$ in Figure 19 at high *T*, and so on for $\Delta$. Nevertheless, this divergence is more modest for the nucleons.

On the other hand, as with the p-dependence corrections, the abandon of the static approximation leads to limited modifications of the Mott temperatures. They are slightly increased with the modeled baryons, except for $\Sigma$.

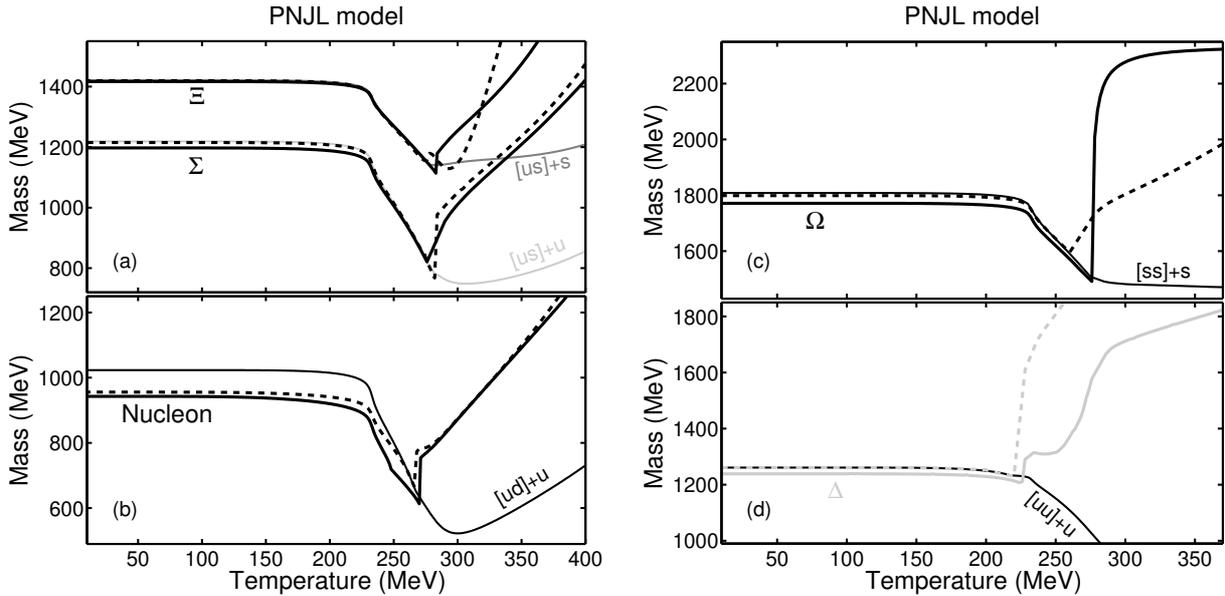

**Figure 19. Masses of (a, b) octet and (c, d) decuplet baryons function of the temperature, at zero density (ISO parameter set). The baryon solid curves do not use the static approximation; the dotted ones employ it. The p-dependence corrections are applied.**

The Figure 20 focuses on the results according to the baryonic density. The Refs. [47, 53] have reported that the static approximation leads to a too great decrease in the nucleon mass according to $\rho_B$, in its stable regime. In parallel, they have estimated a much slower decrease without this approximation. The Figure 20(a) tend to confirm their predictions, at least qualitatively. Indeed, the modifications are weaker as expected in these Refs.

Concerning $\Xi$, the previous subsections reveal a brutal fall of its mass according the baryonic density when it becomes unstable. Without the static approximation, this behavior totally disappears. In Figure 20(a), its mass increases slowly according to $\rho_B$ in this regime. This last statement is also particularly true with $\Sigma$, and with the nucleon: its local maximum vanishes. As a whole, the abandon of the static approximation leads to remove the brutal mass variations observed at the stable-unstable transition according to the density. Concerning $\Delta$, a large gap is found between the two curves at high $\rho_B$. The explanations given above for the temperatures may be extended to the densities to understand this divergence.

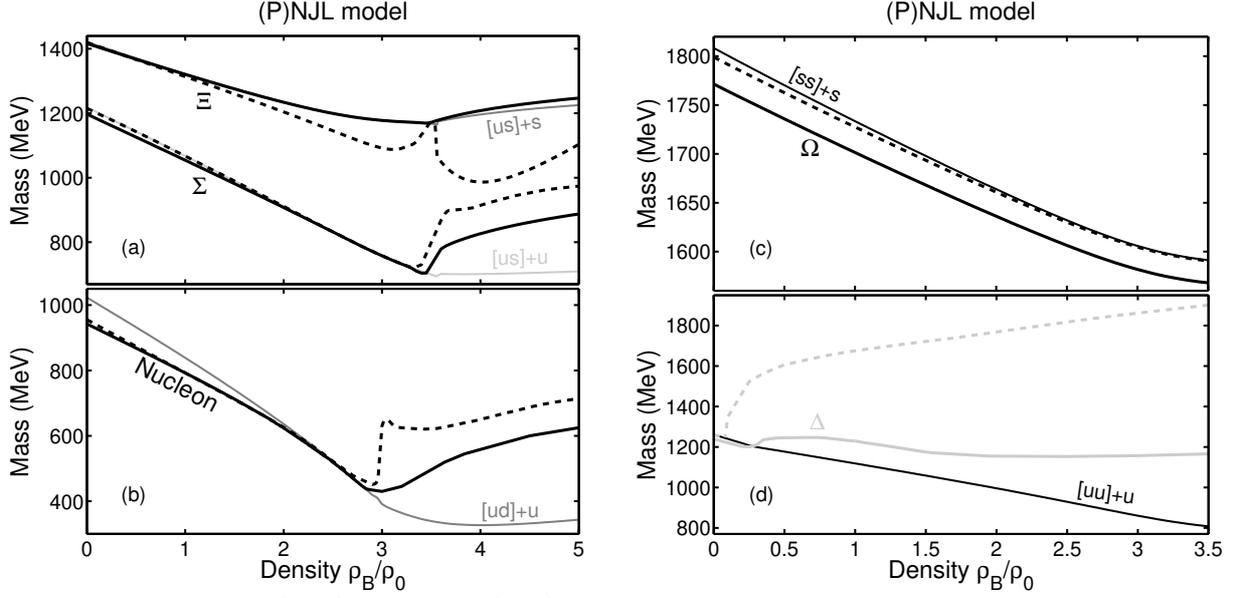

Figure 20. Masses of (a, b) octet and (c, d) decuplet baryons function of the baryonic density, at $T = 10$ MeV (ISO parameter set). The baryon solid curves do not use the static approximation; the dotted ones employ it. The p-dependence corrections are considered.

### D. Octet baryons with scalar and axial flavor components.

The inclusion of the axial flavor components into the octet baryons description is not new in a quark model approach. However, such an evolution has not yet been done previously in a PNJL description at finite temperatures and baryonic densities. Even if minor modifications are expected at $T = 0$ and $\rho_B = 0$, it should be relevant to see its effects elsewhere…

Firstly, at zero temperature and zero density, the masses of the nucleons are clearly underestimated when the p-dependence corrections are not applied, and overestimated otherwise. This remark is valid for the ISO and NISO parameter sets, TABLE VIII and TABLE IX. On the other hand, the proton-neutron mass inversion problem only concerns their scalar flavor components. Consequently, the inclusion of the axial one may intervene in this phenomenon. According to TABLE IX, this hypothesis is only verified with the p-dependence results. In fact, the evolution presented in this subsection has a great sensibility as regards to the used parameter set. The mass inversion can occur (or not) when minor modifications are applied to the parameter set. Concerning the other octet baryons, TABLE VIII, TABLE IX and Appendix E indicate that the gap leaded by this evolution can be rather limited. This is particularly the case with $\Xi^0$ and $\Xi^-$ with p-dependence corrections and the NISO parameter set.

**TABLE VIII** Masses of the octet baryons at $T = 0$ and $\rho_B = 0$, in MeV (ISO parameter set). The row "No p" does not include p-dependence corrections; the "p" row includes them.

|      | Nucleon | $\Lambda$ | $\Sigma$ | $\Xi$  |
|------|---------|-----------|----------|--------|
| No p | 893.0   | 1160.1    | 1192.7   | 1380.5 |
| p    | 1014.6  | 1208.6    | 1218.8   | 1417.1 |

**TABLE IX.** Same legend as **TABLE VIII** with the NISO parameter set.

|      | Proton | Neutron | $\Lambda$ | $\Sigma^+$ | $\Sigma^0$ | $\Sigma^-$ | $\Xi^0$ | $\Xi^-$ |
|------|--------|---------|-----------|------------|------------|------------|---------|---------|
| No p | 890.2  | 886.7   | 1120.8    | 1144.8     | 1146.2     | 1147.7     | 1307.9  | 1310.7  |
| p    | 1006.2 | 1007.4  | 1168.3    | 1173.0     | 1175.8     | 1177.6     | 1339.9  | 1342.4  |

Because of the inclusion of the axial flavor component, all the treated baryons admit several stable-unstable transitions according to the temperature or the density, Figure 21 and Figure 22. With the isospin symmetry, the nucleon has two transitions, and the other baryons three transitions. In these results, a transition is frequently characterized by a brutal increase in the baryon mass. This variation is sometimes so brutal that it looks like a discontinuity. Nevertheless, a brutal increase can also be observed when one of its diquark constituents becomes unstable. This is the case, e.g., for the nucleon at $T \approx 267$ MeV for the two plotted curves. Indeed, this corresponds to the Mott temperature of the scalar $[ud]$ diquark.

In addition, the use of the p-dependence corrections has an influence on the presence/absence of a brutal increase. According to the temperature, the Mott transitions of the nucleon, $\Sigma$ and $\Xi$ are smoother without p-dependence corrections. This affirmation stays true according to the density with the nucleon and $\Sigma$. On the other hand, in Figure 22, the p-dependence corrections prevent the brutal increase in mass found outside the stable-unstable baryon transition, notably for the nucleon and $\Lambda$.

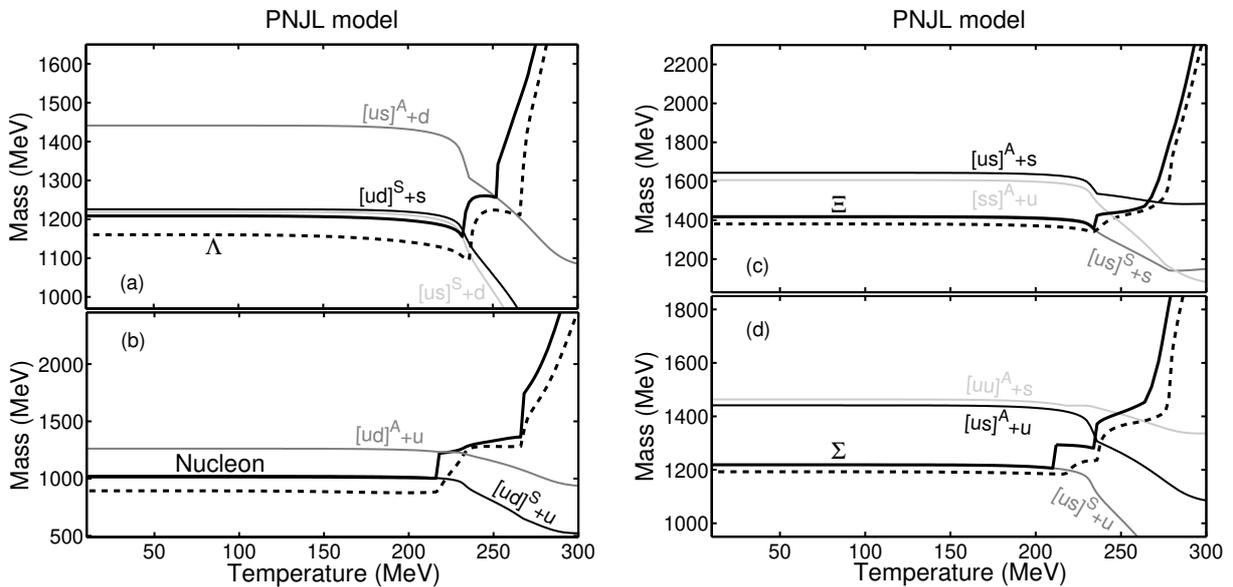

**Figure 21.** Masses of the (a) $\Lambda$, (b) nucleon, (c) $\Xi$ and (d) $\Sigma$ octet baryons function of the temperature, with scalar and axial flavor components, at zero density (ISO parameter set). The baryon solid curves consider p-dependence corrections; the dotted ones do not employ them.

Whatever the treated octet baryon, its first critical temperature and critical density are drastically reduced in comparison with the results of the other subsections. When the p-dependence is taken into account, the binding energies in the stable regimes are weak, whatever the baryon. Furthermore, the axial flavor component leads to higher masses at high temperatures. More precisely, when a baryon and its diquark constituents become unstable, the baryon mass strongly diverges. The maximum temperature of Figure 21 does not exceed $300$ MeV. This behavior is more limited according to the density, notably without the p-dependence corrections. On the one hand, the nucleon mass increases rather slowly. On the other hand, a strong increase is visible for $\Xi$.

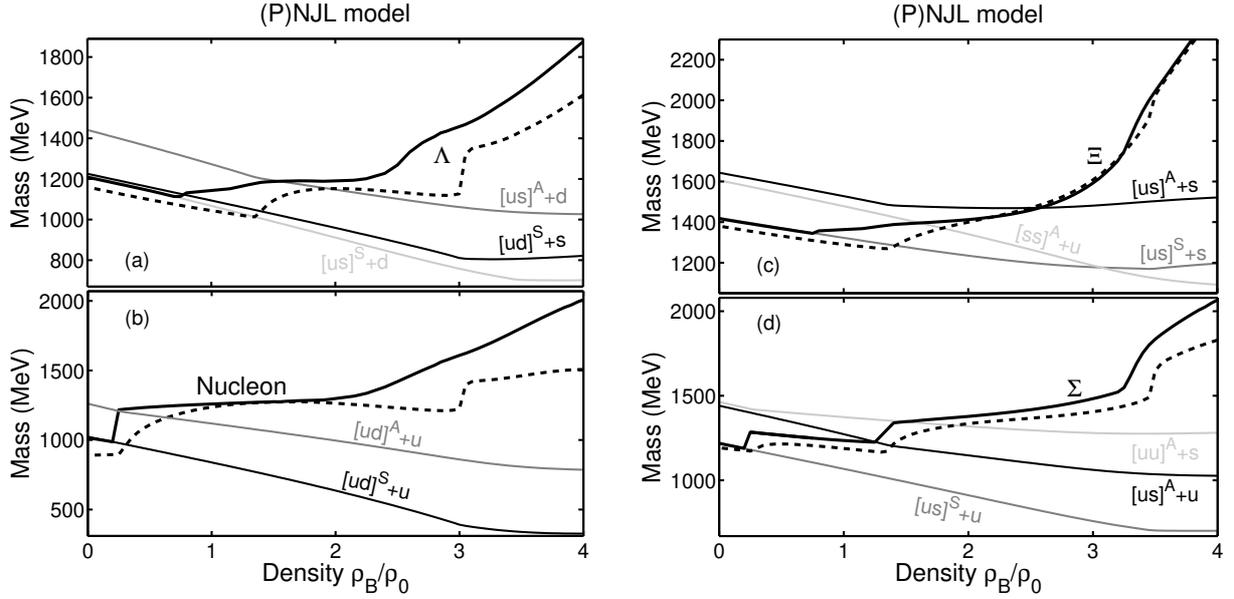

**Figure 22.** Masses of the (a) $\Lambda$, (b) nucleon, (c) $\Xi$ and (d) $\Sigma$ octet baryons function of the baryonic density, with scalar and axial flavor components, at $T=10$ MeV (ISO parameter set). The baryon solid curves include p-dependence corrections; the dotted ones do not consider them.

The quasi-constant masses according to the temperature, in the baryon stable regime, are in agreement with previous results, from this paper or from other ones, e.g., [53, 55, 56, 62, 63]. In contrast, the (P)NJL literature describes a regular decrease in the nucleon mass according to the baryonic density, at least in its stable regime [47]. In the Figure 22(b), the density range in which this statement is verified seems too reduced. This disagreement is explained by the great sensibility of the axial $[ud]$ diquark according to $\rho_B$. This constitutes the main limitation of this evolution. These remarks can be extended to the other octet baryons, even if the axial $[qs]$ diquarks are less sensitive to the baryonic density.

### E. Results involving the NJL diquark propagators.

Even if the QFT and NJL diquark propagators have the same poles, the use of the NJL ones to describe baryons appears to be *terra incognita*. Indeed, far from the poles, their behavior is rather different, Figure 7. This should have non-trivial consequences on the results. Throughout this subsection, the presented calculations include p-dependence corrections. In addition, the octet baryons are exclusively described via their scalar flavor component.

**TABLE X.** Masses (in MeV) of the octet and decuplet baryons at $T=0$ and $\rho_B=0$, including p-dependence corrections and with an NJL diquark propagator, with the ISO parameter set. The "S.A." row stands for static approximation.

|        | Nucleon | $\Lambda$ | $\Sigma$ | $\Xi$ | $\Delta$ | $\Sigma^*$ | $\Xi^*$ | $\Omega$ |
|--------|---------|-----------|----------|-------|----------|------------|---------|----------|
| S.A.   | 944.1   | 1202.5    | 1217.0   | 1423.6 | 1154.8  | 1416.3     | 1643.5  | 1809.7   |
| No S.A.| 974.1   | ✘         | 1154.7   | 1421.6 | 1258.8  | ✘          | ✘       | 1808.6   |

As a whole, the agreement between the found masses at zero temperature and density and the experimental ones stays rather similar in comparison to the previous evolutions, TABLE X, TABLE XI and Appendix E. Nevertheless, with the NISO parameter set, the mass of $\Sigma^\pm$ is clearly underestimated without the static approximation. Concerning the proton-neutron mass inversion evoked above, this problem is only solved here when this approximation is not considered, TABLE XI. Consequently, whatever the used diquark propagator, i.e. QFT or NJL, the abandon of the static approximation constitutes the main solution to fix this anomaly. Concerning the decuplet baryons, the agreement with the experimental data is rather interesting. Indeed, the QFT and NJL axial diquark propagators present notable differences, Figure 7(b). In addition, as done in the Subsec. V.C, the effect of the static approximation on the masses can be commented. This one is very reduced for $\Omega$, which involves a heavy exchanged quark. In contrast, its effect is stronger with $\Delta$, which involves light ones. With the octet baryons, this reasoning does not work with $\Sigma$ and the neutron.

**TABLE XI.** Same legend as **TABLE X** with the NISO parameter set.

|  | Proton | Neutron | $\Lambda$ | $\Sigma^+$ | $\Sigma^0$ | $\Sigma^-$ | $\Xi^0$ | $\Xi^-$ |
|---|---|---|---|---|---|---|---|---|
| S.A. | 933.0 | 930.3 | 1165.5 | 1173.9 | 1175.9 | 1177.9 | 1349.9 | 1353.4 |
| No S.A. | 927.7 | 931.1 | ✘ | 1098.6 | ✘ | 1103.6 | 1340.6 | 1343.0 |

|  | $\Delta^{++}$ | $\Delta^+$ | $\Delta^0$ | $\Delta^-$ | $\Sigma^{*+}$ | $\Sigma^{*0}$ | $\Sigma^{*-}$ | $\Xi^{*0}$ | $\Xi^{*-}$ | $\Omega^-$ |
|---|---|---|---|---|---|---|---|---|---|---|
| S.A. | 1143.2 | 1145.1 | 1147.5 | 1150.4 | 1379.2 | 1382.3 | 1385.7 | 1567.1 | 1568.9 | 1707.3 |
| No S.A. | 1243.0 | ✘ | ✘ | 1251.7 | ✘ | ✘ | ✘ | ✘ | ✘ | 1706.0 |

When the QFT diquark propagators are replaced by NJL ones in the baryon modeling, it was shown in Subsec. IV.E that the diquark coupling constants $g_{qqe}$ disappear in the expression of Eq. (46). Therefore, their variations cannot influence the masses of the baryons. This explains why the increase in these masses after the Mott transition is more regular in Figure 23 and Figure 24 than in the previous results of this paper. This is particularity true for the nucleon according to the temperature, for both curves. Its variations are similar to the ones observed at the Mott transition of a meson or a diquark, Figure 11(a). On the other hand, in Figure 23(c), the two curves of $\Omega$ are very close, until an unexpected divergence at high temperatures. This last behavior should be interpreted as an artifact.

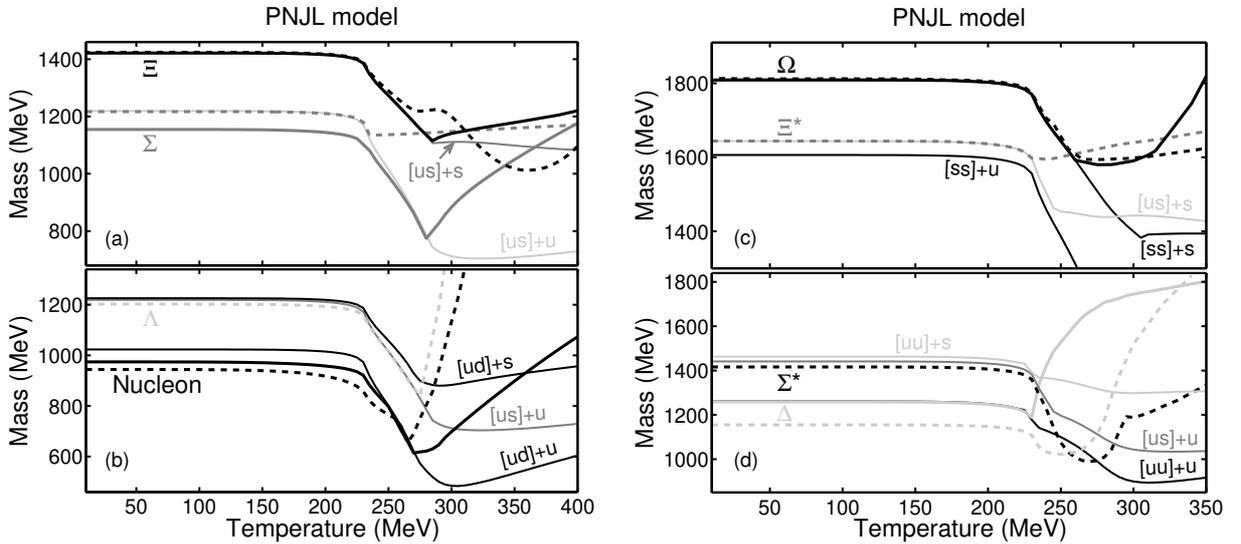

Figure 23. Masses of the (a) $\Sigma$, $\Xi$, (b) nucleon, $\Lambda$, (c) $\Xi^*$, $\Omega$ and (d) $\Sigma^*$, $\Delta$ baryons function of the temperature, at zero density (ISO parameter set). The p-dependence corrections and NJL diquark propagators are considered. The static approximation is used in the baryon dotted curves. The solid ones do not employ it.

With the static approximation, the mass variations of the $\Xi$ in the Figure 23(a) recalls those found in the previous graphs with this baryon. This concerns the minimum visible when its mass is lower than the ones of $[us]+s$. The Mott temperature of the scalar $[us]$ diquark is located at a lower temperature (278 MeV). Therefore, this may go on the sense of the Borromean state previously mentioned [57]. However, this behavior does not survive without the static approximation. Consequently, this leads to consider it as an artifact due to this simplification. This conclusion can also concern the behavior of $\Xi^*$ in Figure 24(c).

According to the density, Figure 24, the $\Omega$ curves and the one of $[ss]+s$ are superimposed. This translates a very low binding energy. A similar behavior is also visible with $\Xi$ and the curve $[us]+s$ without the static approximation. In fact, except for $\Sigma$ and $\Omega$, the baryons masses increase strongly at high densities when this approximation is applied. Concerning the nucleon, the remarks performed in Subsec. V.C about its mass variations at low densities stay valid: the decrease is more rapid with the static approximation. At high densities, the brutal increase when $\rho_B > 3\rho_0$ is due to the instability of the scalar $[ud]$ diquark. This recalls the observations done in Subsec. V.D. In contrast, the mass increase stays moderate in this regime without the static approximation.

This approximation has also a strong influence on the critical temperatures and critical densities of the baryons. In contrast, without it, they are close to the ones found in Subsecs. V.A to V.C with the QFT diquark propagators. In this case, this confirms that the critical temperatures/densities of the baryons are lower than or equal to the ones of their diquark constituents.

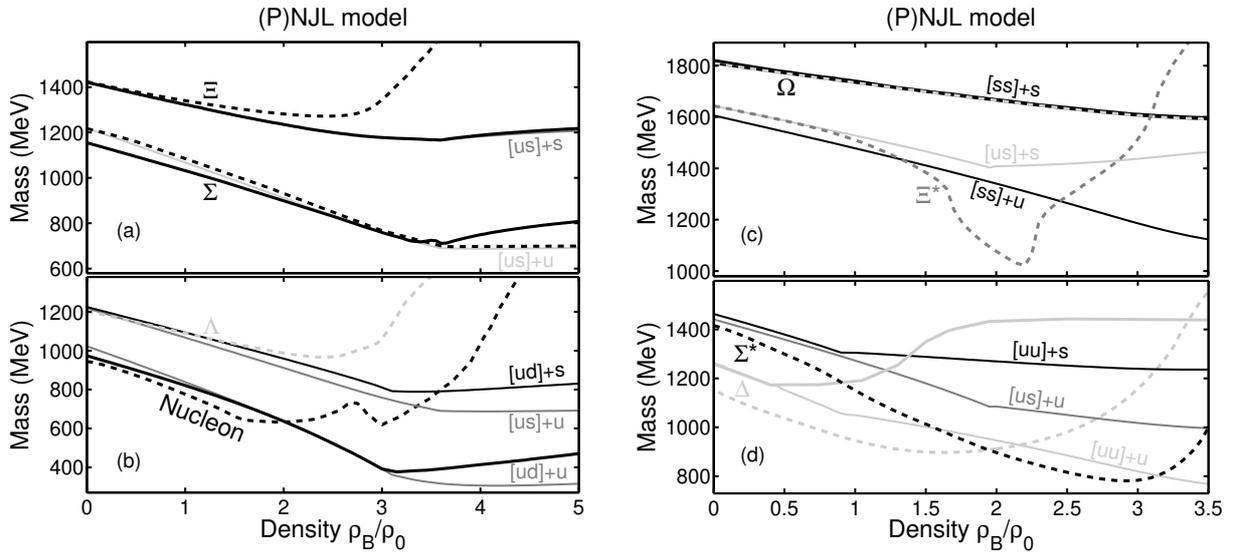

**Figure 24. Same legend as in Figure 23, but according to the baryonic density, at $T=10$ MeV.**

The Figure 25 illustrates the Subsubsec. IV.E.2, related to the poles of the NJL diquark propagators. The results presented Figure 23 and Figure 24 consider $S_D^{NJL}(|p_0|)$. This wants to say that the poles are estimated with the diquark mass. If the absolute values are not used, the negative pole is then obtained with the antidiquark mass. Obviously, there is no difference at $\rho_B = 0$. In addition, $\Omega$ is not impacted at finite baryonic density. Concerning the baryons formed by a scalar or axial $[qs]$ diquark, Figure 25(a, c, d, e), the differences stay rather modest. This remark includes $\Lambda$, even if this baryon is also made by a scalar $[ud]$ diquark. However, the use of $S_D^{NJL}(p_0)$ with $\Xi^*$ strongly reduces the depth of the local minimum at $\rho_B \approx 2.2\,\rho_0$. Another consequence is the interruption of the curves, because the antidiquark or the baryon cannot be modeled for higher densities. As a whole, the observed gap is more important when the static approximation is not included. Moreover, the nucleon is the only baryon modeled exclusively via a scalar $[ud]$ diquark. This explains the notable difference between its curves found via $S_D^{NJL}(|p_0|)$ and $S_D^{NJL}(p_0)$. The axial $\overline{[ud]}$ antidiquark cannot be modeled at positive baryonic density. Therefore, $\Delta$ is missing in Figure 25.

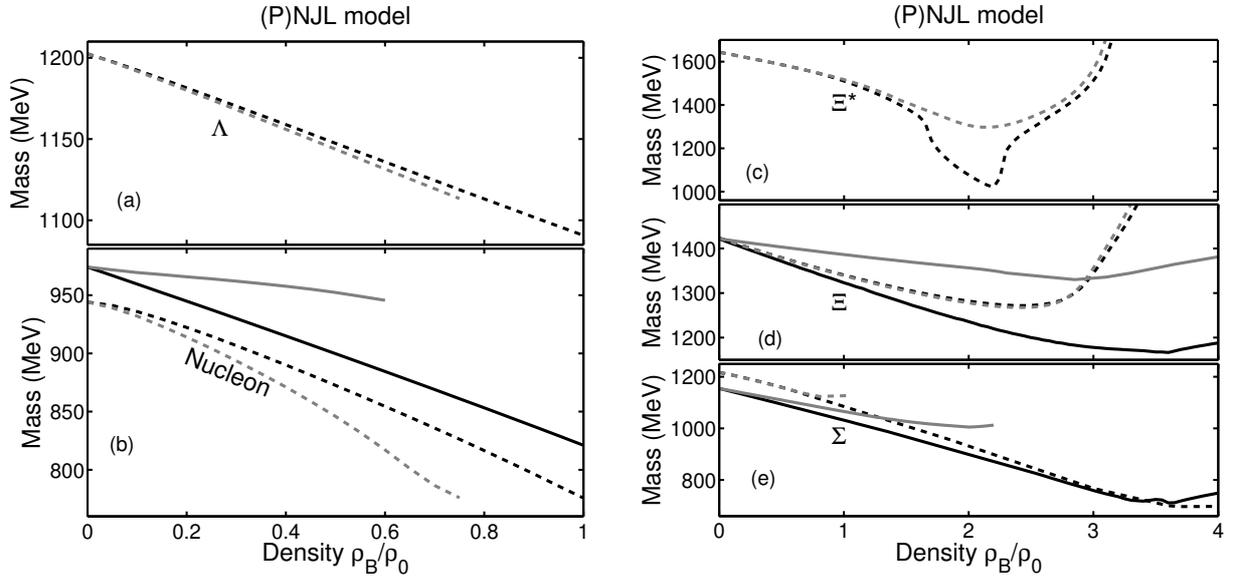

**Figure 25. Differences between the two solutions proposed in Subsubsec. IV.E.2 for the (a) $\Lambda$, (b) nucleon, (c) $\Xi^*$, (d) $\Xi$ and (e) $\Sigma$ baryons (ISO parameter set). In the back curves, $|p_0|$ is used in the $S_D^{NJL}$ diquark propagators, as in the previous graphs of this subsection. The gray curves consider $p_0$ as argument of $S_D^{NJL}$. The dotted curves use the static approximation. The solid ones do not employ it.**

To conclude this subsection, the coupling constant of the nucleon is studied, for each evolution described above in this section, Figure 26. The results for the other baryons are qualitatively comparable. Firstly, according to the temperature, the behavior of the coupling constant is similar whatever the evolution. It is almost constant is the stable regime. At low temperatures and zero density, one reads an initial value of 5.4 in Figure 26(a), 6.1 and 7.2 in (b), 3.9 and 6.1 in (c), and 4.3 in (d). At the Mott transition, it rapidly falls to zero. Then, in the unstable regime, it grows towards high values. These variations look like the ones of the mesons/diquarks coupling constants, e.g., in Refs. [42, 53, 55, 56]. However, the values reached at high temperatures are largely stronger with the baryons. When the axial flavor component is taken into account, the growing seems to be a divergence at $T > 300$ MeV. This seems to be the only difference with the other evolutions described in this figure.

According to the baryonic density, Figure 27, the previous observations stay valid when the static approximation is employed. Otherwise, a regular decrease is visible in the stable regime, with the QFT and NJL diquark propagators. Moreover, the variations of the coupling constant are very different when the axial flavor component is included. Indeed, it recalls the nucleon mass in Figure 22(b), i.e. strong increases at the level of the stable-unstable transitions.

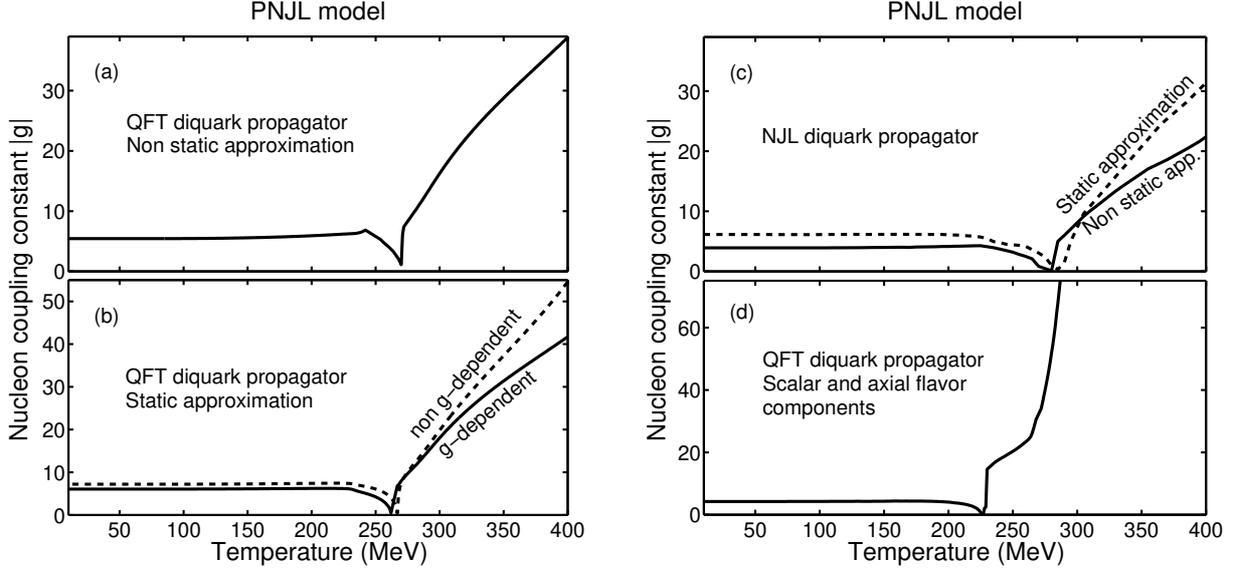

**Figure 26. Coupling constant of the nucleon** $g_{[ud]q}$ **function of the temperature, at zero density, with the ISO parameter set: (a) with a QFT diquark propagator and without the static approximation, (b) with the static approximation, (d) including the axial flavor component and (c) with an NJL diquark propagator. Only the dotted curve in (b) does not apply p-dependence corrections.**

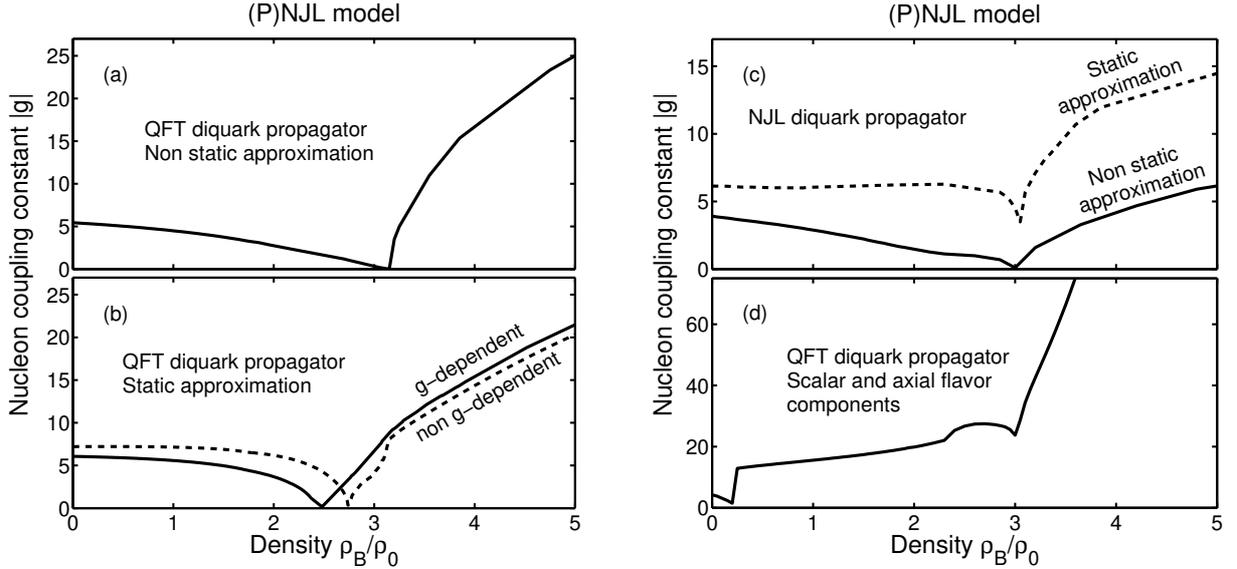

**Figure 27. Same legend as in Figure 26, but with** $g_{[ud]q}$ **as a function of the baryonic density, at** $T = 10$ MeV.

### F. Influence of the color superconductivity.

In the NJL model, the color-superconductivity may intervene at low temperatures when $\rho_B > 0.3\rho_0$, via the 2SC phase. As argued in [38], such a so low density can be explained by the lack of a real quark confinement in the NJL approach. This Ref. also discusses the physical meaning of low baryonic densities, via the quark droplets picture proposed in [95]. Anyway, for example at the standard nuclear density $\rho_0$, a diquark should be confined inside a baryon instead of forming a diquark condensate. However, Figure 28 allows studying the influence of the color-superconductivity on a nucleon. In this investigation, one neglects the mix between the three scalar diquarks $[ud]^{rg}$, $[ud]^{rb}$ and $[ud]^{gb}$, as in [39]. So, one imagines a nucleon made with $[ud]^{rg}$, another with $[ud]^{rb}$, and a third with $[ud]^{gb}$. Obviously, when $\rho_B < 0.3\rho_0$, their masses are degenerate. Otherwise, a partial lifting in degeneracy occurs between the nucleon made with $[ud]^{rg}$ ("with $[ud]^{rg}$" curve) and the two others ("with $[ud]^{rb\,gb}$" curve).

In Figure 28, the "NQ phase" curve refers to a nucleon modeled without the color-superconductivity. For growing densities until $3\rho_0$, the mass gaps between this one and the three mentioned nucleons increase progressively, but stay rather modest. In the vicinity of $\rho_0$, it may be considered as relatively negligible. This confirms that the omission of the color-superconductivity in the baryon description could be considered as an admissible approximation in such cases. Nevertheless, this affirmation becomes incorrect in the cold and very dense matter, i.e. in the conditions of compact stars where $\rho_B$ can exceed several times the standard nuclear density.

Moreover, the nucleon made with $[ud]^{rg}$ is the heaviest, with or without the static approximation. Since a heavy particle can be synonymous of an unstable one, these results seem to be consistent. Indeed, when the 2SC phase appears, the $[ud]^{rg}$ diquarks tend to create the corresponding diquark condensate. If such a diquark is inside a nucleon, this one should undergo a diminution of its stability. In other words, even in the absence of confinement, a competition occurs in this description between the nucleon cohesion and the diquark condensate. Such a result may be compared to [96], in which a three quark clustering is used to model baryons. In addition, in the 2SC phase, the stable-unstable transition of $[ud]^{rg}$ is a crossover according to the density. Its associated nucleon seems to have the same type of transition. Concerning the $[ud]^{rb}$ and $[ud]^{gb}$, they are indirectly impacted by the 2SC phase [39]. Therefore, the effects on the two nucleons formed by these diquarks are weak. Their curve stays close to the "NQ phase" one.

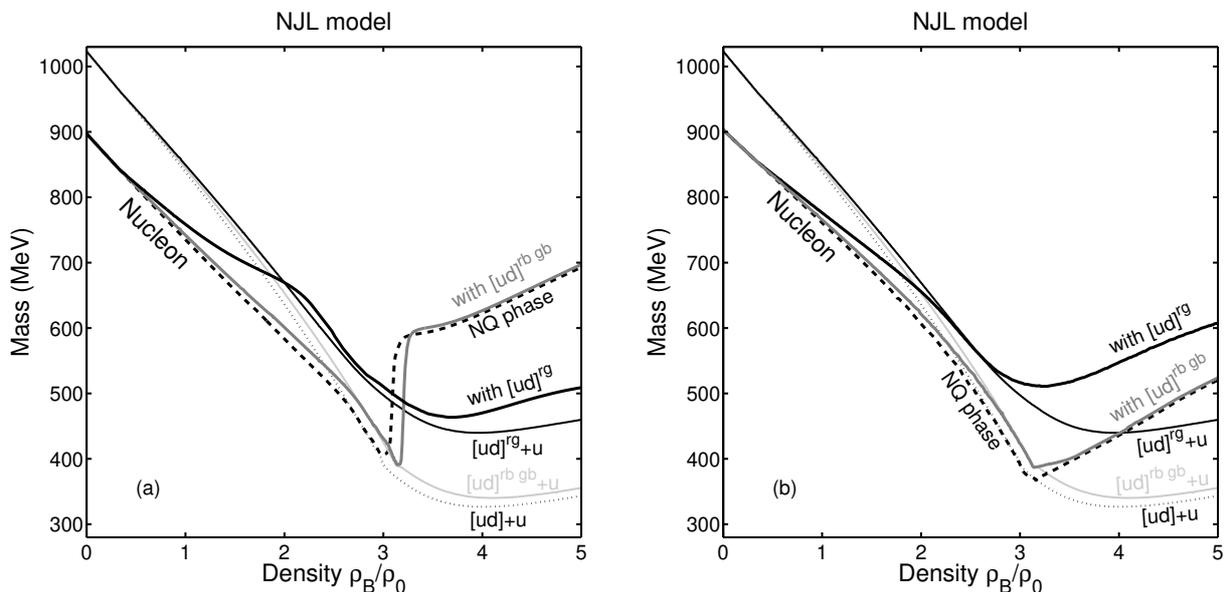

**Figure 28.** Mass of the nucleon according to the baryonic density, (a) with or (b) without the static approximation, at $T = 10$ MeV **(ISO parameter set). Calculations not including color superconductivity correspond to the dotted curves.**

When the density exceeds $3\rho_0$, this scenario is not modified by the results found without the static approximation, Figure 28(b). Nevertheless, this is not the case when this approximation is applied,

Figure 28(a). In this graph, the increase in the mass of the nucleon made with $[ud]^{rg}$ seems to be more reduced, in contrast with the strong increase observed with the others. This behavior appears to be another (unwanted) consequence of the static approximation on the nucleons.

## VI. CONCLUSION

Throughout this paper, a modeling of the $SU(3)_f$ baryons in the Polyakov, Nambu and Jona-Lasinio model is detailed, via the resolution of the Bether-Salpeter equation, in the framework of the quark-diquark picture. This work mainly focuses on the study of their masses according to the temperature and the baryonic density. This document proposes several improvements of this description. On the one hand, this gives the occasion to avoid some of the approximations frequently employed in this framework. On the other hand, this permits to estimate concretely the effects of these approximations on the results and on their reliability.

Firstly, this concerns the possibility to explore the unstable regime of the baryons. This allows extending the domain in which they can be studied, i.e. the whole $T, \rho_B$ plane in which the PNJL model is expected to be reliable. More precisely, extreme densities involve high chemical potentials. A chemical potential too close of the cutoff could lead to unreliable results in the numerical integrations [38], e.g., in the Fermi-Dirac statistics. In addition, according to the temperature, the inclusion of the Polyakov loop prevents to explore temperatures greater than $2.5T_0$ (675 MeV in this work), i.e. when the contribution of the transverse gluons is no longer negligible [29, 34, 36]. Concerning the found results, the nucleon becomes unstable at a Mott temperature lower than or equal to the one of its diquark constituent. Consequently, this does not go on the sense of the Borromean state reported in Refs. [57, 60, 97]. However, when the static approximation is used, the behavior of $\Xi$ may suggest further investigations.

Furthermore, the approximation that considers diquarks constituent at rest is sometimes qualified as "crude" upon a physical point of view. This motivates to neutralize it by including the momentum dependence of the masses and coupling constants of these diquarks into the modeling. Concerning the effects on the results, they appear to be moderate. The behavior of the modeled baryons stays qualitatively similar. For example, the critical temperatures and critical densities of the baryons are almost unchanged, except for the critical densities of the decuplet baryons. Nevertheless, these corrections are not useless and present undeniable advantages. They allow increasing the accuracy of the nucleons masses at zero temperature and zero density, which are underestimated otherwise. In addition, they are fully compatible with the other evolutions mentioned in this work, except in the color-superconducting regime.

Moreover, performing the calculations without the static approximation is also an important aspect of this study. Indeed, it has been confirmed numerically that this approximation is responsible of an unphysical result, i.e. a proton heavier than a neutron. Its abandon leads to correct this anomaly, without any modification of the used parameter set. This affirmation has been verified whatever the used diquark propagator, i.e. the structureless or the NJL one. It is somehow possible to solve this problem in another way, e.g., via a tuning of the parameter set or via the inclusion of the axial flavor component of the nucleons. However, the abandon of the static approximation seems to be the most effective one. This approximation has also non-negligible effects when the NJL diquark propagators are employed. It acts for example on the critical temperatures and densities. Moreover, it has been observed another consequence of this simplification on the nucleon behavior at high densities, in a color-superconducting regime. In addition, the accuracy of the results at zero temperature and zero density are interesting when the static approximation is not taken into account. Nevertheless, this improvement is only possible for baryons modeled via one unique state $D+q$, i.e. not for baryons like $\Lambda$. Therefore, a future development of this work should study the extension of this evolution to such baryons.

In parallel, another possibility of improvement has concerned the inclusion of the axial flavor component of the octet baryons. As expected, it has been observed a modest influence at zero temperature and zero density. Nonetheless, the main difficulty leaded by this component concerns the baryons instability encountered at finite densities. Indeed, the baryons become unstable very rapidly when the baryonic density increases, as the axial diquarks made with light quarks, i.e. $[ud]$ and $[qs]$. Consequently, a possible evolution of this description could be to stabilize the baryons in this unstable regime. This could be done by the inclusion of weighting factors, in order to decrease the contribution of the axial diquarks. These factors could be then compared with the numerical coefficients used in [66]. Another possibility is to imagine a Borromean description, i.e., to obtain a stable baryon via unstable diquarks.

The replacement of the structureless diquark propagator by an NJL one has constituted the next step in this paper. An interesting feature of this evolution is the possibility to include the former ones, except the axial flavor component of the octet baryons. Concerning the found results, the main difference is located at the stable-unstable transitions. Indeed, the diquark coupling constant has a strong influence in this regime when the structureless diquark propagator is employed. In contrast, the use of the NJL diquark propagator suppresses it in the equation to be solved. This leads to a more regular increase in the mass, in the baryon unstable regime. Except for this aspect, the results are qualitatively and quantitatively comparable to the previous ones described in this document. Consequently, the use of the structureless diquark propagator, with momentum dependence corrections, stays a possible option to build a PNJL baryon.

This remark is motivated by the fact that the computation time could be a criterion in the choice of the applied evolution, in addition of its physical interest. Firstly, the use of the complex numbers, to explore the baryon unstable regime, can be faster than the native version of [55, 56]. The abandon of the static approximation does not lead to a significant increase in the computation time. In contrast, the inclusion of the axial flavor component of the octet baryons requires handling several diquarks, and so the calculations are slower. The momentum dependence corrections also involve more resources. The version that uses the NJL diquark propagator with momentum dependence is by far the slowest, but stays accessible to modern computers.

Whatever the described evolution, the accuracy of the baryons masses at zero temperature and zero density stays correct when the isospin symmetry is not considered. However, it is not excellent, in comparison with the masses of the pseudo-scalar mesons found in [38, 55] with the same parameter set (NISO). Consequently, a future evolution should concern an enhancement of this ability to reproduce the experimental masses. The challenge could be to continue to treat baryons and mesons with the same parameter set. Anyway, it could also be interesting to test other evolutions, like the $\Lambda-\Sigma^0$ coupling suggested in [62] when the isospin symmetry is not applied. On the other hand, the improvements listed in this paper focus on the baryons. Therefore, it should be relevant to investigate the effects on these particles of various PNJL improvements available in the literature that act directly on the quarks. One can mention, e.g., the eight quark interactions [98] or the EPNJL approach at finite temperatures/densities [40, 41, 89].

Moreover, a method to evaluate the baryon coupling constant has been proposed. Even if this approach is based upon an approximation via the use of the trace, it allows the calculation whatever the presented evolution. This approach should be improved, in particular for the spin-3/2 baryons. Nevertheless, this opens the door to several applications and other evolutions. This could concern, e.g., the study of reactions involving baryons [64, 65]. More precisely, the consequences of the listed evolutions on the cross sections could be investigated. An extension of these calculations to the decuplet baryons could also be relevant. Indeed, this could allow modeling the creation of such particles in the heavy ion colliders. Moreover, the production of $\Delta$ baryons in neutron stars is particularly investigated [99].

In addition, the quark and diquark modeling performed in the color-superconductivity regime in, respectively, [38] and [39] has permitted to propose a simple modeling of the nucleon submitted to the influence of the 2SC phase. As expected, the found results allow observing a competition between the binding of the $[ud]-q$ couple inside a nucleon and the formation of the $[ud]$ condensate, even in the absence of confinement. The effect on the nucleon increases slowly and progressively with the baryonic density, as soon as the 2SC begins to occur. If this effect stays limited at the vicinity of the standard nuclear density, it becomes important at moderate and high baryonic densities.

Nevertheless, this approach suffers from several limitations. Firstly, it must only be understood as a first step to study baryons in these conditions. A complete modeling of the PNJL baryons in the color-superconductivity regime is still to be done in the $T,\rho_B$ plane, in particular in the quark-diquark picture. In addition, certain of the improvements listed in this paper have not been applied in this color-superconducting description, like the momentum dependence corrections. Therefore, it appears useful to complete the diquarks description of [39] in order to include axial diquarks and this momentum dependence.

These latter updates seem to be non-trivial, but worthwhile. Indeed, such investigations may present important applications to the study of neutron stars. In these conditions, the color superconductivity is expected to intervene in a non-negligible way [100], e.g., via the Meissner effect [101]. Clearly, notably in the case of the magnetars [102, 103], the magnetic field constitutes an additional external parameter that should be taken into account in the modeling [104, 105].

**APPENDIX A: DESCRIPTION OF THE OCTET AND DECUPLET BARYONS.**

*1. Proton and neutron.*

The flavor wavefunction of the proton is written as:

$$|p\rangle = \frac{1}{\sqrt{2}}\begin{pmatrix} u[ud] \\ \sqrt{1/3}\,u[ud] - \sqrt{2/3}\,d[uu] \end{pmatrix}. \tag{74}$$

The first row translates the scalar flavor component, Figure 29, with a scalar diquark. The second row indicates the axial flavor component, Figure 30, involving axial diquarks [51, 106].

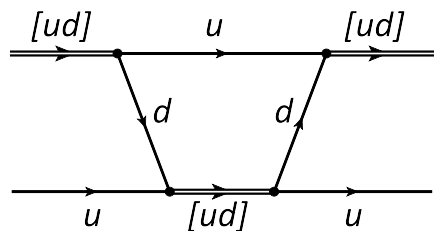

**Figure 29. Scalar flavor component of the proton.**

The black dots translate vertices with scalar interactions. The scalar flavor component of the neutron is found with the exchange $u \leftrightarrow d$.

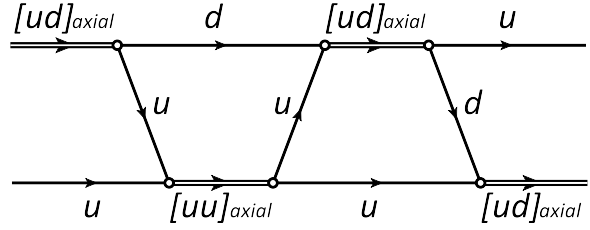

**Figure 30. Axial flavor component of the proton.**

The white dots translate vertices with axial interactions. The axial flavor component of the neutron is found with the exchange $u \leftrightarrow d$. So, the flavor wavefunction of the neutron is expressed as:

$$|n\rangle = \frac{1}{\sqrt{2}} \begin{pmatrix} d[ud] \\ -\sqrt{1/3}\, d[ud] + \sqrt{2/3}\, u[dd] \end{pmatrix}. \tag{75}$$

### 2. Baryon $\Lambda$.

The flavor wavefunction of $\Lambda$ is [106]:

$$|\Lambda\rangle = \frac{1}{\sqrt{2}} \begin{pmatrix} \frac{1}{\sqrt{6}}(2s[ud] - u[ds] + d[us]) \\ \frac{1}{\sqrt{2}}(u[ds] - d[us]) \end{pmatrix}, \tag{76}$$

with the same writing convention as with the nucleons. Different signs are observed in [55, 56, 62, 63] for the scalar component. This has an incidence on the writing of (45), but not on the results. The scalar and axial flavor components are illustrated, respectively, by Figure 31 and Figure 32.

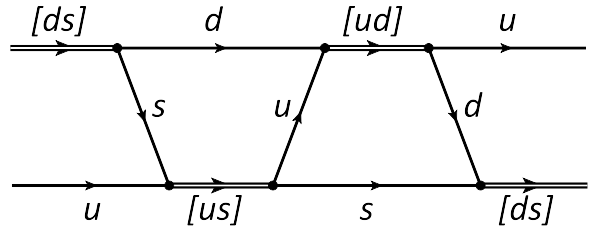

**Figure 31. Scalar flavor component of $\Lambda$.**

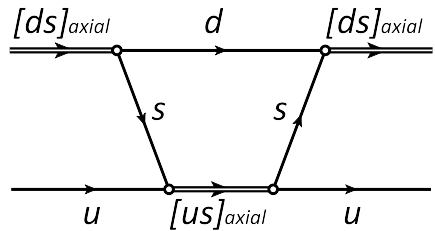

**Figure 32. Axial flavor component of $\Lambda$.**

### 3. Baryons $\Sigma^0$, $\Sigma^\pm$, $\Sigma^{*0}$ and $\Sigma^{*\pm}$.

The flavor wavefunction of $\Sigma^0$ is written on the form [56, 63, 106]:

$$|\Sigma^0\rangle = \frac{1}{\sqrt{2}} \begin{pmatrix} \frac{1}{\sqrt{2}}(u[ds] + d[us]) \\ \frac{1}{\sqrt{6}}(u[ds] + d[us] - 2s[ud]) \end{pmatrix}. \tag{77}$$

Its scalar and axial components are visible, respectively, in Figure 33 and Figure 34. The sign difference in the scalar component found in the quoted references has no influence on the results.

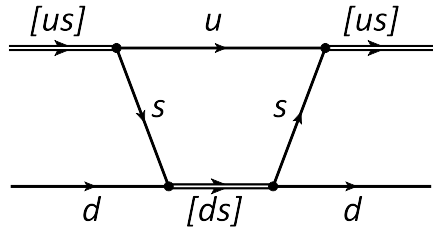

**Figure 33. Scalar flavor component of $\Sigma^0$.**

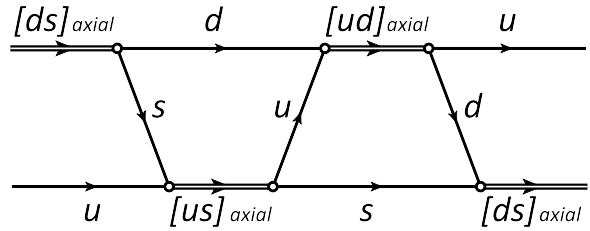

**Figure 34. Axial flavor component of $\Sigma^0$.**

The flavor wavefunction of the decuplet baryon $\Sigma^{*0}$ is expressed with axial diquarks as:

$$\left|\Sigma^{*0}\right\rangle = \frac{1}{\sqrt{3}}\left(u[ds] + d[us] + s[ud]\right) . \tag{78}$$

The Figure 34 can also describe this flavor component.

The flavor wavefunction of $\Sigma^+$ is [51, 106]

$$\left|\Sigma^+\right\rangle = \frac{1}{\sqrt{2}}\begin{pmatrix} u[us] \\ \sqrt{1/3}\,u[us] - \sqrt{2/3}\,s[uu] \end{pmatrix} . \tag{79}$$

Its scalar and axial components are presented Figure 35 and Figure 36.

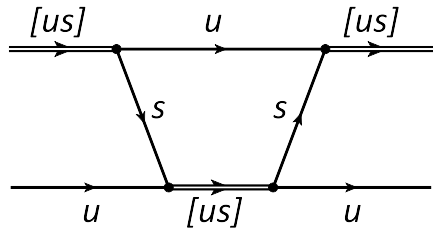

**Figure 35. Scalar flavor component of $\Sigma^+$.**

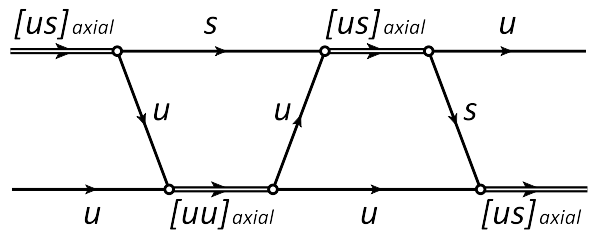

**Figure 36. Axial flavor component of $\Sigma^+$.**

The Figure 36 can also describe the flavor component of $\Sigma^{*+}$, written as [63, 106]:

$$\left|\Sigma^{*+}\right\rangle = \sqrt{1/3}\,s[uu] + \sqrt{2/3}\,u[us]. \tag{80}$$

The components of $\Sigma^-$ and $\Sigma^{*-}$ are obtained with the replacement $u \to d$, with leads to:

$$\left|\Sigma^-\right\rangle = \frac{1}{\sqrt{2}}\begin{pmatrix} d[ds] \\ \sqrt{1/3}\,d[ds] - \sqrt{2/3}\,s[dd] \end{pmatrix}, \tag{81}$$

$$\left|\Sigma^{*-}\right\rangle = \sqrt{1/3}\,s[dd] + \sqrt{2/3}\,d[ds]. \tag{82}$$

### 4. Baryons $\Xi^0$, $\Xi^-$, $\Xi^{*0}$ and $\Xi^{*-}$.

The flavor wavefunction of $\Xi^0$ is [106]:

$$\left|\Xi^0\right\rangle = \frac{1}{\sqrt{2}}\begin{pmatrix} s[us] \\ \sqrt{1/3}\,s[us] - \sqrt{2/3}\,u[ss] \end{pmatrix}, \tag{83}$$

and is illustrated Figure 37 and Figure 38.

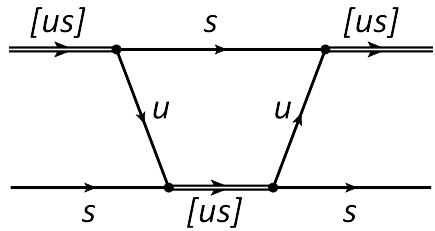

**Figure 37. Scalar flavor component of $\Xi^0$.**

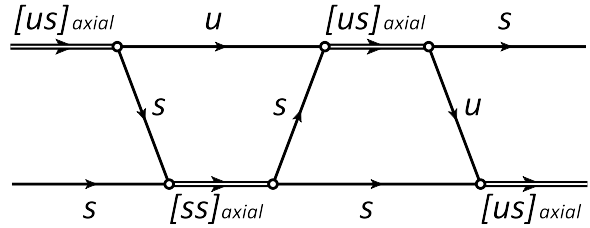

**Figure 38. Axial flavor component of $\Xi^0$.**

The Figure 38 is also usable to describe $\Xi^{*0}$, whose flavor wavefunction is written as [63, 106]:

$$\left|\Xi^{*0}\right\rangle = \sqrt{1/3}\,u[ss] + \sqrt{2/3}\,s[us] \tag{84}$$

The components of $\Xi^-$ and $\Xi^{*-}$ are found with the replacement $u \to d$ in, (83), (84), Figure 37 and Figure 38, i.e.

$$\left|\Xi^-\right\rangle = \frac{1}{\sqrt{2}}\begin{pmatrix} s[ds] \\ \sqrt{1/3}\,s[ds] - \sqrt{2/3}\,d[ss] \end{pmatrix}, \tag{85}$$

$$\left|\Xi^{*-}\right\rangle = \sqrt{1/3}\,d[ss] + \sqrt{2/3}\,s[ds]. \tag{86}$$

### 5. Baryons $\Delta^{++}$, $\Delta^+$, $\Delta^0$ and $\Delta^-$.

The flavor wavefunctions of the $\Delta$ baryons are [51, 63, 106]:

$$\begin{aligned}
\left|\Delta^{++}\right\rangle &= u[uu] \\
\left|\Delta^{+}\right\rangle &= \sqrt{1/3}\,d[uu] + \sqrt{2/3}\,u[ud] \\
\left|\Delta^{0}\right\rangle &= \sqrt{1/3}\,u[dd] + \sqrt{2/3}\,d[ud] \\
\left|\Delta^{-}\right\rangle &= d[dd]
\end{aligned} \qquad (87)$$

where all the mentioned diquarks are axial ones. The Figure 39 and Figure 40 shows the flavor component of, respectively, $\Delta^{++}$ and $\Delta^{+}$.

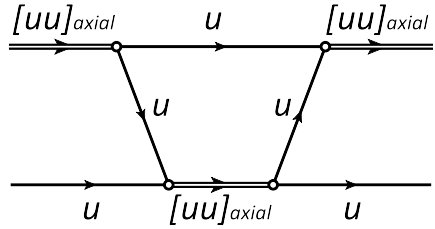

**Figure 39. Axial flavor component of $\Delta^{++}$.**

The $\Delta^{-}$ is found with the substitution $u \to d$.

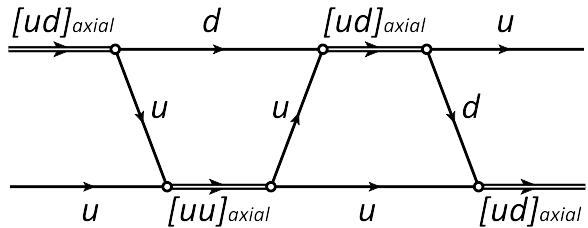

**Figure 40. Axial flavor component of $\Delta^{+}$.**

The $\Delta^{0}$ is obtained with the exchange $u \leftrightarrow d$ in the quarks and diquarks.

### *6. Baryon $\Omega^{-}$.*

The flavor wavefunction of $\Omega^{-}$ is expressed via the axial $[ss]$ diquark [55, 56, 62, 63, 106]:

$$\left|\Omega^{-}\right\rangle = s[ss], \qquad (88)$$

as illustrated via Figure 41.

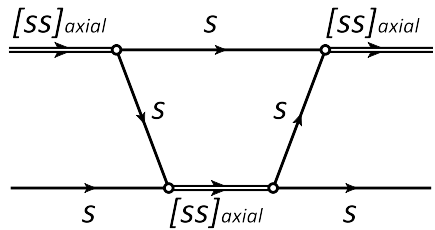

**Figure 41. Axial flavor component of $\Omega^{-}$.**

# APPENDIX B: CALCULATIONS OF THE BARYON LOOP FUNCTION.

## 1. Matsubara sums.

A complex function $\psi: \mathbb{C} \to \mathbb{C}$ is considered. One admits hereafter that $\psi$ has only simple poles $z_e$, and the $i\omega_n$ never belong to them. The evaluation of a Matsubara sum $\sum_n \psi(i\omega_n)$ upon the Matsubara frequencies $i\omega_n$ is performed via the two formulas:

$$\frac{1}{\beta}\sum_n \psi\left(i\omega_n^{BE}\right) = -\sum_e f^{BE}(z_e) Res(\psi, z_e) , \tag{89}$$

if $\omega_n$ is bosonic, i.e. $\omega_n^{BE} = 2n\pi/\beta$, where $n \in \mathbb{Z}$, and:

$$\frac{1}{\beta}\sum_n \psi\left(i\omega_n^{FD}\right) = \sum_e f^{FD}(z_e) Res(\psi, z_e) , \tag{90}$$

if $\omega_n$ is fermionic, i.e. $\omega_n^{FD} = (2n+1)\pi/\beta$. Also, $f^{BE}(z) = \dfrac{1}{e^{\beta z}-1}$ and $f^{FD}(z) = \dfrac{1}{e^{\beta z}+1}$, are, respectively, the Bose-Einstein and Fermi-Dirac statistics. Moreover, if $\psi(z)$ can be written of the form $P(z)/Q(z)$, the evaluation of the residue at $z_e$ can be done with:

$$Res(\psi, z_e) = \frac{P(z_e)}{\left.\dfrac{\partial Q}{\partial z}\right|_{z=z_e}} . \tag{91}$$

## 2. Static approximation and with a QFT diquark propagator.

One begins with the expression extracted from the first term, Eq. (38):

$$\pi^{(1)} = \frac{1}{\beta}\sum_n S_D(i\omega_n, \vec{p}) S_q^C(i\omega_n - i\nu_m, \vec{p}-\vec{k})$$

$$= \frac{1}{\beta}\sum_n \frac{\gamma_0(i\omega_n - i\nu_m - \mu_q) - \vec{\gamma}\cdot(\vec{p}-\vec{k}) + m_q I_{d4}}{\left[(i\omega_n + \mu_D)^2 - E_D^2\right]\left[(i\omega_n - i\nu_m - \mu_q)^2 - E_q^2\right]} . \tag{92}$$

The poles are:

$$\begin{cases} i\omega_n = -\mu_D + E_D \\ i\omega_n = -\mu_D - E_D \\ i\omega_n = \mu_q + E_q + i\nu_m \\ i\omega_n = \mu_q - E_q + i\nu_m \end{cases} . \tag{93}$$

The Eq. (89) allows obtaining:

$$\pi^{(1)} = \left[\gamma_0(-\lambda + E_D) - \vec{\gamma}\cdot(\vec{p}-\vec{k}) + m_q I_{d4}\right]\frac{-f^{BE}(-\mu_D + E_D)}{2E_D\left[(\lambda - E_D)^2 - E_q^2\right]}$$

$$+ \left[\gamma_0(-\lambda - E_D) - \vec{\gamma}\cdot(\vec{p}-\vec{k}) + m_q I_{d4}\right]\frac{f^{BE}(-\mu_D - E_D)}{2E_D\left[(\lambda + E_D)^2 - E_q^2\right]}$$

$$+ \left[\gamma_0(E_q) - \vec{\gamma}\cdot(\vec{p}-\vec{k}) + m_q I_{d4}\right]\frac{-f^{BE}(\mu_q + E_q + i\nu_m)}{2E_q\left[(\lambda + E_q)^2 - E_D^2\right]} \quad , \quad (94)$$

$$+ \left[\gamma_0(-E_q) - \vec{\gamma}\cdot(\vec{p}-\vec{k}) + m_q I_{d4}\right]\frac{f^{BE}(\mu_q - E_q + i\nu_m)}{2E_q\left[(\lambda - E_q)^2 - E_D^2\right]}$$

where $\lambda = i\nu_m + \mu_q + \mu_D$, $E_D^2 = \vec{p}^2 + m_D^2$ and $E_q^2 = (\vec{p}-\vec{k})^2 + m_q^2$.

The Matsubara frequency $\nu_m = (2m+1)\pi/\beta$ is fermonic, with $m \in \mathbb{Z}$. Consequently, $-f^{BE}(\mu_q + E_q + i\nu_m) = f^{FD}(\mu_q + E_q)$ and $f^{BE}(\mu_q - E_q + i\nu_m) = -f^{FD}(\mu_q - E_q)$. In addition, with the relations $f^{FD}(-z) = 1 - f^{FD}(z)$ and $-f^{BE}(-z) = 1 + f^{BE}(z)$, Eq. (94) becomes:

$$\pi^{(1)} = \left[\gamma_0(-\lambda + E_D) - \vec{\gamma}\cdot(\vec{p}-\vec{k}) + m_q I_{d4}\right]\frac{1 + f^{BE}(\mu_D - E_D)}{2E_D\left[(\lambda - E_D)^2 - E_q^2\right]}$$

$$+ \left[\gamma_0(-\lambda - E_D) - \vec{\gamma}\cdot(\vec{p}-\vec{k}) + m_q I_{d4}\right]\frac{-1 - f^{BE}(\mu_D + E_D)}{2E_D\left[(\lambda + E_D)^2 - E_q^2\right]}$$

$$+ \left[\gamma_0(E_q) - \vec{\gamma}\cdot(\vec{p}-\vec{k}) + m_q I_{d4}\right]\frac{1 - f^{FD}(-\mu_q - E_q)}{2E_q\left[(\lambda + E_q)^2 - E_D^2\right]} \quad . \quad (95)$$

$$+ \left[\gamma_0(-E_q) - \vec{\gamma}\cdot(\vec{p}-\vec{k}) + m_q I_{d4}\right]\frac{-1 + f^{FD}(-\mu_q + E_q)}{2E_q\left[(\lambda - E_q)^2 - E_D^2\right]}$$

One has the two following properties:

$$\frac{1}{2E_D\left[(\lambda - E_D)^2 - E_q^2\right]} - \frac{1}{2E_D\left[(\lambda + E_D)^2 - E_q^2\right]}$$
$$+ \frac{1}{2E_q\left[(\lambda + E_q)^2 - E_D^2\right]} - \frac{1}{2E_q\left[(\lambda - E_q)^2 - E_D^2\right]} = 0 \quad (96)$$

and

$$\frac{E_D - \lambda}{2E_D\left[(\lambda - E_D)^2 - E_q^2\right]} - \frac{-E_D - \lambda}{2E_D\left[(\lambda + E_D)^2 - E_q^2\right]}$$
$$+ \frac{E_q}{2E_q\left[(\lambda + E_q)^2 - E_D^2\right]} - \frac{-E_q}{2E_q\left[(\lambda - E_q)^2 - E_D^2\right]} = 0 \quad , \quad (97)$$

which allow removing the $\pm 1$ visible in front on the FD/BE distributions in Eq. (95).

Then, one focuses on the second term, Eq. (41), via

$$\pi^{(2)} = \frac{1}{\beta} \sum_n S_q(i\omega_n, \vec{p}) S_D^C(i\omega_n - i\nu_m, \vec{p} - \vec{k})$$

$$= \frac{1}{\beta} \sum_n \frac{\gamma_0(i\omega_n + \mu_q) - \vec{\gamma} \cdot \vec{p} + m_q I_{d4}}{\left[(i\omega_n + \mu_q)^2 - E_q^2\right]\left[(i\omega_n - i\nu_m - \mu_D)^2 - E_D^2\right]} \quad . \tag{98}$$

The poles are:

$$\begin{cases} i\omega_n = -\mu_q + E_q \\ i\omega_n = -\mu_q - E_q \\ i\omega_n = \mu_D + E_D + i\nu_m \\ i\omega_n = \mu_D - E_D + i\nu_m \end{cases} \quad . \tag{99}$$

The use of Eq. (90) gives:

$$\pi^{(2)} = \left[\gamma_0(E_q) - \vec{\gamma} \cdot \vec{p} + m_q I_{d4}\right] \frac{f^{FD}(-\mu_q + E_q)}{2E_q\left[(\lambda - E_q)^2 - E_D^2\right]}$$

$$+ \left[\gamma_0(-E_q) - \vec{\gamma} \cdot \vec{p} + m_q I_{d4}\right] \frac{-f^{FD}(-\mu_q - E_q)}{2E_q\left[(\lambda + E_q)^2 - E_D^2\right]}$$

$$+ \left[\gamma_0(\lambda + E_D) - \vec{\gamma} \cdot \vec{p} + m_q I_{d4}\right] \frac{f^{FD}(\mu_D + E_D + i\nu_m)}{2E_D\left[(\lambda + E_D)^2 - E_q^2\right]}$$

$$+ \left[\gamma_0(\lambda - E_D) - \vec{\gamma} \cdot \vec{p} + m_q I_{d4}\right] \frac{-f^{FD}(\mu_D - E_D + i\nu_m)}{2E_D\left[(\lambda - E_D)^2 - E_q^2\right]} \quad , \tag{100}$$

with $E_D^2 = (\vec{p} - \vec{k})^2 + m_D^2$ and $E_q^2 = \vec{p}^2 + m_q^2$. Furthermore, $\nu_m = (2m+1)\pi/\beta$ stays fermonic. Therefore, $f^{FD}(\mu_D + E_D + i\nu_m) = -f^{BE}(\mu_D + E_D)$ and $-f^{FD}(\mu_D - E_D + i\nu_m) = f^{BE}(\mu_D - E_D)$.

One rewrites both $\pi^{(1)}$ and $\pi^{(2)}$:

$$\pi^{(1)} = \left[\gamma_0(-E_q) - \vec{\gamma} \cdot (\vec{p} - \vec{k}) + m_q I_{d4}\right] \frac{f^{FD}(-\mu_q + E_q)}{2E_q\left[(\lambda - E_q)^2 - E_D^2\right]}$$

$$+ \left[\gamma_0(E_q) - \vec{\gamma} \cdot (\vec{p} - \vec{k}) + m_q I_{d4}\right] \frac{-f^{FD}(-\mu_q - E_q)}{2E_q\left[(\lambda + E_q)^2 - E_D^2\right]}$$

$$+ \left[\gamma_0(-\lambda - E_D) - \vec{\gamma} \cdot (\vec{p} - \vec{k}) + m_q I_{d4}\right] \frac{-f^{BE}(\mu_D + E_D)}{2E_D\left[(\lambda + E_D)^2 - E_q^2\right]}$$

$$+ \left[\gamma_0(-\lambda + E_D) - \vec{\gamma} \cdot (\vec{p} - \vec{k}) + m_q I_{d4}\right] \frac{f^{BE}(\mu_D - E_D)}{2E_D\left[(\lambda - E_D)^2 - E_q^2\right]} \quad , \tag{101}$$

with $E_D^2 = \vec{p}^2 + m_D^2$ and $E_q^2 = (\vec{p} - \vec{k})^2 + m_q^2$, and

$$\pi^{(2)} = \left[\gamma_0(E_q) - \vec{\gamma}\cdot\vec{p} + m_q I_{d4}\right]\frac{f^{FD}(-\mu_q + E_q)}{2E_q\left[(\lambda - E_q)^2 - E_D^2\right]}$$

$$+ \left[\gamma_0(-E_q) - \vec{\gamma}\cdot\vec{p} + m_q I_{d4}\right]\frac{-f^{FD}(-\mu_q - E_q)}{2E_q\left[(\lambda + E_q)^2 - E_D^2\right]}$$

$$+ \left[\gamma_0(\lambda + E_D) - \vec{\gamma}\cdot\vec{p} + m_q I_{d4}\right]\frac{-f^{BE}(\mu_D + E_D)}{2E_D\left[(\lambda + E_D)^2 - E_q^2\right]}$$

$$+ \left[\gamma_0(\lambda - E_D) - \vec{\gamma}\cdot\vec{p} + m_q I_{d4}\right]\frac{f^{BE}(\mu_D - E_D)}{2E_D\left[(\lambda - E_D)^2 - E_q^2\right]} \quad , \tag{102}$$

with $E_D^2 = (\vec{p}-\vec{k})^2 + m_D^2$ and $E_q^2 = \vec{p}^2 + m_q^2$.

As in Ref. [86], the substitution $\vec{p} \to \vec{k} - \vec{p}$ is applied in the first and second lines of Eq. (101), and in the third and fourth lines of Eq. (102). As a consequence, the definition of $E_D$ becomes the same in the first line of $\pi^{(1)}$ and $\pi^{(2)}$, and so on for $E_q$... In the writing of $(\pi^{(1)} + \pi^{(2)})/2$, the $\gamma_0$ and $\vec{\gamma}$ terms disappear. Only the $m_q I_{d4}$ terms survive. This leads to the expression of the baryon loop function:

$$\Pi_B = \int \frac{d^3 p}{(2\pi)^3} \sum_{j=1}^{4} F_j \quad , \tag{103}$$

where

$$F_1 = m_q I_{d4} \frac{f^{FD}(-\mu_q + E_q)}{2E_q\left[(\lambda - E_q)^2 - E_D^2\right]}, \tag{104}$$

$$F_2 = m_q I_{d4} \frac{-f^{FD}(-\mu_q - E_q)}{2E_q\left[(\lambda + E_q)^2 - E_D^2\right]}, \tag{105}$$

with $E_D^2 = (\vec{p}-\vec{k})^2 + m_D^2$ and $E_q^2 = \vec{p}^2 + m_q^2$,

$$F_3 = m_q I_{d4} \frac{-f^{BE}(\mu_D + E_D)}{2E_D\left[(\lambda + E_D)^2 - E_q^2\right]}, \tag{106}$$

$$F_4 = m_q I_{d4} \frac{f^{BE}(\mu_D - E_D)}{2E_D\left[(\lambda - E_D)^2 - E_q^2\right]}, \tag{107}$$

with $E_D^2 = \vec{p}^2 + m_D^2$ and $E_q^2 = (\vec{p}-\vec{k})^2 + m_q^2$.

### 3. Without the static approximation and with a QFT diquark propagator.

This configuration leads to perform the calculation of:

$$\pi^{(1)} = \frac{1}{\beta}\sum_n S_D(i\omega_n, \vec{p}) i\gamma_5 S_{qe}(2i\omega_n - i\nu_m, 2\vec{p}-\vec{k}) i\gamma_5 S_q^C(i\omega_n - i\nu_m, \vec{p}-\vec{k}). \tag{108}$$

Its poles are visible in Eq. (93) with, in addition,

$$\begin{cases} i\omega_n = (-\mu_{qe} + E_{qe} + i\nu_m)/2 \\ i\omega_n = (-\mu_{qe} - E_{qe} + i\nu_m)/2 \end{cases}. \tag{109}$$

One has also to treat:

$$\pi^{(2)} = \frac{1}{\beta}\sum_n S_q(i\omega_n, \vec{p})i\gamma_5 S_{qe}^C(2i\omega_n - i\nu_m, 2\vec{p} - \vec{k})i\gamma_5 S_D^C(i\omega_n - i\nu_m, \vec{p} - \vec{k}), \tag{110}$$

whose poles are in Eq. (99) and

$$\begin{cases} i\omega_n = (\mu_{qe} + E_{qe} + i\nu_m)/2 \\ i\omega_n = (\mu_{qe} - E_{qe} + i\nu_m)/2 \end{cases}. \tag{111}$$

The methodology is very similar in comparison to the previous case B2. More precisely, the Eqs. (89) and (90) are firstly employed to perform the Matsubara summation. After this stage, $-1$ appears in front of each FD/BE statistics of $\pi^{(1)}$, which are then simplified. Changes of variable are performed: they allow merging $\pi^{(1)}$ and $\pi^{(2)}$. As with the static approximation, only $I_{d4}$ survive during this stage. Finally, six $F_j$ are obtained, written as:

$$F_1 = I_{d4} \frac{-\left[E_q(\lambda - \Delta_\mu - 2E_q) + m_q m_{qe} + \vec{p}\cdot(2\vec{p} - \vec{k})\right]f^{FD}(-\mu_q + E_q)}{2E_q\left[(\lambda - \Delta_\mu - 2E_q)^2 - E_{qe}^2\right]\left[(\lambda - E_q)^2 - E_D^2\right]}, \tag{112}$$

$$F_2 = I_{d4} \frac{\left[-E_q(\lambda - \Delta_\mu + 2E_q) + m_q m_{qe} + \vec{p}\cdot(2\vec{p} - \vec{k})\right]f^{FD}(-\mu_q - E_q)}{2E_q\left[(\lambda - \Delta_\mu + 2E_q)^2 - E_{qe}^2\right]\left[(\lambda + E_q)^2 - E_D^2\right]}, \tag{113}$$

with $E_D^2 = (\vec{p} - \vec{k})^2 + m_D^2$, $E_{qe}^2 = (2\vec{p} - \vec{k})^2 + m_{qe}^2$ and $E_q^2 = \vec{p}^2 + m_q^2$,

$$F_3 = I_{d4} \frac{-\left[-(\lambda + E_D)(\lambda + \Delta_\mu + 2E_D) + m_q m_{qe} + (\vec{p} - \vec{k})\cdot(2\vec{p} - \vec{k})\right]\left[-f^{BE}(\mu_D + E_D)\right]}{2E_D\left[(\lambda + E_D)^2 - E_q^2\right]\left[(\lambda + \Delta_\mu + 2E_D)^2 - E_{qe}^2\right]}, \tag{114}$$

$$F_4 = I_{d4} \frac{\left[-(\lambda - E_D)(\lambda + \Delta_\mu - 2E_D) + m_q m_{qe} + (\vec{p} - \vec{k})\cdot(2\vec{p} - \vec{k})\right]\left[-f^{BE}(\mu_D - E_D)\right]}{2E_D\left[(\lambda - E_D)^2 - E_q^2\right]\left[(\lambda + \Delta_\mu - 2E_D)^2 - E_{qe}^2\right]}, \tag{115}$$

with $E_D^2 = \vec{p}^2 + m_D^2$, $E_{qe}^2 = (2\vec{p} - \vec{k})^2 + m_{qe}^2$ and $E_q^2 = (\vec{p} - \vec{k})^2 + m_q^2$,

$$F_5 = I_{d4} \frac{-2\left[-(\lambda - \Delta_\mu + E_{qe})E_{qe} + 2m_q m_{qe} + 2\vec{p}\cdot(2\vec{p} - \vec{k})\right]f^{FD}(\mu_{qe} + E_{qe})}{E_{qe}\left[(\lambda - \Delta_\mu + E_{qe})^2 - (2E_q)^2\right]\left[(\lambda + \Delta_\mu - E_{qe})^2 - (2E_D)^2\right]}, \tag{116}$$

$$F_6 = I_{d4} \frac{2\left[(\lambda - \Delta_\mu - E_{qe})E_{qe} + 2m_q m_{qe} + 2\vec{p}\cdot(2\vec{p} - \vec{k})\right]f^{FD}(\mu_{qe} - E_{qe})}{E_{qe}\left[(\lambda - \Delta_\mu - E_{qe})^2 - (2E_q)^2\right]\left[(\lambda + \Delta_\mu + E_{qe})^2 - (2E_D)^2\right]}, \tag{117}$$

with $E_D^2 = \left(\vec{p} + \frac{\vec{k}}{2}\right)^2 + m_D^2$, $E_{qe}^2 = (2\vec{p})^2 + m_{qe}^2$ and $E_q^2 = \left(\vec{p} - \frac{\vec{k}}{2}\right)^2 + m_q^2$.

The shorthand notation $\Delta_\mu = \mu_D - \mu_q - \mu_{qe}$ is used in these relations. For a given baryon, if $D$ and $q$ are constant via the quark exchange, $\Delta_\mu = 0$. This is the case for a one state baryon, like the nucleon described via its scalar flavor component.

## 4. Static approximation and with an NJL diquark propagator.

The use of NJL diquark propagator leads to consider the two following expressions:

$$\pi^{(1)} = \frac{1}{\beta}\sum_n S_D(i\omega_n, \vec{p}) S_q^C(i\omega_n - i\nu_m, \vec{p}-\vec{k})$$

$$= \frac{1}{\beta}\sum_n 4G_{DIQ} \frac{\gamma_0(i\omega_n - i\nu_m - \mu_q) - \vec{\gamma}\cdot(\vec{p}-\vec{k}) + m_q I_{d4}}{\left[1 - 2G_{DIQ}\Pi_D(i\omega_n + \mu_D, \vec{p})\right]\left[(i\omega_n - i\nu_m - \mu_q)^2 - E_q^2\right]}, \quad (118)$$

$$\pi^{(2)} = \frac{1}{\beta}\sum_n S_q(i\omega_n, \vec{p}) S_D^C(i\omega_n - i\nu_m, \vec{p}-\vec{k})$$

$$= \frac{1}{\beta}\sum_n 4G_{DIQ} \frac{\gamma_0(i\omega_n + \mu_q) - \vec{\gamma}\cdot\vec{p} + m_q I_{d4}}{\left[(i\omega_n + \mu_q)^2 - E_q^2\right]\left[1 - 2G_{DIQ}\Pi_D(i\omega_n - i\nu_m - \mu_D, \vec{p}-\vec{k})\right]}, \quad (119)$$

where $\Pi_D$ is the diquark irreducible polarization function. The poles of Eq. (118) are given by Eq. (93), and the ones of Eq. (119) by Eq. (99). The Matsubara summation and the changes of variable are feasible in the same way as previously. Nevertheless, the simplifications observed with the QFT diquark propagator are not possible in these calculations. As a consequence, the $\pi^{(1)}$ results labeled as $F_j$ hereafter and the ones of $\pi^{(2)}$, i.e. the $F_j'$, cannot be merged. This wants to say that the eight found relations do not consider only $I_{d4}$ terms, but also $\gamma_0$ and $\vec{\gamma}$ terms,

$$F_1 = 4G_{DIQ}\left[\gamma_0(-E_q) + \vec{\gamma}\cdot\vec{p} + m_q I_{d4}\right] \frac{-f^{FD}(\mu_q - E_q)}{2E_q\left[1 - 2G_{DIQ}\Pi_D(-E_q + \lambda, \vec{p}-\vec{k})\right]}, \quad (120)$$

$$F_1' = 4G_{DIQ}\left[\gamma_0(E_q) - \vec{\gamma}\cdot\vec{p} + m_q I_{d4}\right] \frac{f^{FD}(-\mu_q + E_q)}{2E_q\left[1 - 2G_{DIQ}\Pi_D(E_q - \lambda, \vec{p}-\vec{k})\right]}, \quad (121)$$

$$F_2 = 4G_{DIQ}\left[\gamma_0(E_q) + \vec{\gamma}\cdot\vec{p} + m_q I_{d4}\right] \frac{f^{FD}(\mu_q + E_q)}{2E_q\left[1 - 2G_{DIQ}\Pi_D(E_q + \lambda, \vec{p}-\vec{k})\right]}, \quad (122)$$

$$F_2' = 4G_{DIQ}\left[\gamma_0(-E_q) - \vec{\gamma}\cdot\vec{p} + m_q I_{d4}\right] \frac{-f^{FD}(-\mu_q - E_q)}{2E_q\left[1 - 2G_{DIQ}\Pi_D(-E_q - \lambda, \vec{p}-\vec{k})\right]}, \quad (123)$$

with $E_q^2 = \vec{p}^2 + m_q^2$, and

$$F_3 = 4G_{DIQ}\left[\gamma_0(-E_D - \lambda) - \vec{\gamma}\cdot(\vec{p}-\vec{k}) + m_q I_{d4}\right] \frac{-f^{BE}(-\mu_D - E_D)}{-2\left[(\lambda + E_D)^2 - E_q^2\right] G_{DIQ} \left.\frac{\partial \Pi_D}{\partial z}\right|_{z=-E_D}}, \quad (124)$$

$$F_3' = 4G_{DIQ}\left[\gamma_0(E_D + \lambda) + \vec{\gamma}\cdot(\vec{p}-\vec{k}) + m_q I_{d4}\right] \frac{-f^{BE}(\mu_D + E_D)}{-2\left[(\lambda + E_D)^2 - E_q^2\right] G_{DIQ} \left.\frac{\partial \Pi_D}{\partial z}\right|_{z=E_D}}, \quad (125)$$

$$F_4 = 4G_{DIQ}\left[\gamma_0(E_D - \lambda) - \vec{\gamma}\cdot(\vec{p}-\vec{k}) + m_q I_{d4}\right] \frac{-f^{BE}(-\mu_D + E_D)}{-2\left[(-\lambda + E_D)^2 - E_q^2\right] G_{DIQ} \left.\frac{\partial \Pi_D}{\partial z}\right|_{z=E_D}}, \quad (126)$$

$$F_4' = 4G_{DIQ}\left[\gamma_0(-E_D+\lambda)+\vec{\gamma}\cdot(\vec{p}-\vec{k})+m_q I_{d4}\right]\frac{-f^{BE}(\mu_D-E_D)}{-2\left[(\lambda-E_D)^2-E_q^2\right]G_{DIQ}\left.\frac{\partial\Pi_D}{\partial z}\right|_{z=-E_D}}, \quad (127)$$

with $E_D^2 = \vec{p}^2 + m_D^2$ and $E_q^2 = (\vec{p}-\vec{k})^2 + m_q^2$. However, by symmetry, the $\vec{\gamma}$ terms give zero during the integration upon $p$.

### 5. Without the static approximation and with an NJL diquark propagator.

This last version of the baryon loop function requires evaluating:

$$\pi^{(1)} = \frac{1}{\beta}\sum_n S_D(i\omega_n,\vec{p})i\gamma_5 S_{qe}(2i\omega_n-iv_m, 2\vec{p}-\vec{k})i\gamma_5 S_q^C(i\omega_n-iv_m,\vec{p}-\vec{k}), \quad (128)$$

$$\pi^{(2)} = \frac{1}{\beta}\sum_n S_q(i\omega_n,\vec{p})i\gamma_5 S_{qe}^C(2i\omega_n-iv_m, 2\vec{p}-\vec{k})i\gamma_5 S_D^C(i\omega_n-iv_m,\vec{p}-\vec{k}), \quad (129)$$

but with

$$S_D(i\omega_n,\vec{p}) = \frac{4G_{DIQ}}{1-2G_{DIQ}\Pi_D(i\omega_n+\mu_D,\vec{p})}, \quad (130)$$

$$S_D^C(i\omega_n-iv_m,\vec{p}-\vec{k}) = \frac{4G_{DIQ}}{1-2G_{DIQ}\Pi_D(i\omega_n-iv_m-\mu_D,\vec{p}-\vec{k})}. \quad (131)$$

The poles of Eq. (128) and Eq. (129) can be found in B3. Concerning the methodology of the calculation, the remarks expressed in the previous case are still valid here. A difference is one takes $\vec{k}=\vec{0}$, in order to obtain more tractable expressions of the $F_j$ and $F_j'$, which are:

$$F_1 = \left\{\gamma_0\left[-E_q m_{qe}-(\lambda-\Delta_\mu-2E_q)m_q\right]+I_{d4}\left[E_q(\lambda-\Delta_\mu-2E_q)+m_q m_{qe}+2p^2\right]\right\}$$
$$\times 4G_{DIQ}\frac{1-f^{FD}(-\mu_q+E_q)}{2E_q\left[(\lambda-\Delta_\mu-2E_q)^2-E_{qe}^2\right]\left[1-2G_{DIQ}\Pi_D(-E_q+\lambda,\vec{p})\right]}, \quad (132)$$

$$F_1' = \left\{\gamma_0\left[E_q m_{qe}+(\lambda-\Delta_\mu-2E_q)m_q\right]+I_{d4}\left[E_q(\lambda-\Delta_\mu-2E_q)+m_q m_{qe}+2p^2\right]\right\}$$
$$\times 4G_{DIQ}\frac{-f^{FD}(-\mu_q+E_q)}{2E_q\left[(\lambda-\Delta_\mu-2E_q)^2-E_{qe}^2\right]\left[1-2G_{DIQ}\Pi_D(E_q-\lambda,\vec{p})\right]}, \quad (133)$$

$$F_2 = \left\{\gamma_0\left[E_q m_{qe}-(\lambda-\Delta_\mu+2E_q)m_q\right]+I_{d4}\left[-E_q(\lambda-\Delta_\mu+2E_q)+m_q m_{qe}+2p^2\right]\right\}$$
$$\times 4G_{DIQ}\frac{-1+f^{FD}(-\mu_q-E_q)}{2E_q\left[(\lambda-\Delta_\mu+2E_q)^2-E_{qe}^2\right]\left[1-2G_{DIQ}\Pi_D(E_q+\lambda,\vec{p})\right]}, \quad (134)$$

$$F_2' = \left\{\gamma_0\left[-E_q m_{qe}+(\lambda-\Delta_\mu+2E_q)m_q\right]+I_{d4}\left[-E_q(\lambda-\Delta_\mu+2E_q)+m_q m_{qe}+2p^2\right]\right\}$$
$$\times 4G_{DIQ}\frac{f^{FD}(-\mu_q-E_q)}{2E_q\left[(\lambda-\Delta_\mu+2E_q)^2-E_{qe}^2\right]\left[1-2G_{DIQ}\Pi_D(-E_q-\lambda,\vec{p})\right]}, \quad (135)$$

$$F_3 = \left\{\gamma_0\left[-(\lambda+E_D)m_{qe}+(\lambda+\Delta_\mu+2E_D)m_q\right]+I_{d4}\left[-(\lambda+E_D)(\lambda+\Delta_\mu+2E_D)+m_q m_{qe}+2p^2\right]\right\}$$
$$\times 4G_{DIQ}\frac{-1-f^{BE}(\mu_D+E_D)}{-2\left[(\lambda+E_D)^2-E_q^2\right]\left[(\lambda+\Delta_\mu+2E_D)^2-E_{qe}^2\right]G_{DIQ}\left.\frac{\partial\Pi_D}{\partial z}\right|_{z=-E_D}}, \quad(136)$$

$$F_3' = \left\{\gamma_0\left[(\lambda+E_D)m_{qe}-(\lambda+\Delta_\mu+2E_D)m_q\right]+I_{d4}\left[-(\lambda+E_D)(\lambda+\Delta_\mu+2E_D)+m_q m_{qe}+2p^2\right]\right\}$$
$$\times 4G_{DIQ}\frac{f^{BE}(\mu_D+E_D)}{-2\left[(\lambda+E_D)^2-E_q^2\right]\left[(\lambda+\Delta_\mu+2E_D)^2-E_{qe}^2\right]G_{DIQ}\left.\frac{\partial\Pi_D}{\partial z}\right|_{z=E_D}}, \quad(137)$$

$$F_4 = \left\{\gamma_0\left[-(\lambda-E_D)m_{qe}+(\lambda+\Delta_\mu-2E_D)m_q\right]+I_{d4}\left[-(\lambda-E_D)(\lambda+\Delta_\mu-2E_D)+m_q m_{qe}+2p^2\right]\right\}$$
$$\times 4G_{DIQ}\frac{-1-f^{BE}(\mu_D-E_D)}{-2\left[(\lambda-E_D)^2-E_q^2\right]\left[(\lambda+\Delta_\mu-2E_D)^2-E_{qe}^2\right]G_{DIQ}\left.\frac{\partial\Pi_D}{\partial z}\right|_{z=E_D}}, \quad(138)$$

$$F_4' = \left\{\gamma_0\left[(\lambda-E_D)m_{qe}-(\lambda+\Delta_\mu-2E_D)m_q\right]+I_{d4}\left[-(\lambda-E_D)(\lambda+\Delta_\mu-2E_D)+m_q m_{qe}+2p^2\right]\right\}$$
$$\times 4G_{DIQ}\frac{f^{BE}(\mu_D-E_D)}{-2\left[(\lambda-E_D)^2-E_q^2\right]\left[(\lambda+\Delta_\mu-2E_D)^2-E_{qe}^2\right]G_{DIQ}\left.\frac{\partial\Pi_D}{\partial z}\right|_{z=-E_D}}, \quad(139)$$

$$F_5 = \left\{\gamma_0\left[2E_{qe}m_q-(\lambda-\Delta_\mu+E_{qe})m_{qe}\right]+I_{d4}\left[-E_{qe}(\lambda-\Delta_\mu+E_{qe})+2m_q m_{qe}+4p^2\right]\right\}$$
$$\times 4G_{DIQ}\frac{1-f^{FD}(\mu_{qe}+E_{qe})}{2E_{qe}\left[(\lambda-\Delta_\mu+E_{qe})^2-(2E_q)^2\right]\left[1-2G_{DIQ}\Pi_D\left(\frac{\lambda+\Delta_\mu-E_{qe}}{2},\vec{p}\right)\right]}, \quad(140)$$

$$F_5' = \left\{\gamma_0\left[-2E_{qe}m_q+(\lambda-\Delta_\mu+E_{qe})m_{qe}\right]+I_{d4}\left[-E_{qe}(\lambda-\Delta_\mu+E_{qe})+2m_q m_{qe}+4p^2\right]\right\}$$
$$\times 4G_{DIQ}\frac{-f^{FD}(\mu_{qe}+E_{qe})}{2E_{qe}\left[(\lambda-\Delta_\mu+E_{qe})^2-(2E_q)^2\right]\left[1-2G_{DIQ}\Pi_D\left(\frac{-\lambda-\Delta_\mu+E_{qe}}{2},\vec{p}\right)\right]}, \quad(141)$$

$$F_6 = \left\{\gamma_0\left[-2E_{qe}m_q-(\lambda-\Delta_\mu-E_{qe})m_{qe}\right]+I_{d4}\left[E_{qe}(\lambda-\Delta_\mu-E_{qe})+2m_q m_{qe}+4p^2\right]\right\}$$
$$\times 4G_{DIQ}\frac{-1+f^{FD}(\mu_{qe}-E_{qe})}{2E_{qe}\left[(\lambda-\Delta_\mu-E_{qe})^2-(2E_q)^2\right]\left[1-2G_{DIQ}\Pi_D\left(\frac{\lambda+\Delta_\mu+E_{qe}}{2},\vec{p}\right)\right]}, \quad(142)$$

$$F_6' = \left\{\gamma_0\left[2E_{qe}m_q+(\lambda-\Delta_\mu-E_{qe})m_{qe}\right]+I_{d4}\left[E_{qe}(\lambda-\Delta_\mu-E_{qe})+2m_q m_{qe}+4p^2\right]\right\}$$
$$\times 4G_{DIQ}\frac{f^{FD}(\mu_{qe}-E_{qe})}{2E_{qe}\left[(\lambda-\Delta_\mu-E_{qe})^2-(2E_q)^2\right]\left[1-2G_{DIQ}\Pi_D\left(\frac{-\lambda-\Delta_\mu-E_{qe}}{2},\vec{p}\right)\right]}. \quad(143)$$

with $E_D^2=\vec{p}^2+m_D^2$, $E_{qe}^2=(2\vec{p})^2+m_{qe}^2$ and $E_q^2=\vec{p}^2+m_q^2$.

# APPENDIX C: AXIAL FLAVOR COMPONENT IN THE OCTET BARYONS DESCRIPTION.

## 1. Focus on the scalar-axial terms.

In the octet baryons modeling proposed in Subsec. IV.D, scalar-axial couplings appear. This is due to the $Z_{1,2}$, $Z_{1,3}$, $Z_{2,1}$ and $Z_{3,1}$ matrix elements in Eq. (60). Couplings of this kind are visible, e.g., in [107, 108]. In contrast with the scalar/scalar and axial/axial $Z_{i,j}$ matrix elements, they carry $\gamma$ matrices that cannot be simplified directly. For the proton, the determinant in Eq. (46) leads to handle terms like $Z_{2,1} \Pi_u^S Z_{1,2} \Pi_d^A$. This one translates the coupling between the $[ud]^S + u$ and $[uu]^A + d$ states. Starting from Eqs. (38) and (41), $Z_{2,1} \Pi_u^S$ is firstly written with:

$$Z_{2,1} \Pi_B^{(1)}\left(i\nu_m, \vec{k}\right) = \frac{1}{\beta} \sum_n \int \frac{d^3 p}{(2\pi)^3} S_{[ud]^S}\left(i\omega_n, \vec{p}\right) \left(-2\sqrt{2} \frac{\gamma^5 g_{ud}^S \gamma^\mu g_{uu}^A}{m_u}\right) S_u^C\left(i\omega_n - i\nu_m, \vec{p} - \vec{k}\right) \quad (144)$$

and

$$Z_{2,1} \Pi_B^{(2)}\left(i\nu_m, \vec{k}\right) = \frac{1}{\beta} \sum_n \int \frac{d^3 p}{(2\pi)^3} S_u\left(i\omega_n, \vec{p}\right) \left(-2\sqrt{2} \frac{\gamma^\mu g_{uu}^A \gamma^5 g_{ud}^S}{m_u}\right) S_{[ud]^S}^C\left(i\omega_n - i\nu_m, \vec{p} - \vec{k}\right). \quad (145)$$

The calculation is the same as in Appendix B.2, until Eqs. (101) and (102). Indeed, the presence of the $\gamma$ matrices prevents to perform the simplifications described later in B.2. Nevertheless, after some manipulations, it is still possible to obtain a form compatible with Eq. (103), but with the rewriting of Eq. (104):

$$F_1^S = \frac{f^{FD}\left(-\mu_q + E_q\right)}{2E_q\left[\left(\lambda - E_q\right)^2 - E_D^2\right]} \gamma^0 E_q \gamma^a \gamma^5, \quad (146)$$

with $a = 1, 2, 3$, and so on with the three others $F_j$, Eqs. (105) to (107). Concerning the $Z_{1,2} \Pi_d^A$ part, one obtains, in the same way:

$$F_1^A = \gamma_5 \gamma_a \gamma_0 E_q \frac{f^{FD}\left(-\mu_q + E_q\right)}{2E_q\left[\left(\lambda - E_q\right)^2 - E_D^2\right]}, \quad (147)$$

and so on with the three others $F_j$. Consequently, the $Z_{2,1} \Pi_u^S Z_{1,2} \Pi_d^A$ term may be expressed in the simplified form:

$$Z_{2,1} \Pi_u^S Z_{1,2} \Pi_d^A = \overline{\Pi}_u^S \overline{\Pi}_d^A \, 3\left(2\sqrt{2} \frac{g_{uu}^A g_{ud}^S}{m_u}\right)^2, \quad (148)$$

where $\overline{\Pi}_u^S$ and $\overline{\Pi}_d^A$ are defined Eqs. (103) to (107), but with the replacements of $m_q I_{d4}$ by $E_q I_{d4}$, $-E_q I_{d4}$, $(E_D + \lambda) I_{d4}$ and $(-E_D + \lambda) I_{d4}$ in, respectively, Eqs. (104) to (107). Such a simplified writing of $Z_{2,1} \Pi_u^S Z_{1,2} \Pi_d^A$ cannot be found when the NJL diquark propagator is employed.

## 2. Description of the $\Lambda$ and $\Sigma^0$ baryons.

When the isospin symmetry is considered, $\Lambda$ and $\Sigma^0$ are modeled via $3 \times 3$ matrices [66]. In contrast, without this symmetry, $5 \times 5$ matrices are required. With the same notations of Eq. (60), they are written for the $\Lambda$ baryon as:

$$Z = \begin{bmatrix} 0 & -\dfrac{g^S_{ds}\,g^S_{us}}{m_s} & 2\dfrac{g^S_{ds}\,g^S_{ud}}{m_d} & 0 & -\dfrac{\gamma_5 g^S_{ds}\,\gamma_\mu g^A_{us}}{m_s} \\ -\dfrac{g^S_{us}\,g^S_{ds}}{m_s} & 0 & 2\dfrac{g^S_{us}\,g^S_{ud}}{m_u} & -\dfrac{\gamma_5 g^S_{us}\,\gamma_\mu g^A_{ds}}{m_s} & 0 \\ 2\dfrac{g^S_{ud}\,g^S_{ds}}{m_d} & 2\dfrac{g^S_{ud}\,g^S_{us}}{m_u} & 0 & 2\dfrac{\gamma_5 g^S_{ud}\,\gamma_\mu g^A_{ds}}{m_d} & 2\dfrac{\gamma_5 g^S_{ud}\,\gamma_\mu g^A_{us}}{m_u} \\ 0 & -\dfrac{\gamma^\mu g^A_{ds}\,\gamma^5 g^S_{us}}{m_s} & 2\dfrac{\gamma^\mu g^A_{ds}\,\gamma^5 g^S_{ud}}{m_d} & 0 & -\dfrac{4 g^A_{ds}\,g^A_{us}}{m_s} \\ -\dfrac{\gamma^\mu g^A_{us}\,\gamma^5 g^S_{ds}}{m_s} & 0 & 2\dfrac{\gamma^\mu g^A_{us}\,\gamma^5 g^S_{ud}}{m_u} & -\dfrac{4 g^A_{us}\,g^A_{ds}}{m_s} & 0 \end{bmatrix} \quad (149)$$

and

$$\Pi_B = \mathrm{diag}\left([ds]^S + u,\ [us]^S + d,\ [ud]^S + s,\ [ds]^A + u,\ [us]^A + d\right). \tag{150}$$

With $\Sigma^0$, one has:

$$Z = \begin{bmatrix} 0 & -2\dfrac{g^S_{ds}\,g^S_{us}}{m_s} & 0 & -\dfrac{\gamma_5 g^S_{ds}\,\gamma_\mu g^A_{us}}{m_s} & -2\dfrac{\gamma_5 g^S_{ds}\,\gamma_\mu g^A_{ud}}{m_d} \\ -\dfrac{g^S_{us}\,g^S_{ds}}{m_s} & 0 & -\dfrac{\gamma_5 g^S_{us}\,\gamma_\mu g^A_{ds}}{m_s} & 0 & -2\dfrac{\gamma_5 g^S_{us}\,\gamma_\mu g^A_{ud}}{m_u} \\ 0 & -\dfrac{\gamma^\mu g^A_{ds}\,\gamma^5 g^S_{us}}{m_s} & 0 & -\dfrac{4 g^A_{ds}\,g^A_{us}}{m_s} & -2\dfrac{4 g^A_{ds}\,g^A_{ud}}{m_d} \\ -\dfrac{\gamma^\mu g^A_{us}\,\gamma^5 g^S_{ds}}{m_s} & 0 & -\dfrac{4 g^A_{us}\,g^A_{ds}}{m_s} & 0 & -2\dfrac{4 g^A_{us}\,g^A_{ud}}{m_u} \\ -2\dfrac{\gamma^\mu g^A_{ud}\,\gamma^5 g^S_{ds}}{m_d} & -2\dfrac{\gamma^\mu g^A_{ud}\,\gamma^5 g^S_{us}}{m_u} & -2\dfrac{4 g^A_{ud}\,g^A_{ds}}{m_d} & -2\dfrac{4 g^A_{ud}\,g^A_{us}}{m_u} & 0 \end{bmatrix} \quad (151)$$

and

$$\Pi_B = \mathrm{diag}\left([ds]^S + u,\ [us]^S + d,\ [ds]^A + u,\ [us]^A + d,\ [ud]^A + s\right). \tag{152}$$

### APPENDIX D: INTEROPERABILITY OF THE EVOLUTIONS MENTIONED IN THIS PAPER.

The various evolutions of the baryons modeling cannot always be applied together. Consequently, the symmetric interoperability matrix TABLE XII proposes to sum up all the possible configurations. The gray squares materialize its main diagonal. The ✓ symbol translates that the two evolutions can work together. In contrast, the ✗ symbol indicates the opposite, because this would lead to a too great complexity of the associated calculations.

**TABLE XII.** Interoperability matrix.

|                                    | A. Complex numbers | B. Momentum dependence | C. Without static approximation | D. Axial flavor component | E. NJL diquark propagator | F. Color superconductivity |
|------------------------------------|:---:|:---:|:---:|:---:|:---:|:---:|
| A. Complex numbers                 |     | ✓   | ✓   | ✓   | ✓   | ✓   |
| B. Momentum dependence             | ✓   |     | ✓   | ✓   | ✓   | ✗   |
| C. Without static approximation    | ✓   | ✓   |     | ✗   | ✓   | ✓   |
| D. Axial flavor component          | ✓   | ✓   | ✗   |     | ✗   | ✗   |
| E. NJL diquark propagator          | ✓   | ✓   | ✓   | ✗   |     | ✓   |
| F. Color superconductivity         | ✓   | ✗   | ✓   | ✗   | ✓   |     |

In the color-superconducting regime, the NJL diquark propagator can be used to model baryons. However, this evolution is more accurate when momentum dependence corrections are applied. Therefore, the calculations have not been performed in this E-F configuration.

## APPENDIX E: COMPARED PERFORMANCES OF EACH EVOLUTION.

The TABLE XIII shows a comparison of the ability of each treated evolution to reproduce the baryons experimental masses at zero temperature and zero baryonic density. The color superconductivity is not included in these calculations, since this phenomenon does not intervene at zero density. In this table, (p), (no p), (S.A.), (no S.A.) stand for, respectively, momentum dependence corrections, no p-dependence corrections, static approximation, no static approximation.

**TABLE XIII.** Percent errors between the NISO parameter set results and the experimental values, in %.

|                                       | Proton | Neutron | $\Lambda$ | $\Sigma^+$ | $\Sigma^0$ | $\Sigma^-$ | $\Xi^0$ | $\Xi^-$ |
|---------------------------------------|:------:|:-------:|:---------:|:----------:|:----------:|:----------:|:-------:|:-------:|
| A. Complex numbers                    | -5.5   | -5.8    | -0.95     | -3.4       | -3.5       | -3.7       | 1.3     | 1.0     |
| B. Momentum dependence                | 0.69   | 0.46    | 2.2       | -2.1       | -2.2       | -2.4       | 1.9     | 1.6     |
| C. Without static approximation (no p)| -5.0   | -4.8    | ✗         | -4.2       | ✗          | -4.4       | 1.3     | 0.92    |
| C. Without static approximation (p)   | -0.79  | -0.57   | ✗         | -3.2       | ✗          | -3.4       | 1.7     | 1.3     |
| D. Axial flavor component (no p)      | -5.1   | -5.6    | 0.46      | -3.7       | -3.9       | -4.2       | -0.53   | -0.83   |
| D. Axial flavor component (p)         | 7.2    | 7.2     | 4.7       | -1.4       | -1.4       | -1.7       | 1.9     | 1.6     |
| E. NJL diquark propagator (S.A.)      | -0.56  | -0.98   | 4.5       | -1.3       | -1.4       | -1.6       | 2.7     | 2.4     |
| E. NJL diquark propagator (no S.A.)   | -1.1   | -0.90   | ✗         | -7.6       | ✗          | -7.8       | 2.0     | 1.6     |

| | $\Delta^{++}$ $\Delta^{+}$ $\Delta^{0}$ $\Delta^{-}$ | $\Sigma^{*+}$ | $\Sigma^{*0}$ | $\Sigma^{*-}$ | $\Xi^{*0}$ | $\Xi^{*-}$ | $\Omega^{-}$ |
|---|---|---|---|---|---|---|---|
| A. Complex numbers | 0.13 to 0.35 | -3.2 | -3.2 | -3.4 | -3.5 | -3.6 | 0.14 |
| B. Momentum dependence | 3.3 to 3.8 | 0.95 | 1.0 | 0.94 | 0.0 | -0.07 | 1.6 |
| C. Without static approximation (no p) | 0.77 to 1.2 | ✗ | ✗ | ✗ | ✗ | ✗ | -0.46 |
| C. Without static approximation (p) | 1.5 to 1.9 | ✗ | ✗ | ✗ | ✗ | ✗ | 0.12 |
| E. NJL diquark propagator (S.A.) | -5.5 to -4.9 | -0.26 | -0.10 | -0.11 | 2.3 | 2.2 | 2.1 |
| E. NJL diquark propagator (no S.A.) | 2.7 to 3.4 | ✗ | ✗ | ✗ | ✗ | ✗ | 2.0 |